\documentclass[aps,pop,twocolumn,showpacs,superscriptaddress,nofootinbib,10pt]{revtex4-1}   

\usepackage[T1]{fontenc}
\usepackage{bm}
\usepackage{multirow}
\usepackage[english]{babel}  
\usepackage{amsmath}
\usepackage{amsfonts}
\usepackage{amssymb} 
\usepackage{textcomp}
\usepackage{graphicx}
\usepackage{epstopdf}
\usepackage{subcaption}
\usepackage{txfonts}
\usepackage{color}
\usepackage{xfrac}
\usepackage{tabularx}
\usepackage[colorlinks]{hyperref}
\hypersetup{colorlinks=true,citecolor=blue,linkcolor=blue}
\usepackage{floatrow}
\usepackage{multirow}
\usepackage{xcolor}

\definecolor{verdino}{rgb}{0, 0.5, 0}
\definecolor{azzu}{rgb}{0.3010, 0.7450, 0.9330}
\definecolor{aran}{RGB}{255,165,0}

\makeatletter
\renewcommand\@author{\ifx\AB@affillist\AB@empty\AB@author\else
      \ifnum\value{affiliation}>\value{Maxaffil}\def\rlap##1{##1}%
    \AB@authlist\\[\affilsep]\behalf\\\AB@affillist
    \else  \AB@authors\fi\fi}
\makeatother

\begin{document}

\title{\bf Study of the impact of fast ions on core turbulence at rational surfaces via global gyrokinetic simulations}

\author{D. Brioschi}
\thanks{Electronic address: \texttt{davide.brioschi@ipp.mpg.de}}
\affiliation{\small\emph{Max-Planck-Institut für Plasmaphysik, 85748 Garching, Germany}}
\author{A. Di Siena}
\affiliation{\small\emph{Max-Planck-Institut für Plasmaphysik, 85748 Garching, Germany}}
\author{R. Bilato}
\affiliation{\small\emph{Max-Planck-Institut für Plasmaphysik, 85748 Garching, Germany}}
\author{A. Bottino}
\affiliation{\small\emph{Max-Planck-Institut für Plasmaphysik, 85748 Garching, Germany}}
\author{\\ T. Hayward-Schneider}
\affiliation{\small\emph{Max-Planck-Institut für Plasmaphysik, 85748 Garching, Germany}}
\author{A. Mishchenko}
\affiliation{\small\emph{Max-Planck-Institut für Plasmaphysik, 17491 Greifswald, Germany}}
\author{E. Poli}
\affiliation{\small\emph{Max-Planck-Institut für Plasmaphysik, 85748 Garching, Germany}}
\author{A. Zocco}
\affiliation{\small\emph{Max-Planck-Institut für Plasmaphysik, 17491 Greifswald, Germany}}
\author{F. Jenko}
\affiliation{\small\emph{Max-Planck-Institut für Plasmaphysik, 85748 Garching, Germany}}
\affiliation{\small\emph{Institute for Fusion Studies, The University of Texas
at Austin, Austin, TX, USA}}

\date{\today}

\begin{abstract}

%In this work, enhancement of zonal modes generation through fast ions dilution is studied via global nonlinear gyrokinetic simulations. The effect fast particles have on the development of shearing structures from turbulence self-interaction is analyzed. Our study takes into account the competition between this mechanism and other fast ions effects, i.e. thermal profiles dilution and quasi-resonant interaction. In the context of this analysis, we detail the process of shearing layers generation via turbulent eddy self-interaction at rational surfaces. We find the fast ions-induced reduction of destabilization threshold for these zonal modes to be a very efficient way to suppress turbulence. Indeed, when the suprathermal particles contribution is maximized, this mechanism leads to the formation of regions where turbulent transport is reduced by 90\% of its original value.  

In this work, the interplay between fast ions and safety factor rational surfaces is studied in a turbulent plasma via global nonlinear gyrokinetic simulations. Initially, the fast particles-induced enhancement of shearing structures from turbulence self-interaction is analyzed. Our study takes into account the competition between this mechanism and other fast ions effects, i.e. thermal profiles dilution and quasi-resonant interaction. We find the fast ions-induced reduction of destabilization threshold for the zonal modes to be a very efficient way to suppress turbulence. Indeed, it leads to the formation of regions where turbulent transport is reduced by 90\% of its original value. Furthermore, an $n=m=1$ fishbone is driven unstable inside the plasma and its interaction with turbulence is studied. We find the beat-driven zonal structure generate by this mode to further reduce turbulence when its presence does not drastically flatten the thermal profiles. 

\end{abstract}

\maketitle

\section{Introduction}\label{intro}
\vspace{-0.5pc}

\begin{figure*}[ht!]
\noindent\begin{subfigure}[t]{0.33\textwidth}
\begin{center}
\includegraphics[width=\textwidth]{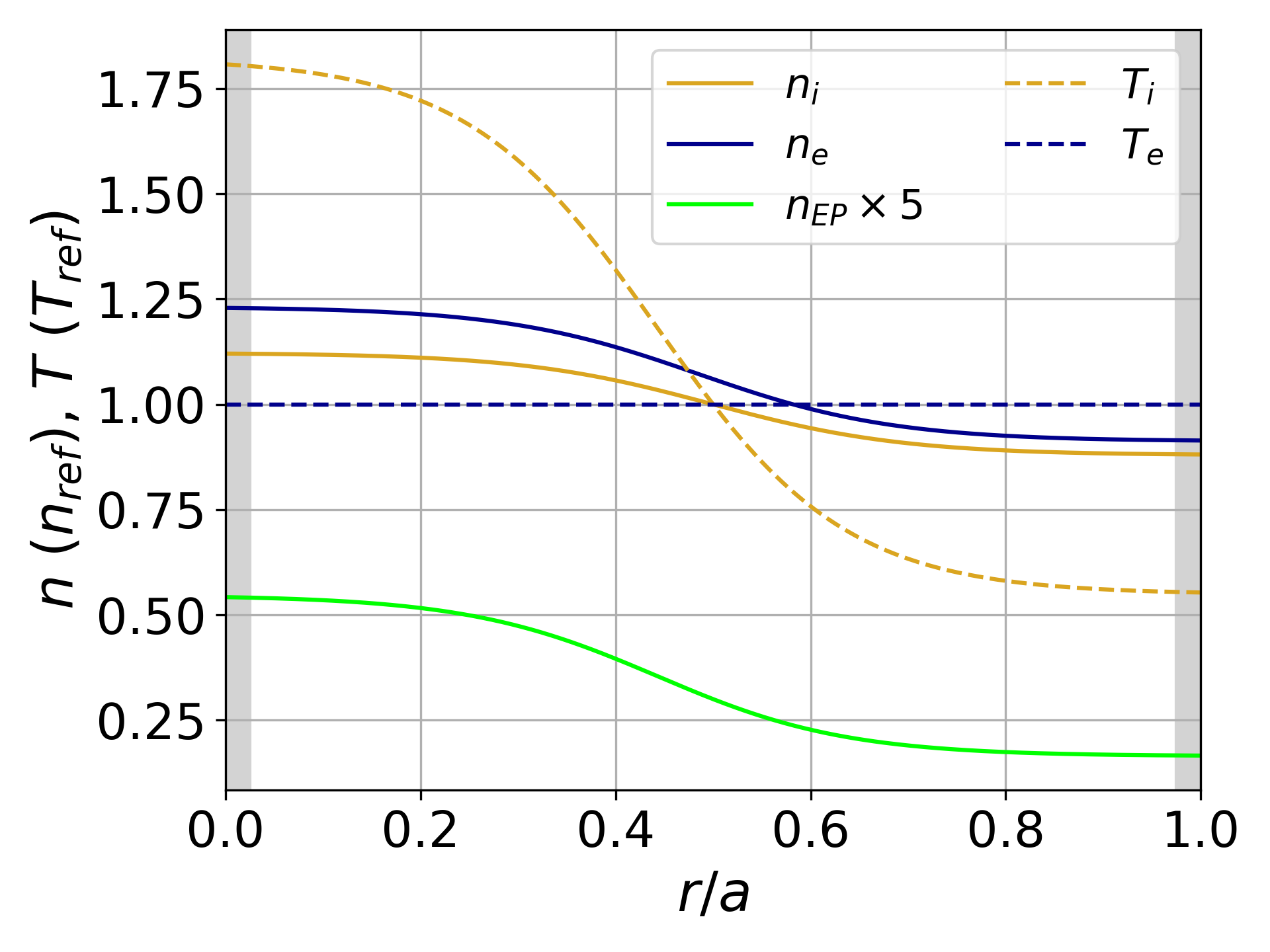}
\caption{}
\end{center}
\end{subfigure}
\noindent\begin{subfigure}[t]{0.33\textwidth}
\begin{center}
\includegraphics[width=\textwidth]{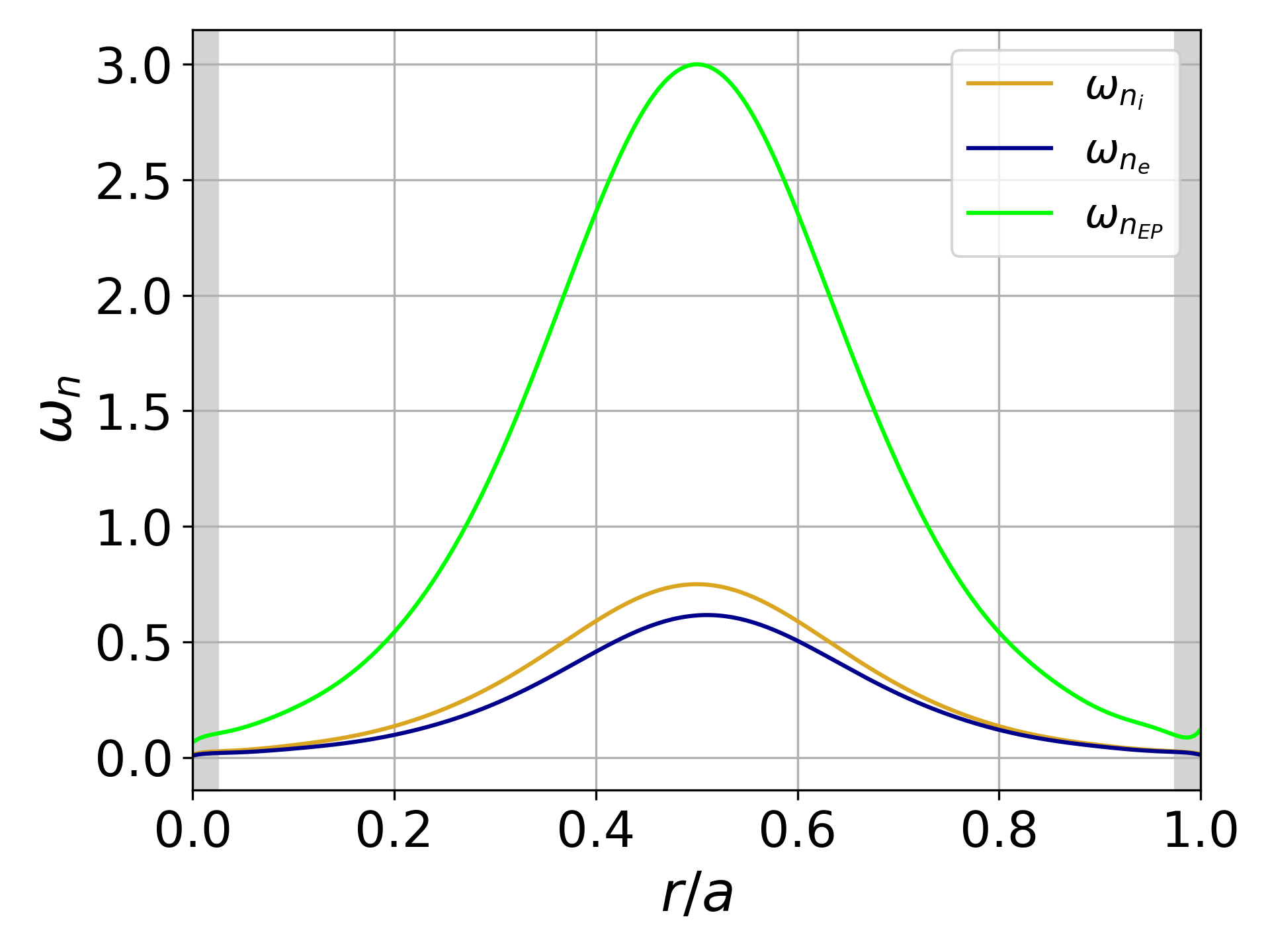}
\caption{}
\end{center}
\end{subfigure}
\noindent\begin{subfigure}[t]{0.33\textwidth}
\begin{center}
\includegraphics[width=\textwidth]{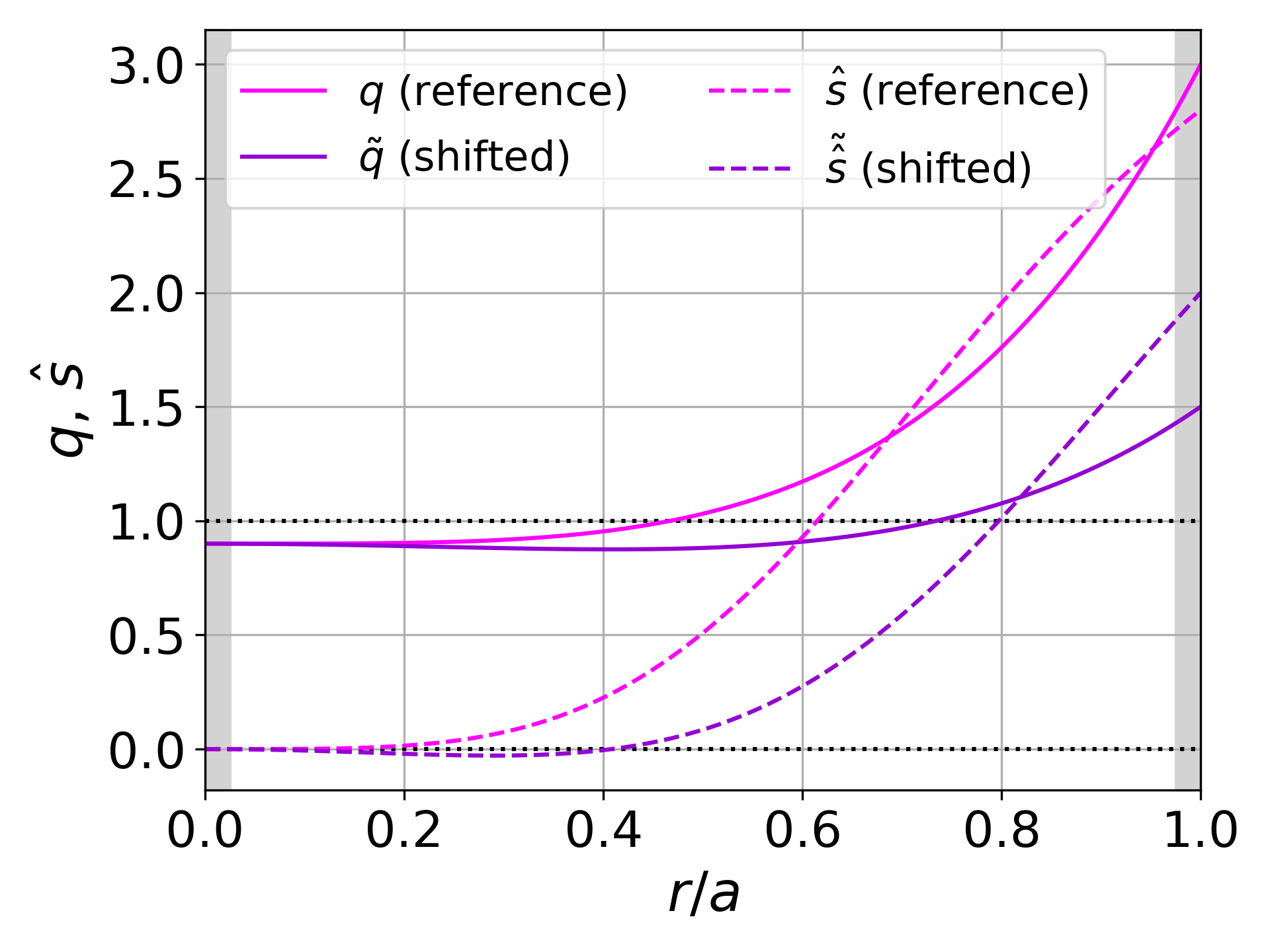}
\caption{}
\end{center}
\end{subfigure}
\caption{(a) Density, temperature and (b) normalized density gradients radial profiles for each plasma species. (c) Safety factor and magnetic shear profiles for the "reference" and "shifted" rational surface (identified by the $\:\tilde{(...)}\:$ superscript) setups. In (a), the fast ions density is rescaled by a factor 5 in order to facilitate visualization.}
\label{figure1}
\end{figure*}

One of the main challenges controlled nuclear fusion research has to face is the development of  methods to efficiently confine the plasma and reach reactor-relevant performances. To this scope, the outward fluxes of particles and heat, which remove particles (fuel) and energy from the plasma, must be minimized. Since these fluxes are driven by the turbulence destabilized by thermal density and temperature gradients, understanding how different plasma scenarios affect the turbulence is of crucial importance. In this work, we investigate how fully-developed plasma turbulence behaves around rational surfaces in the presence of suprathermal particles. The influence the safety factor ($q$) rational surfaces have on turbulence stabilization is a topic that has been explored both experimentally \cite{joffrin,yu,chung,chu} and numerically \cite{waltz,weikl,ajay_1,chen,volcokas_3,digiannatale}. In recent years, a picture started delineating in which the turbulence levels at rational surfaces are reduced due to the generation of $E\times B$ shearing structures \cite{diamond} through turbulence self interaction \cite{weikl,ball_1,ajay_1,ajay_2,volcokas_2,digiannatale}. The process is influenced by the shape of the $q$ profile, whose value can modify the turbulent eddy shape, thus acting on the self-interaction itself. \\
Furthermore, fusion reactor plasmas are expected to contain a non-negligible fraction of alpha particles whose temperature will surpass roughly by a factor of 10$^2$ the thermal one. For this reason, the physics of fast ions (FI) and their impact on turbulence is of great importance. Moreover, the action of heating systems like radio-frequency heating or neutral beam injection (NBI) generates a fraction of suprathermal particles also in modern day experimental plasmas. Fast ions influence plasma turbulence through different channels, which can act at the same time and interact one with the other while affecting the behavior of plasma bulk \cite{lauber,na}. Energetic particles are known to destabilize so-called EP-driven modes \cite{chen_1,chen_zonca}, whose interaction with turbulence can be direct, bringing to a modification of growth rates of the latter \cite{liu_3,yan}, or mediated by the generation of potential zonal structures \cite{chen_zonca_2,qiu,disiena_1}, which are known to shear away turbulent eddies. The presence of fast ions can also impact directly the turbulence characteristics, via quasi-resonant interaction \cite{disiena_2,disiena_3} and through the mechanism of phase-synchronization \cite{ghizzo}. Furthermore, suprathermal ions act as an additional species in the plasma, diluting the thermal ones and modifying their profiles, impacting this way on the turbulence level \cite{tardini,kim}. Fast ions dilution can also modify the threshold and growth rates for the generation and development of $E\times B$ zonal structures, therefore impacting the shearing effect they have on turbulence \cite{hahm,choi}. In this work, we will study via numerical simulations how this mechanism affects the generation of zonal structures at rational surfaces. \\
The interplay between fast ions and turbulence at rational surfaces is of great importance also when rational $q$ values are a fundamental condition for the development of modes. This is the case for fishbone instabilities \cite{mcguire,chen_FB,coppi,zonca_FB}, which can interact with turbulence. These are $n=qm$ modes (with $n$ and $m$ toroidal and poloidal mode numbers) driven by the resonance between fast ions and plasma bulk at rational surface. Indeed, such modes have been recently studied, both experimentally \cite{liu_1,he,ge} and numerically \cite{liu_2,brochard}, in the context of strong turbulent transport reduction. Within this work, the interaction between zonal structures generated by the fishbone and turbulence is analyzed. \\
This paper describes how fast ions dilution affects the characteristic of $E\times B$ zonal structures generated at rational surfaces by interaction of plasma turbulent eddies with themselves. The competition between this effect and other channels through which fast ions influence turbulence is analyzed. In particular, we find a linear interplay between FI quasi-resonant effect and FI dilution, and regimes in which one dominates over the other are delineated. It is shown that the presence of fast ions with certain characteristics can lead to a fundamental reduction of turbulent fluxes at specific radial positions. This effect is maximized when turbulence develops at rational surfaces. Moreover, the destabilization of a fishbone mode in the plasma leads to a further modification of the turbulent fluxes. This is shown to be due to the development of a beat-driven zonal structure from the fishbone. This whole analysis is carried out by mean of global, nonlinear simulations using the gyrokinetic code GENE \cite{jenko,gorler}. \\
The remainder of this paper is organized as follows. In section \ref{sec:sim_det}, details of the numerical setup are discussed. First, the code GENE, which has been used to perform the simulations described, is presented in section \ref{subsec:GENE}. Then, in section \ref{subsec:setup}, the profiles used in the numerical analysis are given, and the simulations setup is described. Section \ref{sec:linear} is dedicated to the linear study of the modes that develop in the considered scenario, along with an analysis of the fast ions dilution. In section \ref{subsec:qr_effect}, a description of the quasi-resonant effect we observe when fast ions are included is given. Section \ref{sec:transp_supp} is dedicate to the description of the turbulent transport suppression observed at rational surfaces in the presence of energetic particles. The zonal structures generation mechanism via self-interaction is analyzed in section \ref{subsec:ZF_generation}. A detailed analysis is performed by selecting a single mode and studying its interaction with the zonal one in section \ref{subsec:n_min_5}. The dilution effect through which fast ions enhance the stabilizing effect due to self-interaction is studied in \ref{subsec:FI_dilution}. Section \ref{sec:FB} is dedicated to the effect the destabilization of a fishbone mode has on the plasma. Linear characteristics of the fishbone are reported in section \ref{subsec:FB_lin}, and the interaction between this mode and turbulence is analyzed in section \ref{subsec:FB_nl}. Conclusions are reported in section \ref{sec:conclusions}. 

\vspace{-0.5pc}
\section{Simulation details}\label{sec:sim_det}
\vspace{-0.5pc}

\subsection{The gyrokinetic code GENE}\label{subsec:GENE}
\vspace{-0.5pc}

\begin{figure*}[ht!]
\noindent\begin{subfigure}[t]{0.33\textwidth}
\begin{center}
\includegraphics[width=\textwidth]{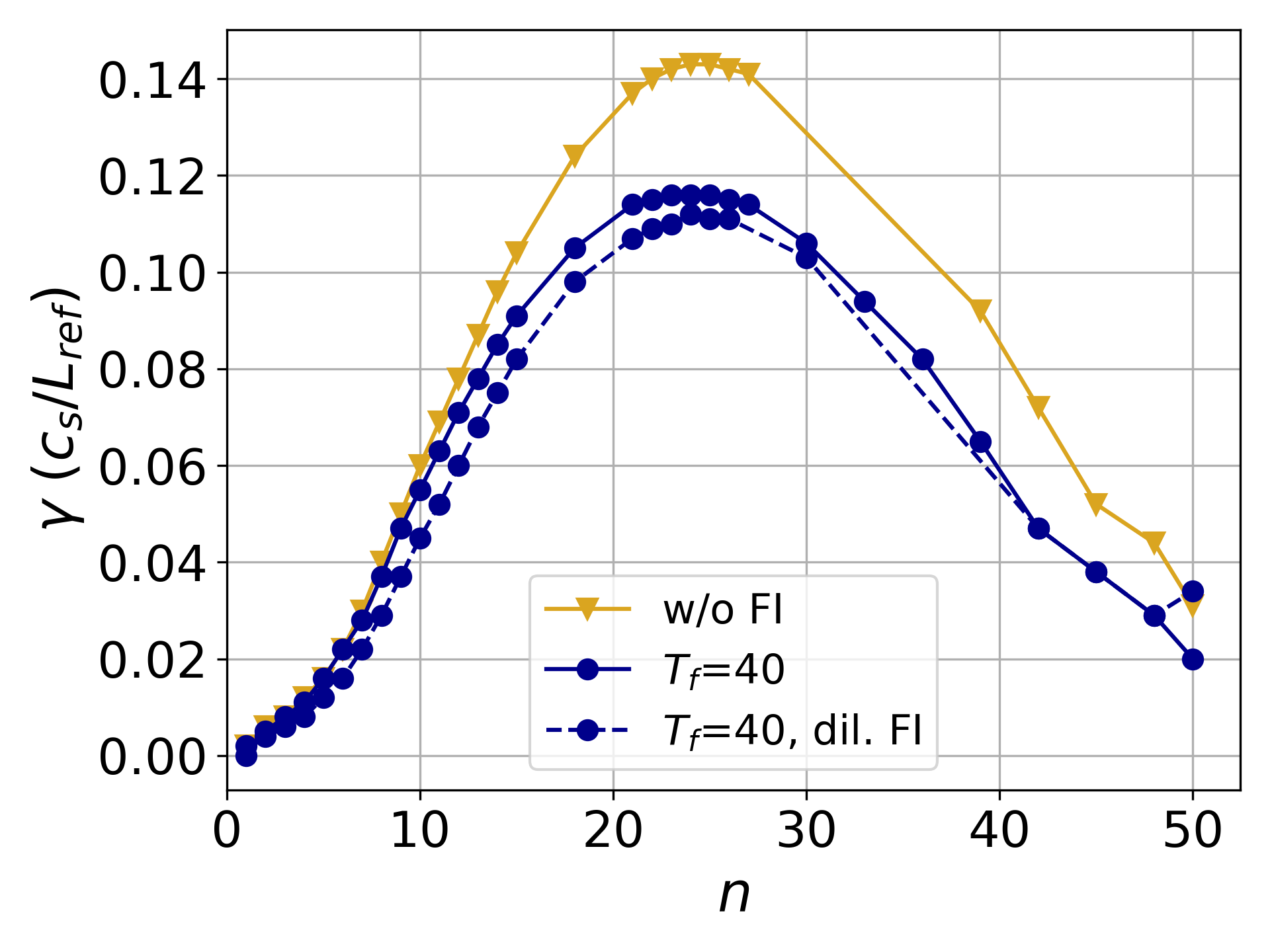}
\caption{}
\end{center}
\end{subfigure}
\noindent\begin{subfigure}[t]{0.33\textwidth}
\begin{center}
\includegraphics[width=\textwidth]{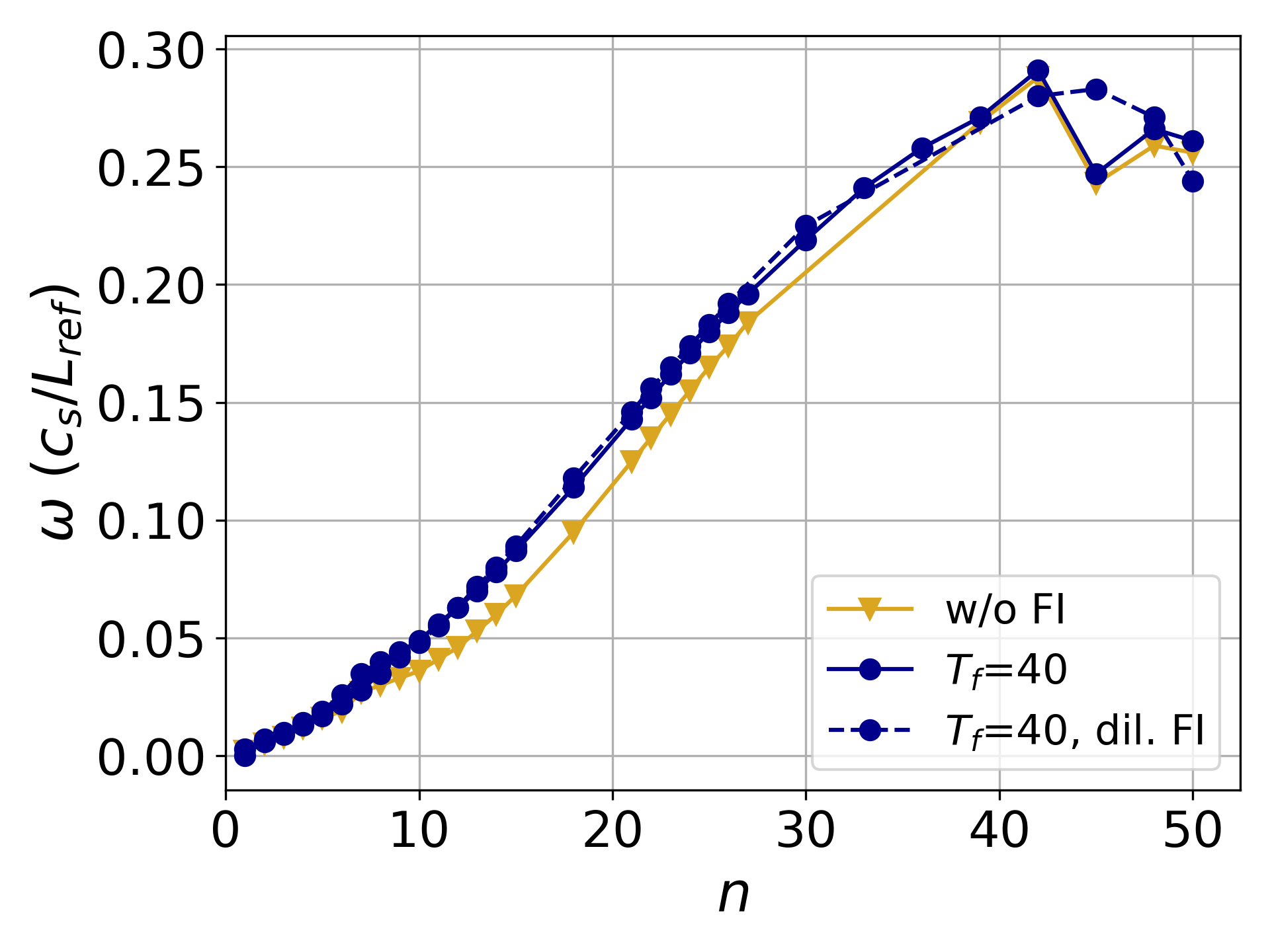}
\caption{}
\end{center}
\end{subfigure}
\caption{Normalized (a) growth rate and (b) frequency radial average for three scenarios. One without fast particles, one with $T_f=40$ and one with $T_f=40$ but fast ions set as a dilution species. In the setup without fast particles, $n_i$ and $\omega_{n_i}$ are raised to 1.06 and 0.75, respectively.}
\label{figure2}
\end{figure*}

\begin{figure*}[ht!]
\noindent\begin{subfigure}[t]{0.33\textwidth}
\begin{center}
\includegraphics[width=\textwidth]{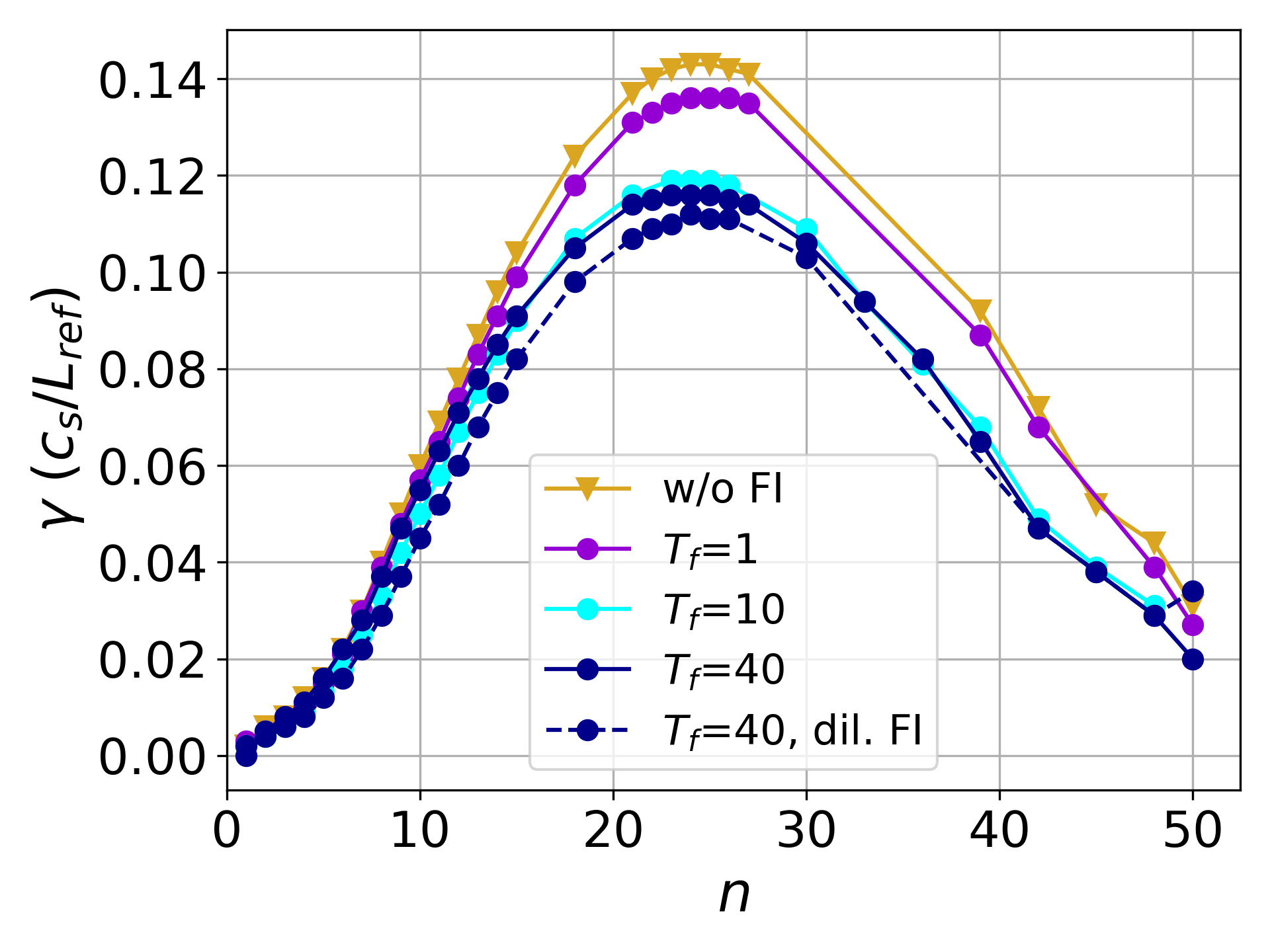}
\caption{}
\end{center}
\end{subfigure}
\noindent\begin{subfigure}[t]{0.33\textwidth}
\begin{center}
\includegraphics[width=\textwidth]{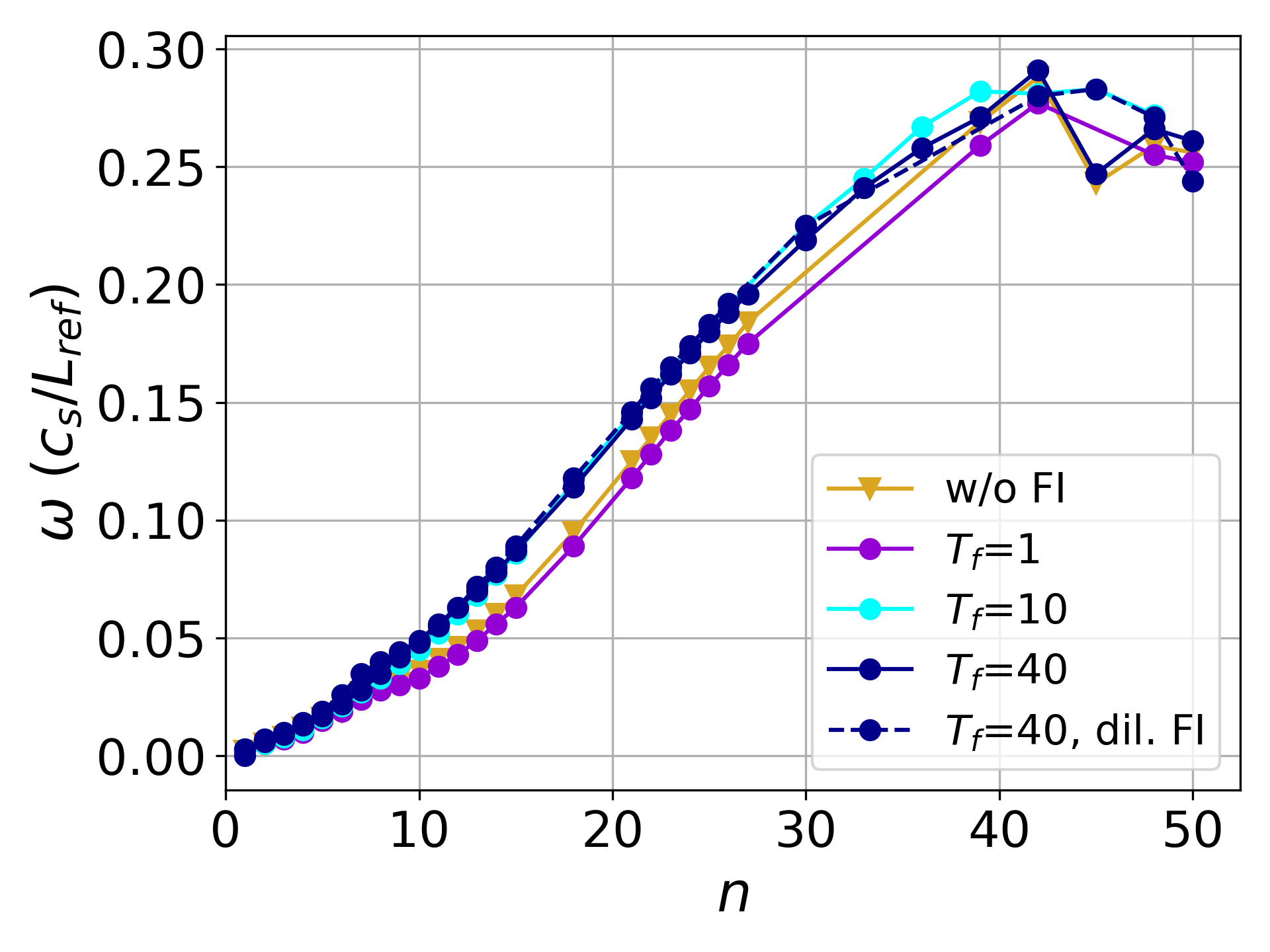}
\caption{}
\end{center}
\end{subfigure}
\caption{Normalized (a) growth rate and (b) frequency radial average for some of the scenarios considered. In particular, one with $T_f=1$ and one with $T_f=10$ are added with respect to the previous figure.}
\label{figure2bis}
\end{figure*}

The Eulerian gyrokinetic code GENE solves the perturbed Vlasov-Maxwell system on a five-dimensional grid for each of the plasma species considered. The field-aligned spatial coordinates in which the code works are $x$, $y$ and $z$. They are defined as $x=r/a=\sqrt{\Phi/\Phi_{max}}$ radial-like coordinate working as a flux-surface label, $y=x_0(q\chi-\varphi)/q(x_0)$ binormal coordinate spanning magnetic field lines and $z=\chi$ moving along a given field line. Here, $\Phi$ is the toroidal magnetic flux through a given flux surface, $x_0$ is the center of the domain and $\chi$ and $\varphi$ are the poloidal and toroidal angles in straight-field line coordinates, respectively. The velocity space is spanned by $v_\parallel$ and $\mu=mv_\perp^2/(2B_0)$, representing the projection of the particle velocity along the magnetic field line and the particle magnetic moment, respectively. In GENE, an equidistant grid is implemented along all directions except the $\mu$ one, where Gauss-Laguerre points are used. The code can be used both in local and global approach. The first paradigm considers only a thin plasma section around a magnetic field line as simulation domain. The second one simulates a whole annulus of the tokamak volume. Periodic boundary conditions are considered for the binormal coordinate $y$ in both approaches, and Dirichlet boundary conditions are used in $x$ for global simulations. This imposes the vanishing of fluctuations at the boundaries of the simulation domain. Inside the code, this is obtained through the application of Krook operators inside the Vlasov equation in the form $\partial_t f=-\gamma_kf$. Furthermore, plasma species and safety factor profiles are linearized around the simulation box center in local simulations, while during global runs the full radial profile is retained. The quantities computed with the code are normalized to reference units given in the following: $m_{ref}=m_H=1$ a.m.u., $n_{ref}=1\times10^{19}$ m$^{-3}$, $T_{ref}=T_e=1$ keV, $L_{ref}=1$ m and $B_{ref}=1$ T. The reference sound speed is defined as $c_s=\sqrt{T_e/m_i}=\sqrt{T_{ref}/m_{ref}}$. 

\vspace{-0.5pc}
\subsection{Plasma parameters and simulation setup}\label{subsec:setup}
\vspace{-0.5pc}

\begin{figure*}[ht!]
\noindent\begin{subfigure}[t]{0.33\textwidth}
\begin{center}
\includegraphics[width=\textwidth]{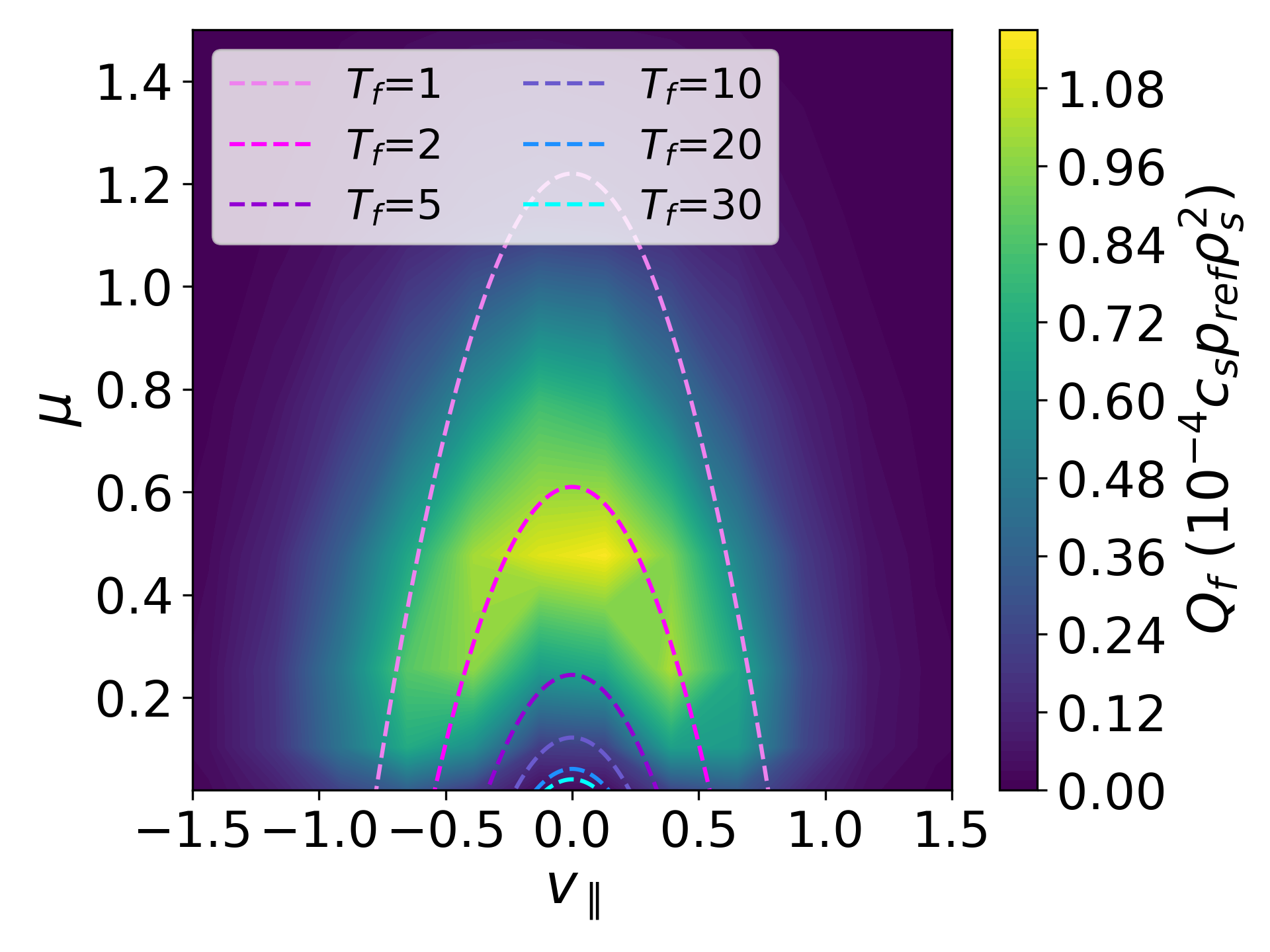}
\caption{}
\end{center}
\end{subfigure}
\noindent\begin{subfigure}[t]{0.33\textwidth}
\begin{center}
\includegraphics[width=\textwidth]{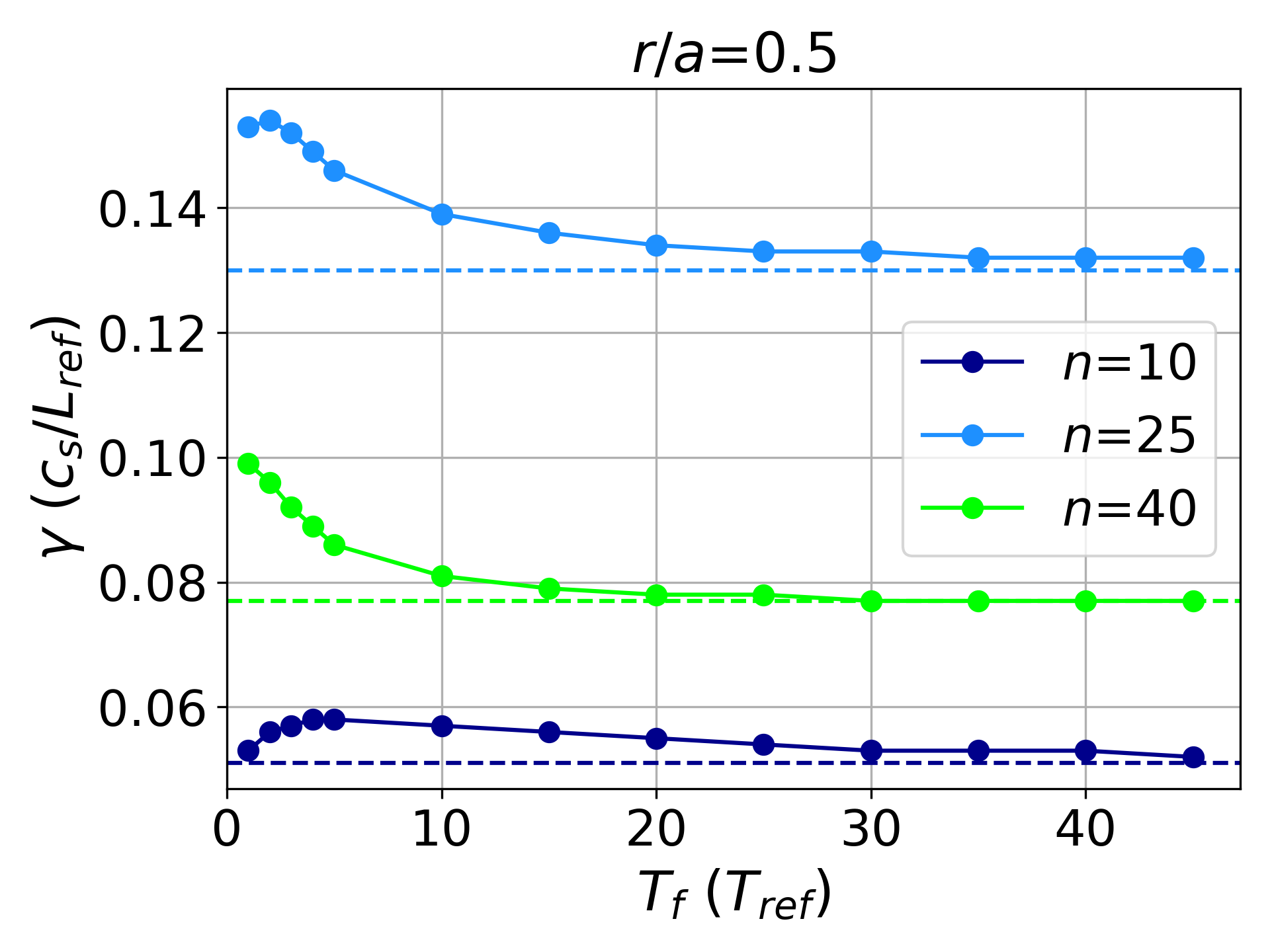}
\caption{}
\end{center}
\end{subfigure}
\caption{(a) Electrostatic heat flux for the fast ions minority at $T_f=10$, $n=25$, $z=0$ in velocity space. Resonance curves are plotted from Eq.(\ref{Eq2}) using $k_y=0.31$ and $\omega_r=0.18$, corresponding to $n=25$ in figure \ref{figure2bis}. (b) Normalized growth rates from local runs at $r/a=0.5$ for the $n=10$, 25, 40 modes at different $T_f$ values. Dashed lines represent $\gamma$ values obtained setting fast ions as a dilution species.}
\label{figure3}
\end{figure*}

The plasma setup considered in this paper is designed to develop high ion temperature gradient (ITG)-driven turbulence in correspondence to the rational surface $q=1$. At the same time, it avoids the destabilization of EP-driven modes when fast particles with temperature $T_f\leq40$ are added to the description. The presence of such modes would alter the picture of zonal structure generation, introducing contributions to the instauration of such phenomena \cite{ishizawa,liu_3,qiu_2}. However, the impact of an $n=1$ EP-driven mode on the setup is also studied with dedicated linear and nonlinear runs by increasing $T_f$. Density profiles for electrons ($n_e$) and fast ions ($n_f$) are given by the relation
\begin{equation} \label{Eq1}
    n=n_0\textrm{exp}\Biggl(-\kappa_n\Delta_n\textrm{tanh}\biggl(\frac{r-r_0}{a\Delta_n}\biggl)\Biggl),
\end{equation}
where $n_0$ is the density at $r_0/a=0.5$, center of the simulation domain, $\kappa_n=-n^{-1}dn/dr$ is the inverse of the characteristic density variation length, and $\Delta_n$ gives a measure of the density gradient peak width. In Eq.(\ref{Eq1}), $a$ is the plasma minor radius. For the simulation setup we have, in GENE units, $n_{0e}=1.06$, $n_{0f}=0.06$, $\kappa_{n_e}=0.75$, $\kappa_{n_f}=3$, $\Delta_{n_e}=\Delta_{n_f}=0.2$. Plasma quasi-neutrality holds through ions, whose density is computed as $n_i=n_e-n_f$. Electron and fast ions temperatures are considered flat, with $T_e=1$ in GENE units, while $T_i$ profile is assigned by Eq.(\ref{Eq1}) with $T_{0i}=1$, $\kappa_{T_i}=3$ and $\Delta_{T_i}=0.2$. In the context of this work, fast ions temperature is varied from 1 to 40. An additional scenario with $T_f=180$ is also considered. For simplicity, all species are modeled through an equivalent Maxwellian distribution. In figure \ref{figure1} (a) and (b), density and temperature profiles are reported for the three plasma species, along with the normalized density gradients $\omega_n=L_{ref}/L_n=-L_{ref}n^{-1}dn/dr$. In the global runs, we simulate 95\% of the plasma volume, with the radial domain spanning the interval $r/a\in[0.025,0.975]$. It is important to stress out that the density and ion temperature gradients peak at the center of the simulation domain, i.e. at $r/a=0.5$. All the simulations are carried out in a collisionless hydrogen plasma with reduced mass ratio $m_i/m_e=200$ (used in order to reduce computational costs) and deuterium fast ions, thus with $m_f=2$. In all simulations, the electronic over magnetic pressure ratio is set to $\beta_e=7.5\times10^{-4}$ and $\rho^*=\rho_s/L_{ref}=6\times 10^{-3}$, with $\rho_s=\sqrt{T_em_i}/(eB_0)$ sound Larmor radius, is fixed. In simulations without fast ions, where $n_i$ is increased up to $n_e$, the value of $\beta_e$ is modified in order to keep $\beta_{tot}=\sum_{\sigma}\beta_{\sigma}$ fixed. In this case, $\beta_f$ in the sum is computed using $n_f=0.06$ and $T_f=40$. We consider a simplified geometry with circular concentric magnetic flux surfaces. The minor radius is $a=1$, while the major radius is $R=10$. The analytical safety factor $q$ profiles we are considering are reported in figure \ref{figure1} (c), along with the respective magnetic shear $\hat{s}=(r/q)dq/dr$. The reference $q$ is assigned by $q(r/a)=0.9+2.1(r/a)^4$, and has two low order rational surfaces, the first one at $r/a=0.47$ ($q=1$) and the second one at $r/a=0.85$ ($q=2$). Another safety factor profile, labeled $\tilde{q}$, is considered. This safety factor profile is similar to the first one but presents a shifted $\tilde{q}=1$ rational surface at $r/a=0.74$. Its analytical expression is $\tilde{q}=0.9-0.3(r/a)^2+0.9(r/a)^4$. Both profiles are characterized by very low values of the magnetic shear between $r/a=0$ and $r/a\approx 0.3$, with $\hat{s}(0.2)=1.5\times10^{-2}$ and $\tilde{\hat{s}}(0.2)=-2.1\times10^{-2}$. Furthermore, it is worth noticing that the position of the gradients maximum ($r/a=0.5$) coincides with the $q=1$ one ($r/a=0.47$). \\
After performing linear scans in all directions of the grid, we select a numerical resolution for the global runs which is $(n_x,n_{k_y},n_z,n_{v_\parallel},n_\mu)=(256,48,24,24,16)$. The grid setup is also tested nonlinearly by performing runs with $n_{v_\parallel}=48$ for the scenario without fast ions and the one with $T_f=40$. Since $n_{k_y}=48$, in the nonlinear simulation we are including all the modes with toroidal mode number $n\in\{0,1,...,47\}$. Simulation domain dimensions along the velocity space directions are $l_{v_\parallel,min}=-3$, $l_{v_\parallel,max}=3$, $l_{\mu,min}=0$ and $l_{\mu,max}=13$ in $c_s$ units. Lower ($l$) and upper ($u$) buffer regions in which fluctuations are damped near the plasma border represent the $2.5\%$ of the simulation domain each. The value of Krook operators which carry out the damping in these regions is $\gamma_{kl}=\gamma_{ku}=1$ in $c_s/L_{ref}$ units. Krook operators are also used to minimize the modifications of temperature and density profiles due to heat ($h$) and particle ($p$) fluxes, respectively. For the two channels, the values we use are $\gamma_{kh}=0.03$ (smaller than the maximum linear growth rate observed) and $\gamma_{kp}=0.1$. 

\vspace{-0.5pc}
\section{Modes characterization and fast ions linear physics}\label{sec:linear}
\vspace{-0.5pc}

\begin{figure*}[ht!]
\noindent\begin{subfigure}[t]{0.33\textwidth}
\begin{center}
\includegraphics[width=\textwidth]{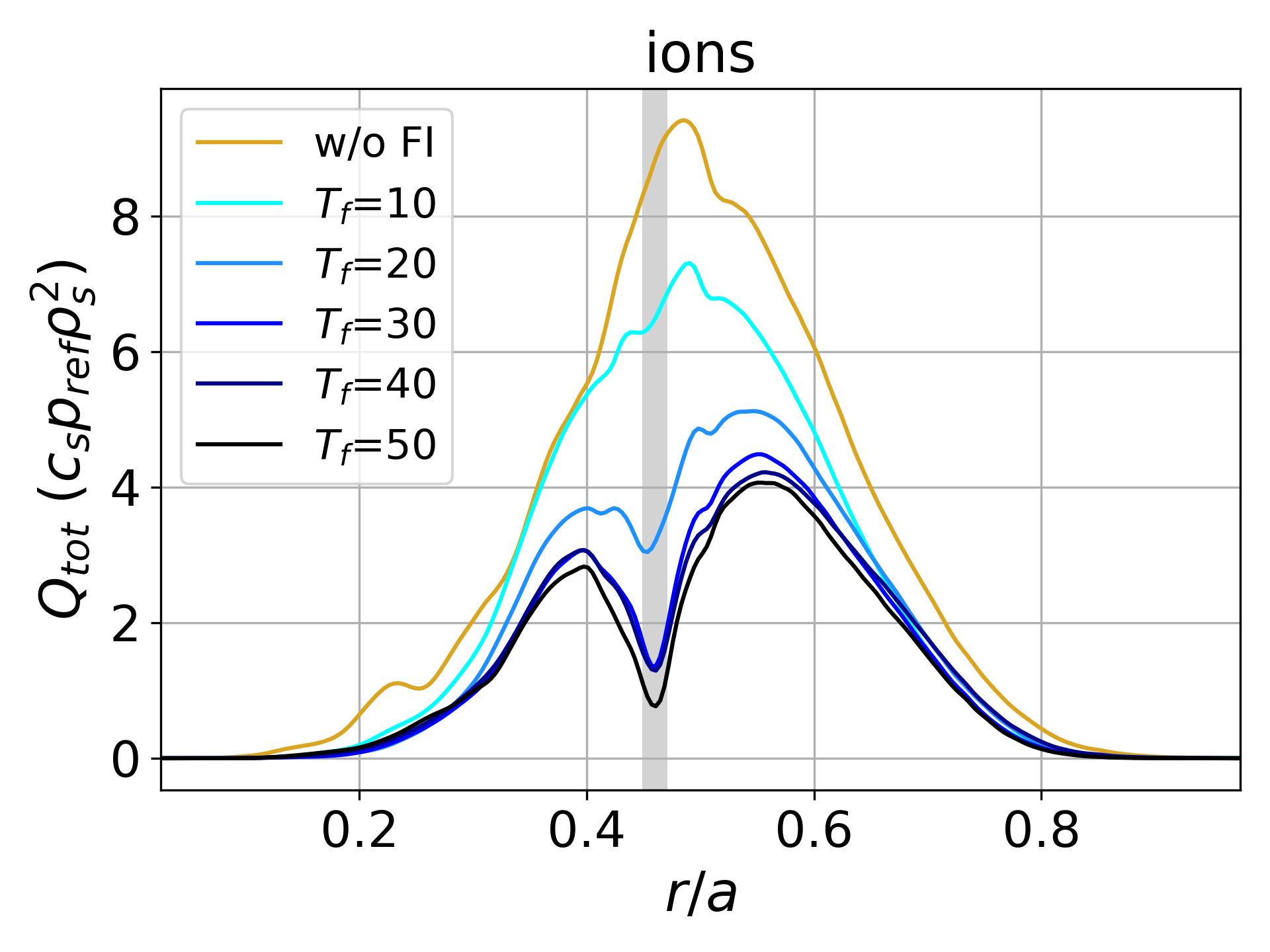}
\caption{}
\end{center}
\end{subfigure}
\noindent\begin{subfigure}[t]{0.33\textwidth}
\begin{center}
\includegraphics[width=\textwidth]{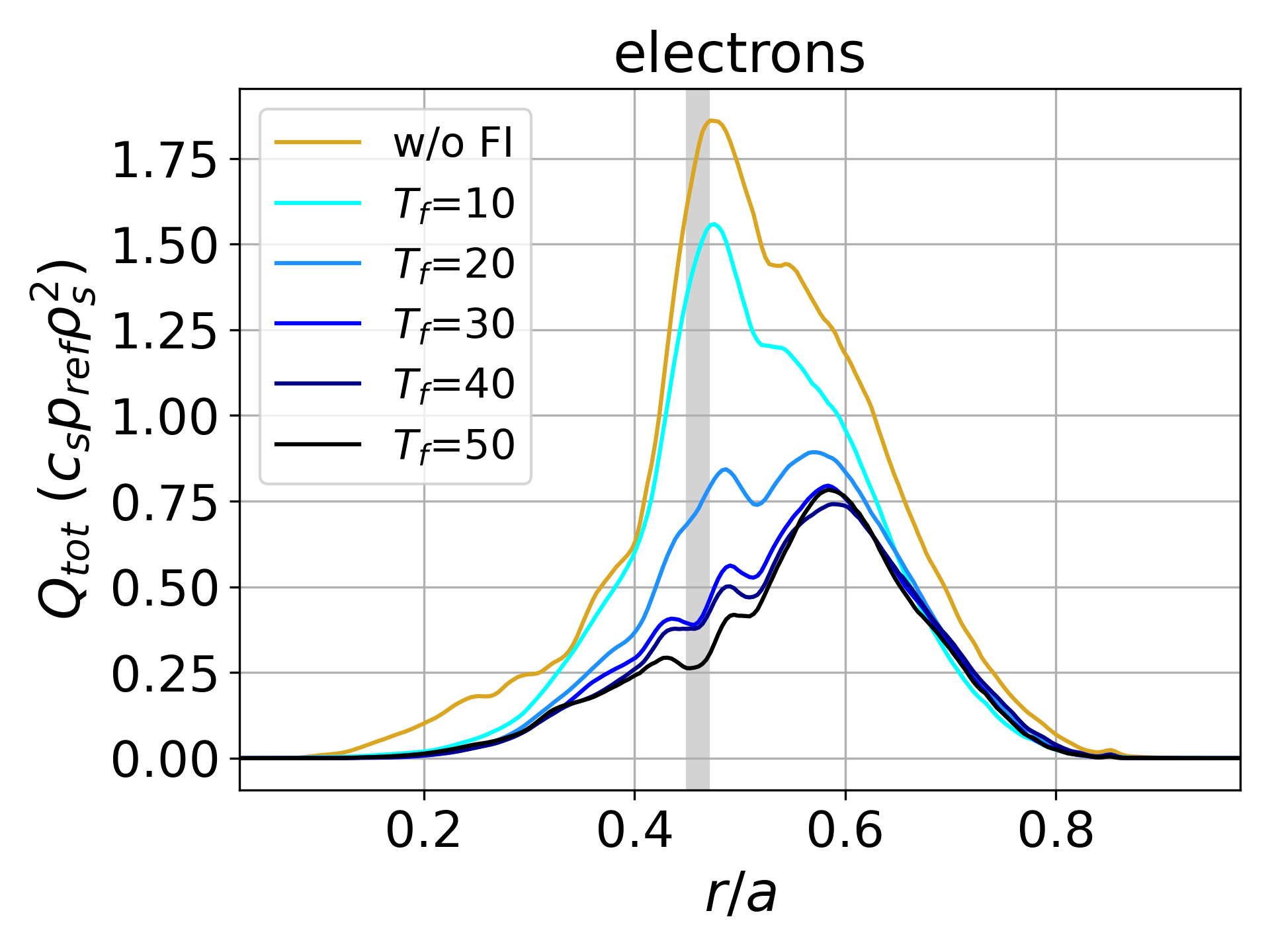}
\caption{}
\end{center}
\end{subfigure}
\noindent\begin{subfigure}[t]{0.33\textwidth}
\begin{center}
\includegraphics[width=\textwidth]{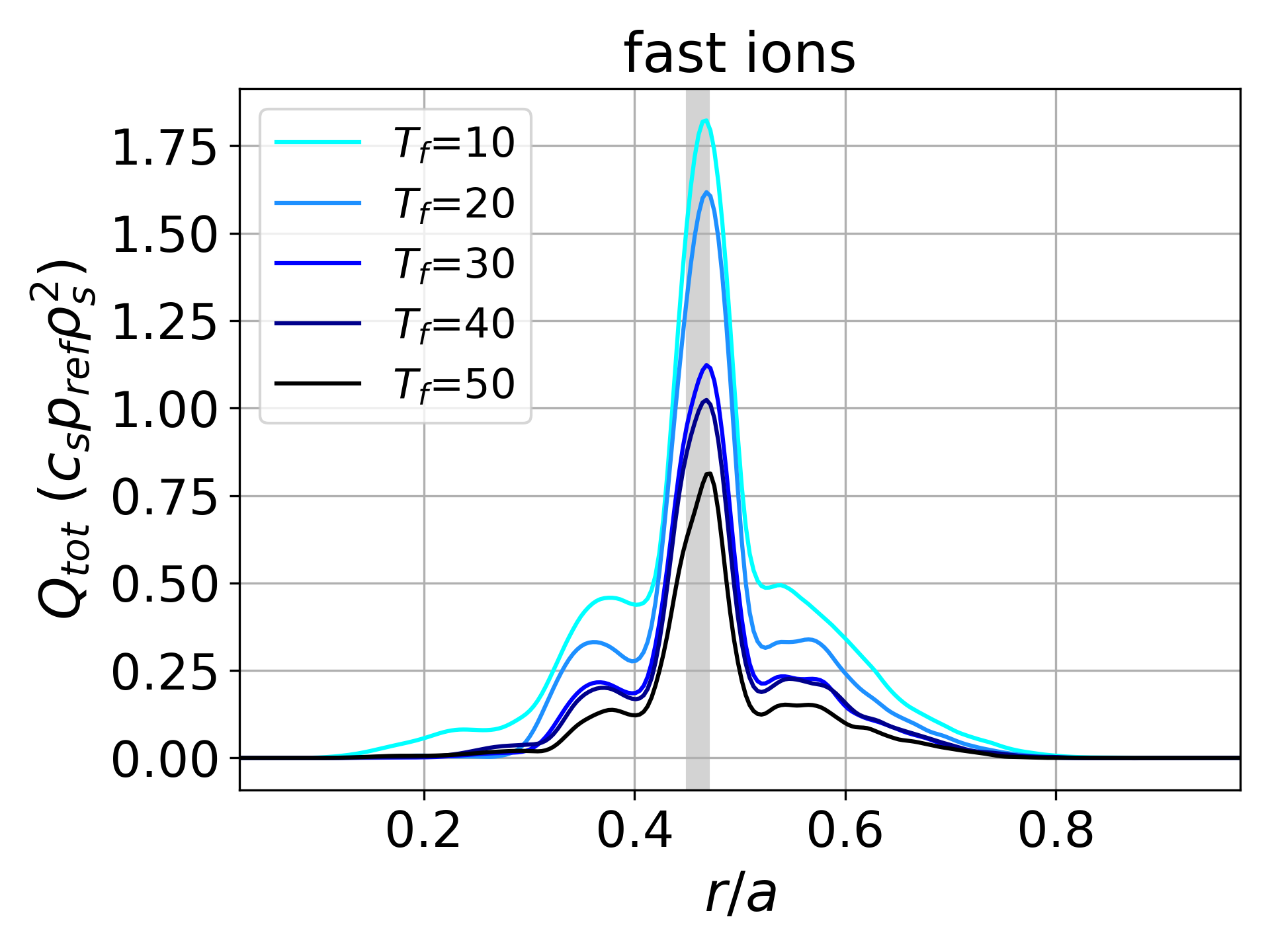}
\caption{}
\end{center}
\end{subfigure}
\caption{Turbulent total (electrostatic + electromagnetic) heat fluxes for (a) ions, (b) electrons and (c) fast ions species for different $T_f$ values. All the profiles are averaged over a time interval in which turbulence is saturated. The shaded region represents an interval of amplitude $r/a=0.01$ around the $q=1$ position.}
\label{figure4}
\end{figure*}

%\begin{figure*}[ht!]
%\noindent\begin{subfigure}[t]{0.33\textwidth}
%\begin{center}
%caption{}
%\end{center}
%\end{subfigure}
%\noindent\begin{subfigure}[t]{0.33\textwidth}
%\begin{center}
%\includegraphics[width=\textwidth]{figure5b.png}
%\caption{}
%\end{center}
%\end{subfigure}
%\caption{Normalized (a) growth rate and (b) frequency at $r/a=0.47$ for some of the scenarios considered. In the setup without fast particles, $n_i$ and $\omega_{n_i}$ are raised to 1.06 and 0.75, respectively.}
%\label{figure4bis}
%\end{figure*}

\begin{figure*}[ht!]
\noindent\begin{subfigure}[t]{0.33\textwidth}
\begin{center}
\includegraphics[width=\textwidth]{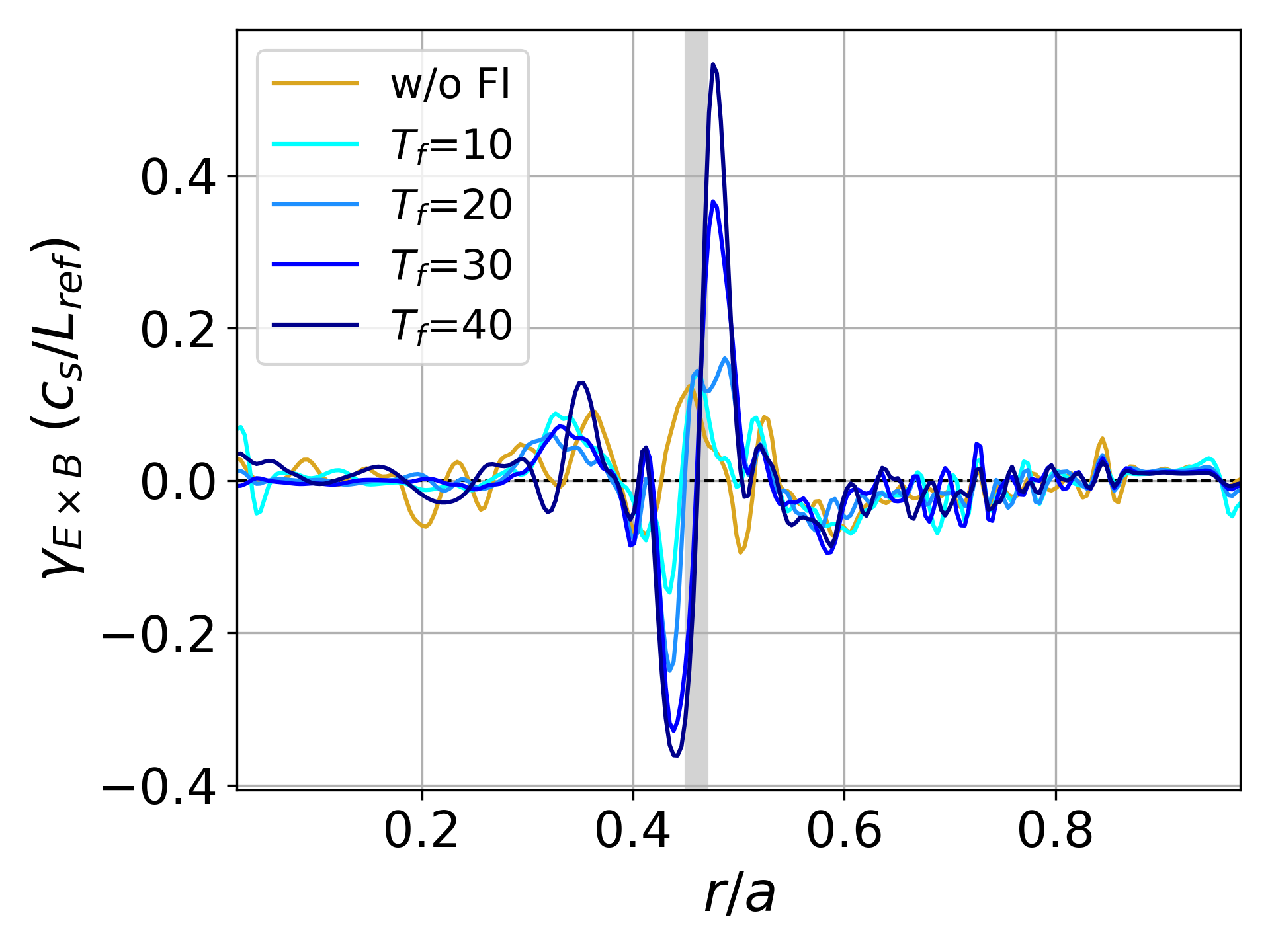}
\end{center}
\end{subfigure}
\caption{Shearing rate $\gamma_{E\times B}$ radial profiles for different $T_f$ values. The quantity is averaged over a time interval in which turbulence is saturated. The shaded region represents an interval of amplitude $r/a=0.01$ around the $q=1$ position.}
\label{figure5}
\end{figure*}

Linear global simulations are performed to characterize the instabilities that develop in the setup, along with the influence energetic particles has on them. Figure \ref{figure2} shows the radially-averaged growth rates and frequencies obtained without and with fast ions at temperatures $T_f=40$ with solid lines. We see that the instability $\gamma$ peaks around $n=25$. Recalling that in GENE a positive frequency corresponds to a mode rotating in the ion diamagnetic direction, we conclude that the spectra of our setup is ITG-dominated. By including fast particles (from yellow to solid blue curve), a reduction of growth rates -- i.e. a linear stabilization of the mode -- is evident. This reduction of growth rates leads to $\gamma$ values 20\% lower than the initial ones. We argue this feature can be explained in terms of the fast ions dilution effect on the thermal plasma species \cite{tardini,kim}. Indeed, when fast ions are included in the plasma setup, thermal species profiles are modified in order to respect the quasi-neutrality condition $\sum_\sigma Z_\sigma n_\sigma=0$. The modification changes the amount of particles participating in the mode dynamics, affecting the linear dispersion relation and leading in turn to different growth rates between the cases with and without fast ions. We test this conclusion by performing a simulation with fast particles considered as a dilution species inside the plasma. Therefore, the contribution given by this species is retained in the code only inside the computation of the polarization density. Results for this scenario are plotted with a dashed line in figure \ref{figure2}. We see that they coincide with the growth rates and frequencies obtained for the $T_f=40$ case.

\vspace{-0.5pc}
\subsection{Fast ions quasi-resonant effect} \label{subsec:qr_effect}
\vspace{-0.5pc}

In order to explore the influence of fast ions parameters on linear modes, we scan different values of the temperature $T_f$. Results obtained for two representative cases, i.e. $T_f=1$ and $T_f=10$, are reported in figure \ref{figure2bis}. By looking at figure \ref{figure2bis} (a), we observe an increase in growth rates as the value of $T_f$ decreases, thus moving from 40 to 10 to 1. This effect can be explained by the quasi-resonant interaction \cite{disiena_2,disiena_3} between particles and linear instability when the magnetic-drift frequency of the former matches the dominant frequency of the latter, 
\begin{equation} \label{Eq2}
    \omega_r=-\frac{T_f}{q_f}\Biggl(\frac{\mu B_0+2v_\parallel^2}{B_0}\Biggl)\mathcal{K}_yk_y.
\end{equation}
Here, $\omega_r$ is the real frequency of the ITG mode, $q_f$ the fast ions charge, $\mathcal{K}_y=-((\mathbf{B}\times\nabla B_0)\cdot\hat{y})/B_0^2$ the magnetic curvature along $y$ and $k_y$ the binormal wave number of the dominant mode. The quantity is linked to $n$ by $k_y=2\pi n/l_y$, with $l_y$ simulation domain length along $y$. When Eq.(\ref{Eq2}) holds, drift waves and fast particles can interact and exchange energy, resulting in a modification of the instability growth rate. The dependence of this resonant interaction mechanism on $T_f$ is shown in figure \ref{figure3} (a). Here, the velocity space component of the fast ions heat flux is plotted for the $n=25$ mode, along with Eq.(\ref{Eq2}) for different $T_f$ values at $r/a=0.5$. The fast ions flux is maximized for $T_f\approx 2$, and decreases as we move away from this resonance condition, both decreasing or increasing $T_f$. It is also worth noticing that, since here $\omega_{T_f}=0$, from \cite{disiena_2} we have that the fast ions contribution to the mode can only be destabilizing. This leads to an increase of the linear growth rates and fast particles fluxes. The mechanism is illustrated also in figure \ref{figure3} (b). Here, growth rates for some representative $n$ modes are plotted for a range of $T_f$ values. For $n=25$, $\gamma$ is maximized at $T_f=2$, confirming what said above. Moreover, as fast ions temperature increases -- i.e. as we move away from the resonance --, growth rates approach the ones obtained by considering fast ions as a dilution species (dashed lines). This observation further confirms the explanation given for the results in figure \ref{figure2}. What we identify linearly is then a competition between two mechanisms. The dilution effect tends to stabilize the turbulence independently on the fast ions temperature, and is independent on the suprathermal particles temperature. At the same time, the quasi-resonant mechanism enhances the mode destabilization as we approach the value of $T_f$ for which Eq.(\ref{Eq2}) holds. We will observe again this interplay when moving on to analyze the nonlinear development of turbulence in this setup. 

\begin{figure*}[ht!]
\noindent\begin{subfigure}[t]{0.33\textwidth}
\begin{center}
\includegraphics[width=\textwidth]{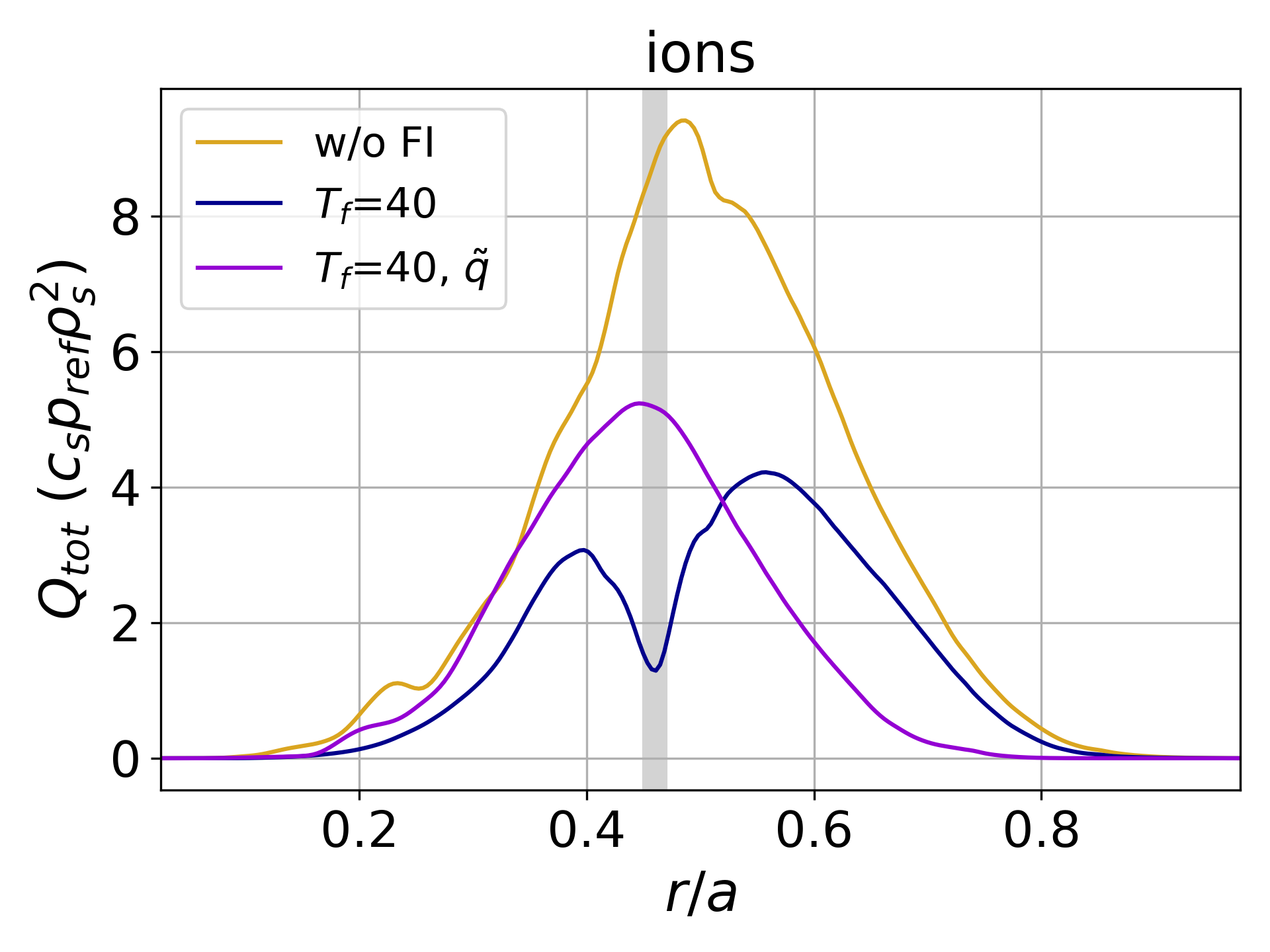}
\caption{}
\end{center}
\end{subfigure}
\noindent\begin{subfigure}[t]{0.33\textwidth}
\begin{center}
\includegraphics[width=\textwidth]{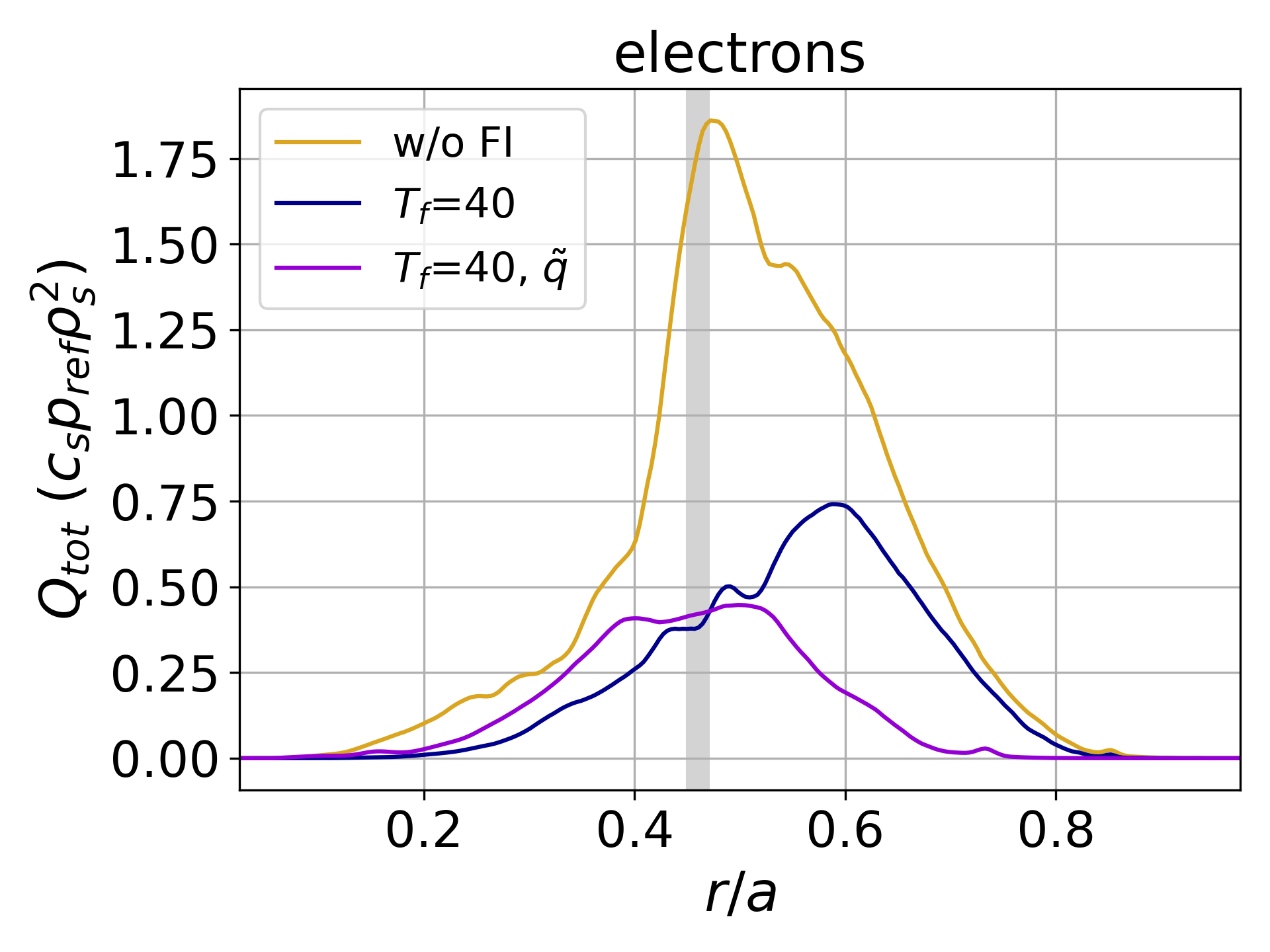}
\caption{}
\end{center}
\end{subfigure}
\noindent\begin{subfigure}[t]{0.33\textwidth}
\begin{center}
\includegraphics[width=\textwidth]{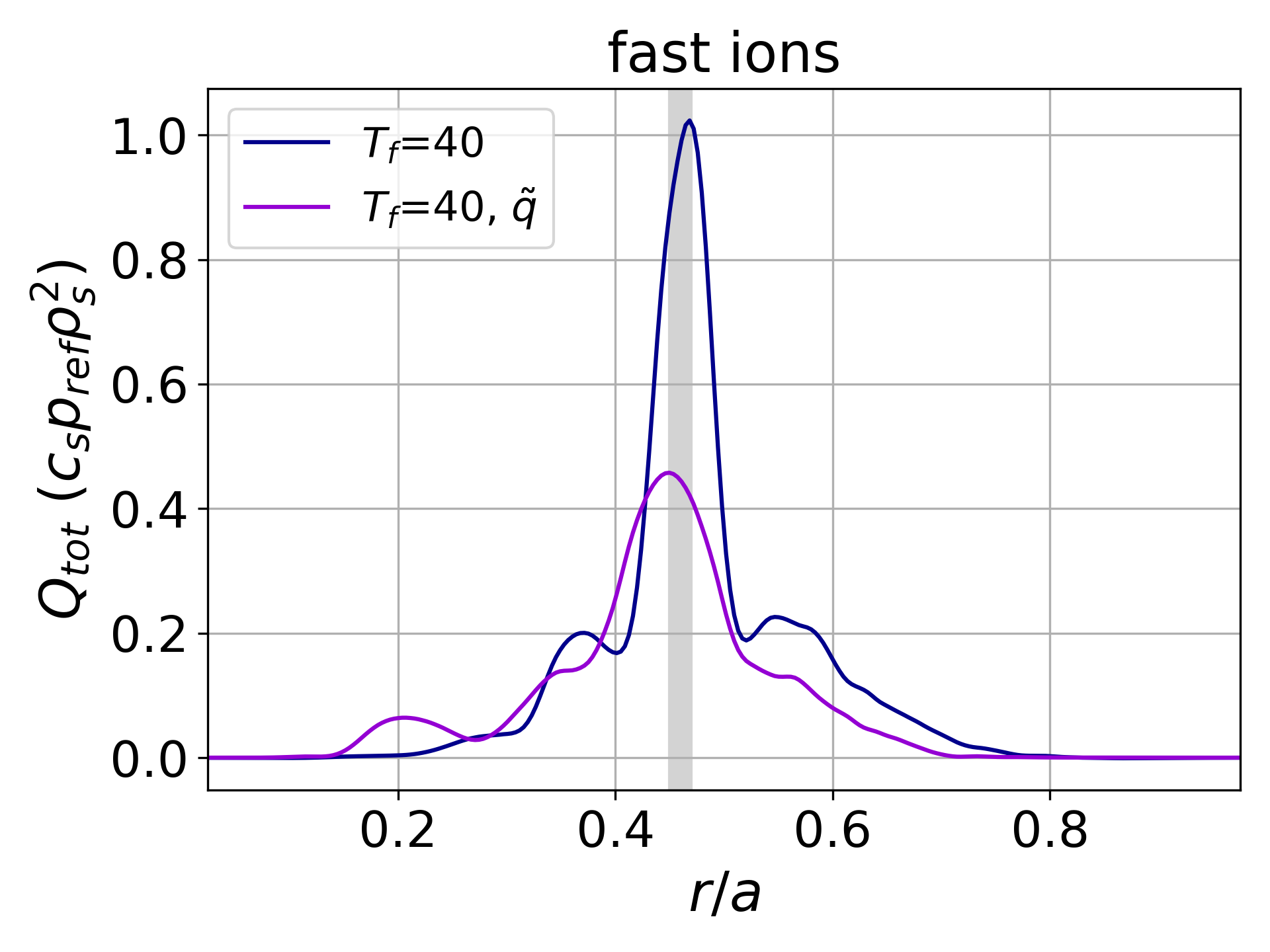}
\caption{}
\end{center}
\end{subfigure}
\caption{Turbulent total heat fluxes for (a) ions, (b) electrons and (c) fast ions species for three scenarios. Reference ones without and with ($T_f=40$) fast ions and one with $T_f=40$ and shifted $\tilde{q}$ profile.}
\label{figure7}
\end{figure*}

\begin{figure*}[ht!]
\noindent\begin{subfigure}[t]{0.33\textwidth}
\begin{center}
\includegraphics[width=\textwidth]{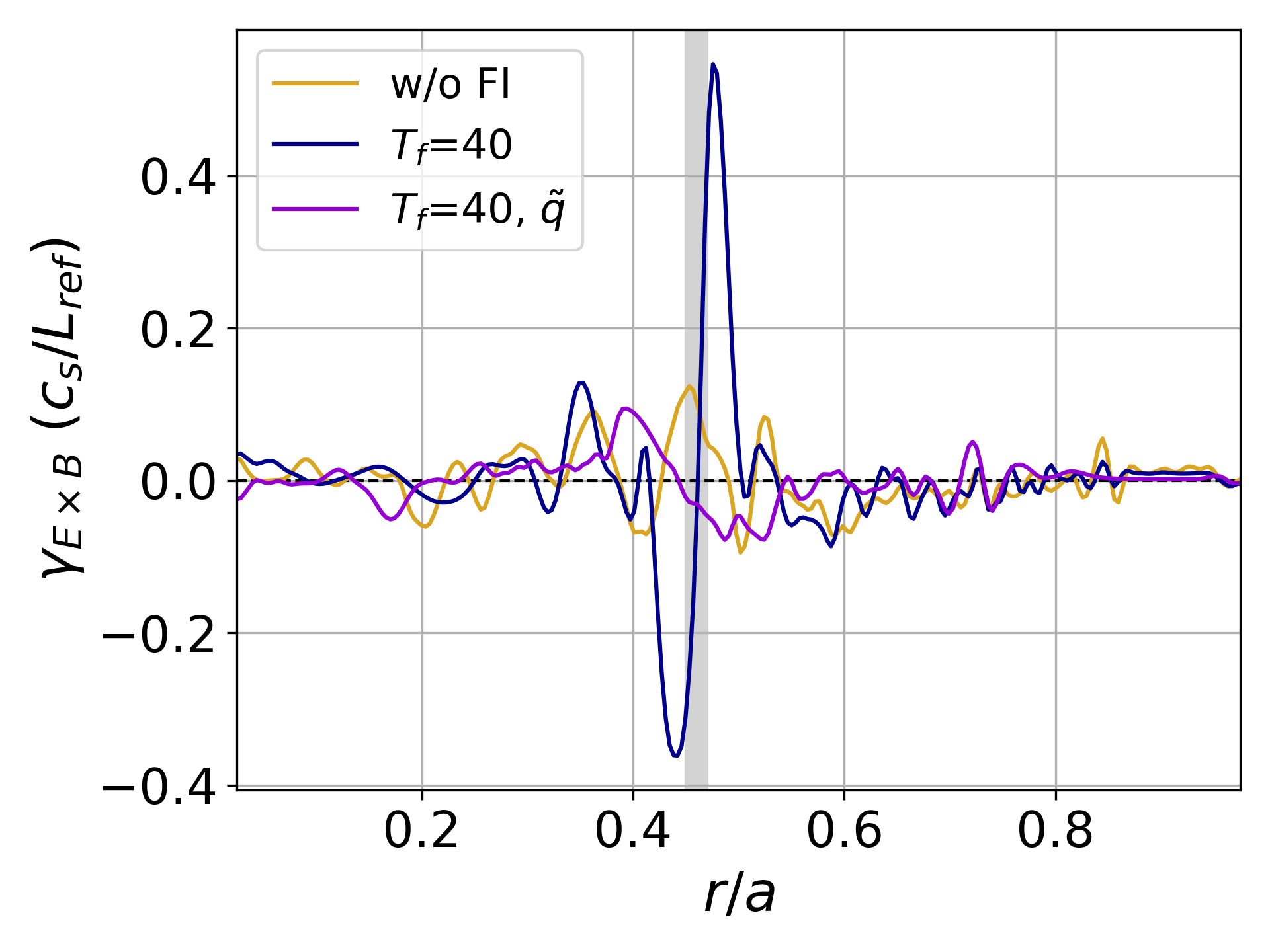}
%\caption{}
\end{center}
\end{subfigure}
\caption{Shearing rate $\gamma_{E\times B}$ radial profile for three scenarios. Reference ones without and with ($T_f=40$) fast ions and one with $T_f=40$ and shifted $\tilde{q}$ profile.}
\label{figure8}
\end{figure*}

\section{Turbulent transport suppression at rational surfaces} \label{sec:transp_supp}
\vspace{-0.5pc}
In the previous section, we have identified two effects the fast ions have on the linear instabilities which develop in the chosen setup. We now proceed to study fully-developed turbulence dynamics through nonlinear simulations. In this runs, the $E\times B$-nonlinearity of the Vlasov equation is retained, allowing for coupling between different $n$ modes. We start by performing global simulations for the case without fast ions. We then include them changing $T_f$ between 10 and 40. Heat fluxes radial profiles obtained for all the species are reported in figure \ref{figure4}. By looking at these results, we see that two different mechanism are acting to regulate the turbulent fluxes. The first one acts along the whole domain and reduces the fluxes almost independently on $r/a$ as $T_f$ increases (see e.g. the transition from yellow to cyan line in figure \ref{figure4} (a)). This stabilization is given by the superposition of dilution and linear quasi-resonant mechanisms described in section \ref{subsec:qr_effect}. Since we already know that the resonance sets around $T_f=2$ while the lowest $T_f$ value we consider here is $T_f=10$, we understand why fast ions fluxes keep decreasing as $T_f$ increases. \\
The second kind of transport reduction we observe is localized around the $q=1$ rational surface (shaded region in figure \ref{figure4}) and intensifies with increasing $T_f$. This lowering of turbulence levels is very strong, and leads to a reduction of the ion heat flux of up to 90\% of the value obtained without fast particles. 
%Given the localized nature of this stabilization, linear local simulations are performed to verify if the linear physics of the system differs from the one found in section \ref{sec:linear} for this specific position. Results for these runs are reported in figure \ref{figure4bis}. We see that the growth rate and frequency trends with varying $T_f$ are the same discussed in the previous section. Thus, the strong transport reduction observed around the rational surface cannot be ascribed to a linear effect. 
To better understand the nature of this stabilization, we study the behavior of the $E\times B$ shearing rate, defined as 
\begin{equation} \label{Eq3}
    \gamma_{E\times B}=-r\partial_r\Biggl(\dfrac{\partial_r\langle\phi\rangle}{rB}\Biggl),
\end{equation}

\begin{figure*}[ht!]
\noindent\begin{subfigure}[t]{0.33\textwidth}
\begin{center}
\includegraphics[width=\textwidth]{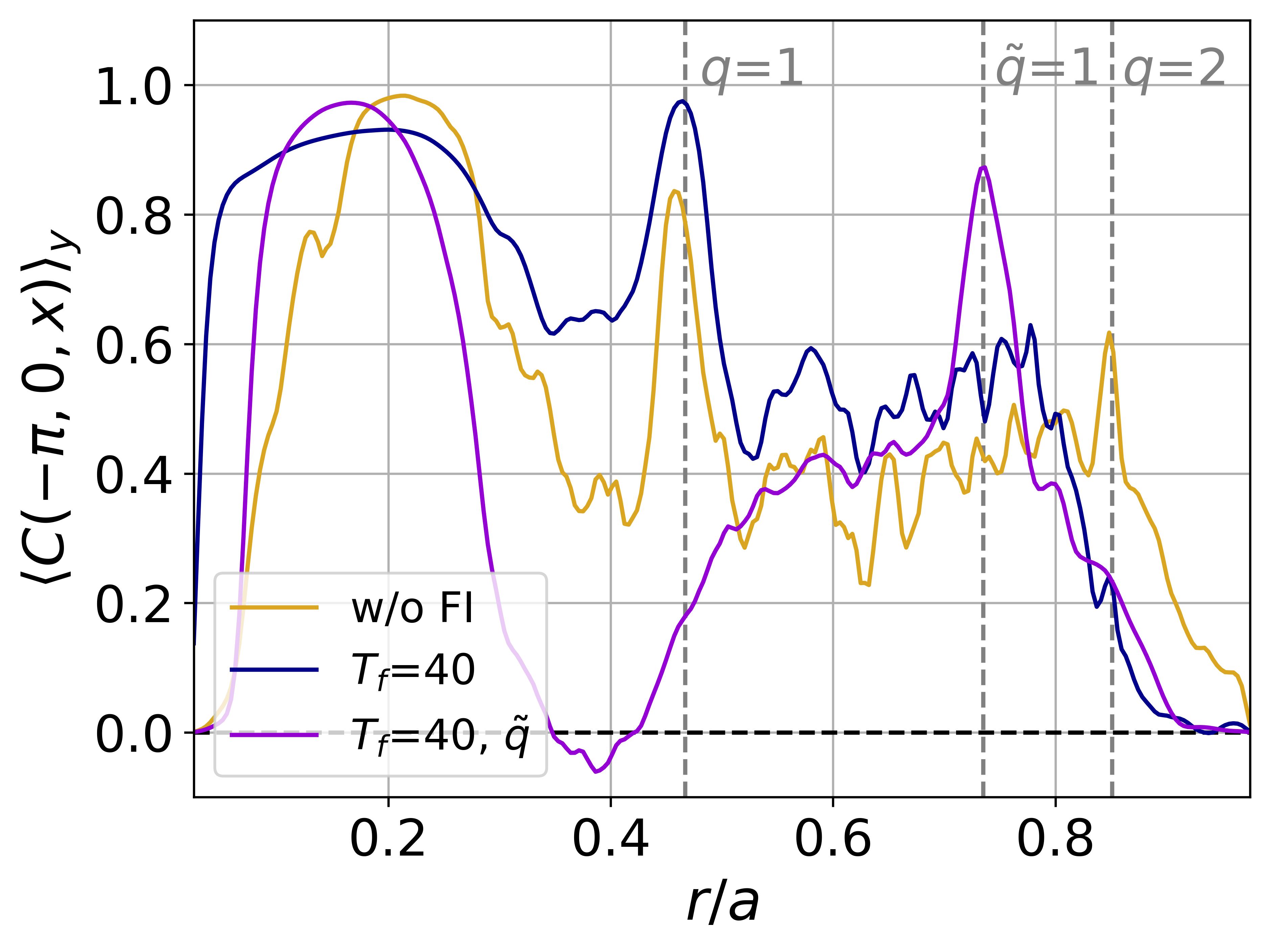}
\end{center}
\end{subfigure}
\caption{Normalized parallel correlation computed from Eq.(\ref{Eq4}) for the three cases compared in this section. Dashed gray lines represent $q=1$ and $q=2$ rational surfaces for the $q$ profile. The dotted gray line represents $\tilde{q}=1$ for the shifted $\tilde{q}$ profile.}
\label{figure9}
\end{figure*}

\noindent with $\langle\:\cdot\:\rangle=\langle_{\textcolor{white}{y}}\langle\:\cdot\:\rangle_y\rangle_z$ the flux-surface average. This quantity gives a measure of the shearing turbulent eddies are ongoing radially \cite{smolyakov}. Radial profiles for this quantity for the considered cases are reported in figure \ref{figure5}. By looking at the plot, we clearly see a monotonic increase of $\gamma_{E\times B}$ amplitude with $T_f$ around the rational surface. By comparing these values with the growth rates reported in figure \ref{figure2bis} (a), we see that the condition $\gamma_{E\times B}>\gamma$ holds when turbulence suppression is observed. Furthermore, the shearing rate is negligible for the reference case without fast particles. We can therefore assert that the turbulence reduction observed at $q=1$ is linked to the generation of a strong local $E\times B$ shearing layer when fast particles are included. Now we have to investigate what is generating this structure and how fast particles participate in its development. In the remaining of this section, we describe how such zonal modes are generated independently on fast particles through a mechanism of self-interaction of the turbulence. Afterwards, the dilution role fast ions play in lowering the generation threshold of this structure is presented. We will see that this allows the shearing layer to efficiently develop and suppress turbulence only under certain conditions. 

\vspace{-0.5pc}
\subsection{Zonal structures generation through eddy self-interaction} \label{subsec:ZF_generation}
\vspace{-0.5pc}

We want to investigate the role the $q=1$ rational surface has on the development of the shearing structure. To do so, we perform a simulation using the $\tilde{q}$ safety factor profile (see figure \ref{figure1} (c)) in the $T_f=40$ setup. We recall that this profiles as a $\tilde{q}=1$ surface at $r/a=0.74$. Thus, in this setup, gradient maxima and first order rational surface do not coincide anymore. Fluxes and shearing rate profiles are reported in figure \ref{figure7} and \ref{figure8}, where they are compared with the ones for the case without fast ions and with $T_f=40$. We observe that by shifting the $q=1$ rational surface from $r/a\approx 0.5$ -- i.e. the position at which thermal species gradients peak -- we do not observe the stabilizing effect anymore (figure \ref{figure7} (a)). Furthermore, the shearing structure completely disappears, as we can see from figure \ref{figure8}. In order to explain this generation of shearing zonal structure at rational surfaces, we invoke the mechanism of turbulence self-interaction \cite{weikl,ball_1,ajay_1,ajay_2,volcokas_2,digiannatale}. In this paradigm, turbulent eddies with toroidal mode number $n$ can efficiently generate an $n=0$ zonal structure by interacting with themselves through three wave-coupling. This requires a superposition of the eddies in real space, which is achieved when a turbulent coherent structure is able to "bite its own tail" after a whole turn around the torus. The condition is fulfilled at low integer rational surfaces. Indeed here, given a reference position $z_0$ along a field line, after a given number of toroidal and poloidal turns an eddy reconnects with itself exactly at $z_0$. The lower the order of the rational surface, the shorter the parallel length the eddy has to span in order to "bite its own tail". Therefore, the probability for it to de-correlate before being able to self-interact is minimized. We see that what we observe in figure \ref{figure8} is the generation of an $n=0$ structure at $q=1$ due to this turbulence self-interaction. To verify this, we compute the parallel correlation of the eddy with itself between two different positions along the field line, namely $z=-\pi$ and $z=0$. This has been previously done in \cite{ball_1} to study a similar phenomenon. The  electrostatic potential correlation between two positions $\mathbf{x_1}$ and $\mathbf{x_2}$ is defined as 
\begin{equation} \label{Eq4}
    C(\mathbf{x_1},\mathbf{x_2})\equiv\dfrac{\langle\delta\phi_{NZ}(\mathbf{x_1})\delta\phi_{NZ}(\mathbf{x_2})\rangle_t}{\sqrt{\langle\delta\phi_{NZ}^2(\mathbf{x_1})\rangle_t\langle\delta\phi_{NZ}^2(\mathbf{x_2})\rangle_t}}.
\end{equation}
Here, $\delta\phi_{NZ}=\phi_{NZ}-\langle\phi_{NZ}\rangle_t$, where $\phi_{NZ}=\phi-\langle\phi\rangle$ is the non-zonal component of the electrostatic potential. Normalized values of this quantity for $\mathbf{x_1}=(x,y,-\pi)$, $\mathbf{x_2}=(x,y,0)$ averaged in $y$ are reported in figure \ref{figure9} for each of the cases studied in figure \ref{figure7}, \ref{figure8}. We see that $\langle C(-\pi,0,x)\rangle_y$ represents the electrostatic potential correlation along a given field line computed on the flux surface labeled by $x$. Focusing only on the $q=1$ rational surface (dashed gray line), we see that the correlation peaks here only for the cases with reference $q$ profile, both with and without fast ions. Moreover, $C$ is much lower for the case with shifted $\tilde{q}$. Notice that, for the case without fast particles, $C$ also has a minor peak at $q=2$ ($r/a=0.85$). These observations confirm our hypothesis on the nature of the shearing structure developing in correspondence to rational surfaces. Furthermore, the presence of fast ions does not influence the correlation of the mode with itself at the rational surface. \\
Figure \ref{figure9} shows high values of the potential parallel correlation in the region $0.1\lesssim r/a\lesssim0.3$ for all the considered cases. 
This observations can be linked to an increase of the eddy size along the parallel direction $z$. As noticed in section \ref{subsec:setup}, both $q$ and $\tilde{q}$ safety factor profiles present an almost-zero shear region inside $r/a\approx 0.3$. Low values of $\hat{s}$ are known to induce eddy elongation along the parallel direction, as shown in \cite{volcokas_3}. Here, the authors derive the scaling $\ell_\parallel\sim\hat{s}^{-1}$ for the eddy parallel length $\ell_\parallel$. Therefore, this leads to an increase in the eddy self-correlation even at non-integer rational surfaces. By looking at figure \ref{figure8}, we see that no shearing layer is generated in this region even if such high values of $C$ are reached. This can be addressed to the difference in turbulence intensity we obtain when comparing $r/a=0.25$ (took as representative position for this region) and $q=1$. Indeed, a linear local analysis shows that, for the $n=25$ mode in the setup without fast ions, $\gamma_{q=1}\sim 10\gamma_{r/a=0.25}$ holds. Nonlinearly, turbulent heat fluxes (figure \ref{figure7}) are much lower at $r/a=0.25$ if compared to the ones achieved at the first order rational surface. The considerations explain why no zonal structure generation is registered inside $r/a\approx0.3$ despite values of the turbulence parallel correlation as high as the ones at $q=1$ are obtained here.

\begin{figure*}[ht!]
\noindent\begin{subfigure}[t]{0.33\textwidth}
\begin{center}
\includegraphics[width=\textwidth]{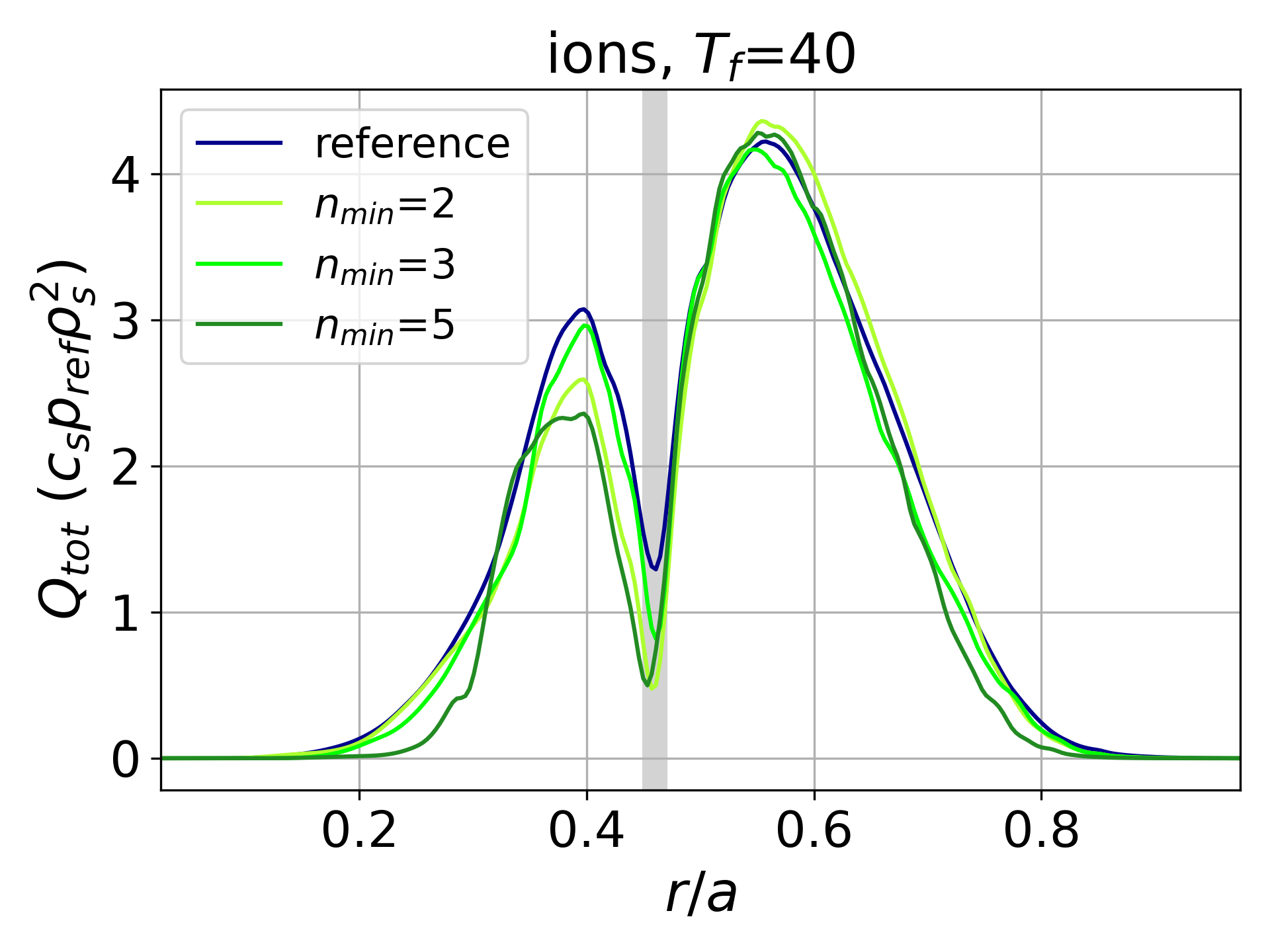}
\caption{}
\end{center}
\end{subfigure}
\noindent\begin{subfigure}[t]{0.33\textwidth}
\begin{center}
\includegraphics[width=\textwidth]{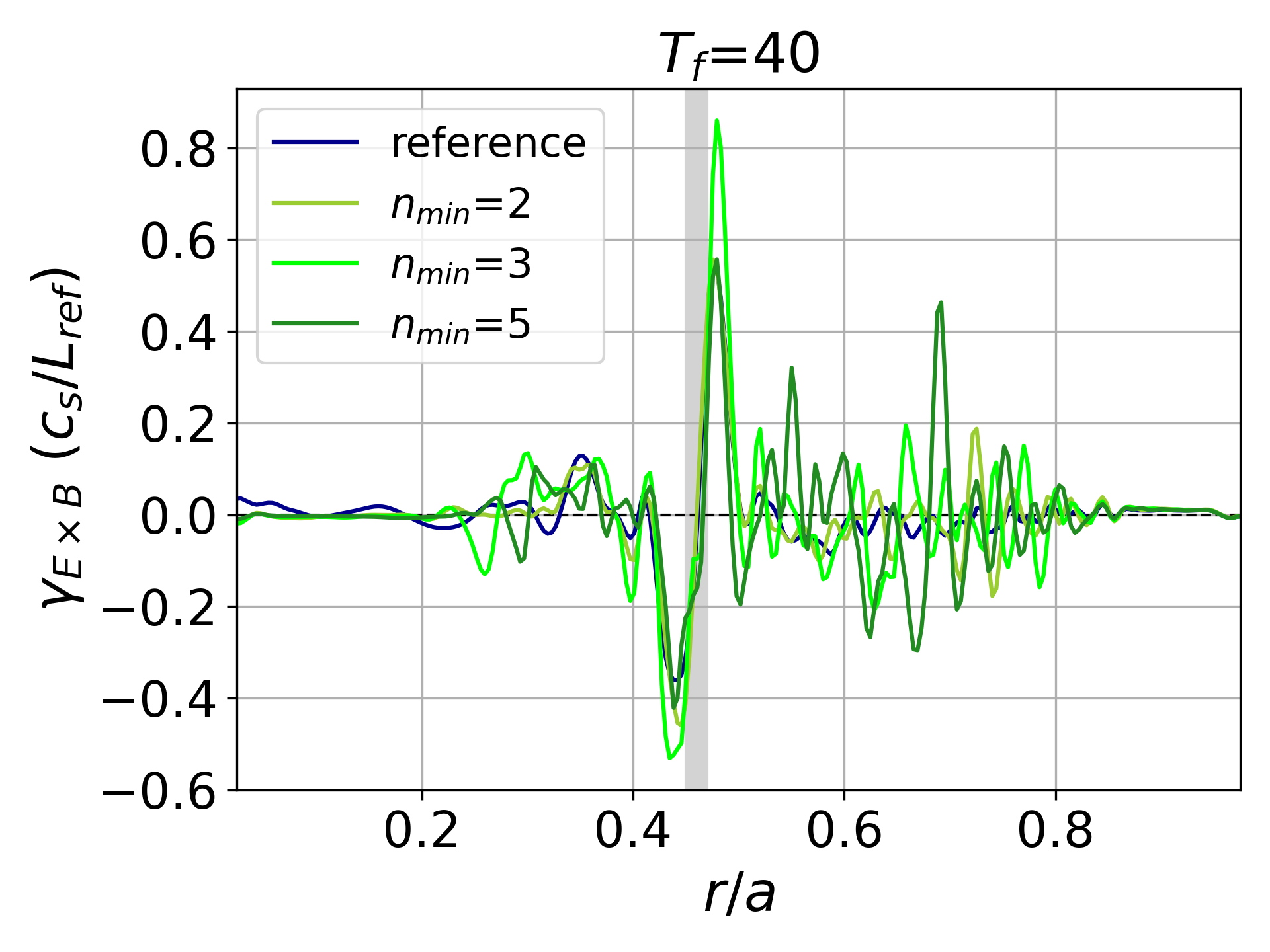}
\caption{}
\end{center}
\end{subfigure}
\noindent\begin{subfigure}[t]{0.33\textwidth}
\begin{center}
\includegraphics[width=\textwidth]{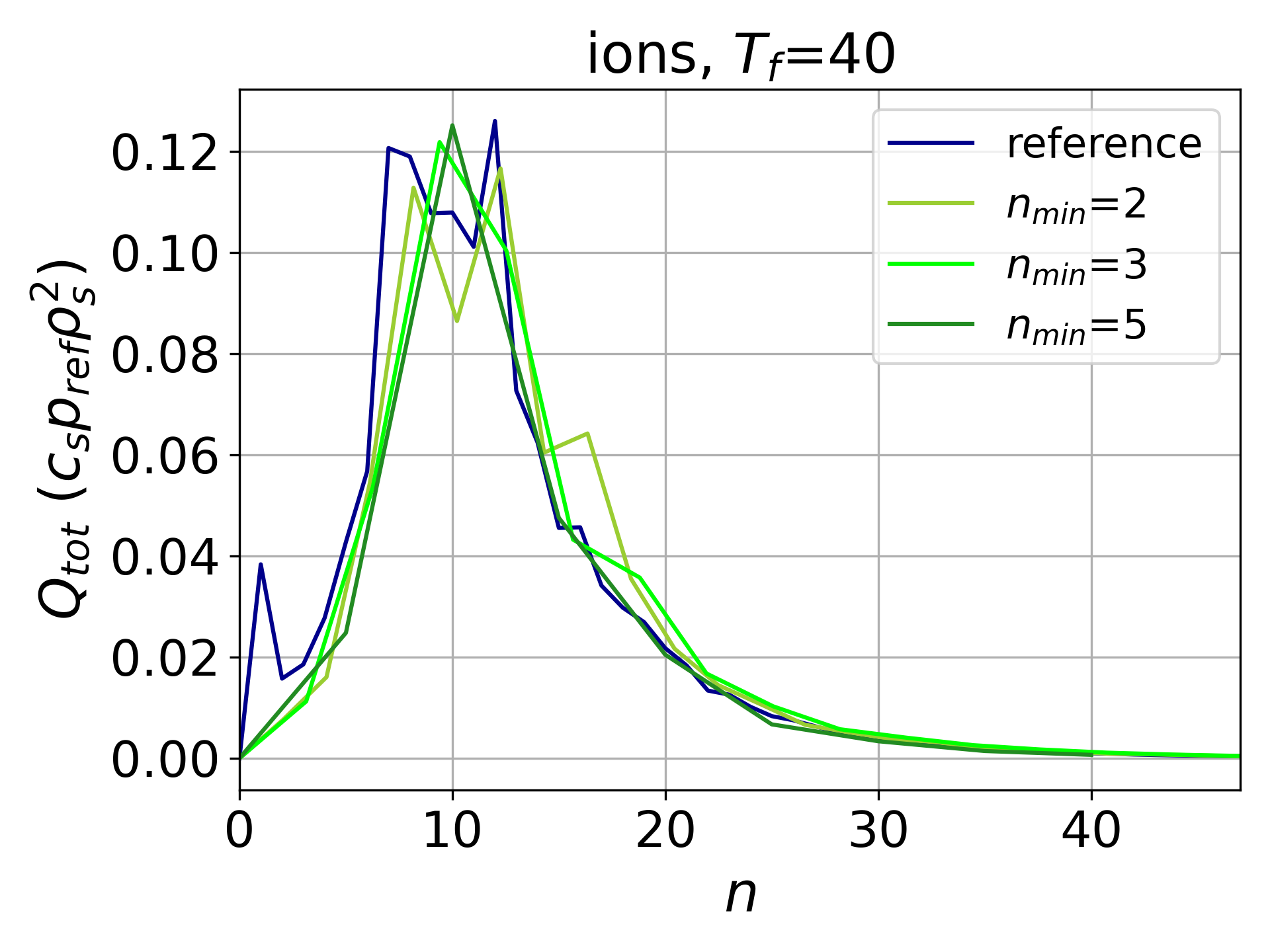}
\caption{}
\end{center}
\end{subfigure}
\caption{(a) Turbulent total ion heat fluxes, (b) shearing rate profiles and (c) ion heat flux spectra for different $n_{min}$ values at $T_f=40$.}
\label{figure6}
\end{figure*}

\begin{figure*}[ht!]
\noindent\begin{subfigure}[t]{0.33\textwidth}
\begin{center}
\includegraphics[width=\textwidth]{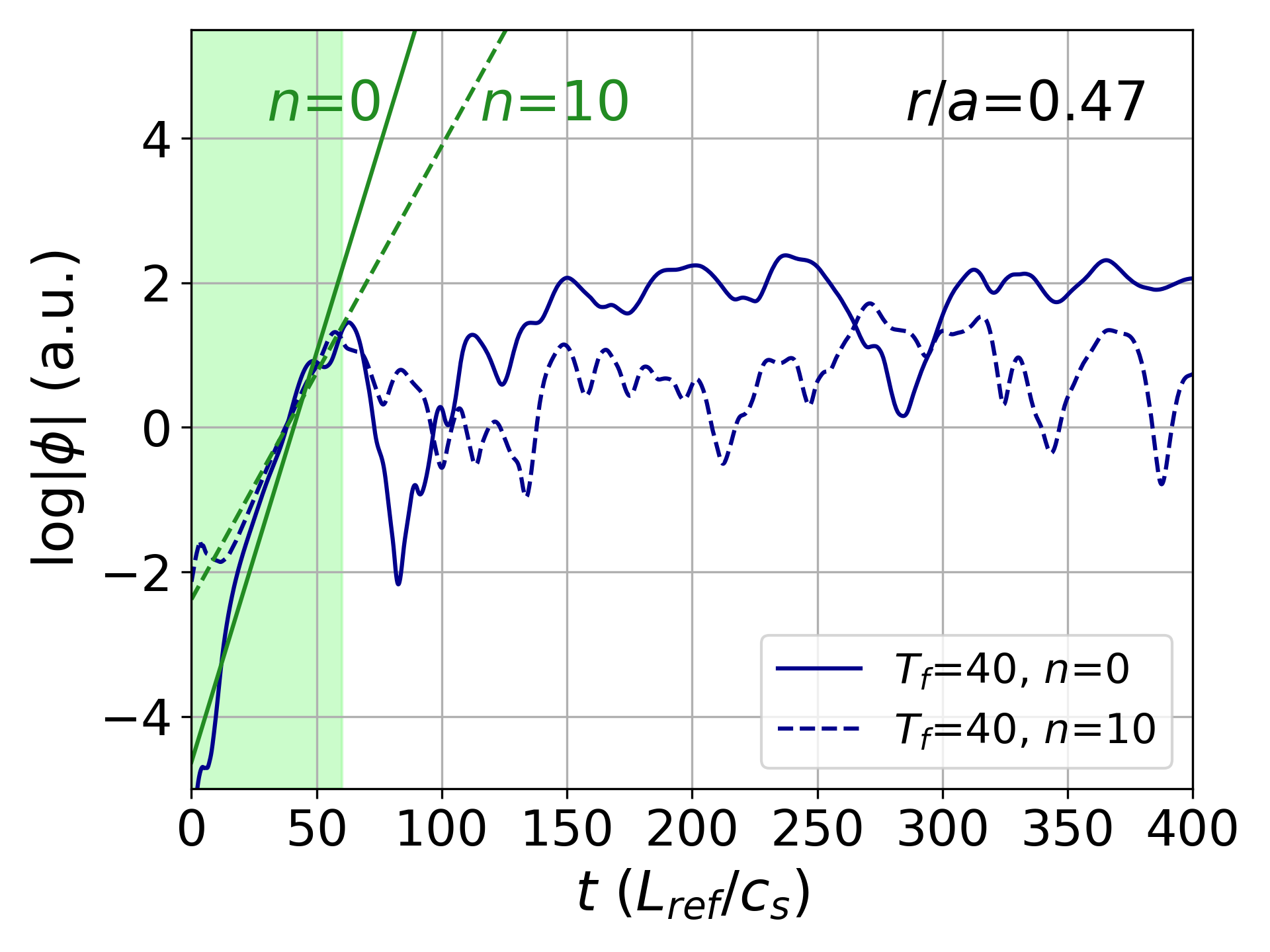}
\end{center}
\end{subfigure}
\caption{Time traces of $\textrm{log}|\phi|$ for $T_f=40$ at $r/a=0.47$ ($q=1$). The shaded region represents the linear growths phase of $|\phi|_{n=0}$ and $|\phi|_{n=10}$. Here, time traces are fitted with a linear regression. Results of the fit are reported as green lines.}
\label{figure6bis}
\end{figure*}

\vspace{-0.5pc}
\subsection{Analysis of the self-interaction mechanism for \textit{n}=10} \label{subsec:n_min_5}
\vspace{-0.5pc}

A study of the interaction of the single-$n$ mode with itself to generate the $n=0$ structure is now performed. Before doing so, some remarks are necessary. The picture of self-interacting turbulence generating $E\times B$ structures described above holds independently on the considered eddy mode number $n$. Indeed, on low integer rational surfaces eddies at each $n$ interact with themselves to generate the shearing structure. We want to study this mechanism for only one of the many $n$s involved in the layer generation. Since the reference setup with FI has 47 $n\neq 0$ modes, a reduction of the number of modes included in the description is needed to select the $n$ to focus on. For this purpose, we perform runs with different $n_{min}$ values in the $T_f=40$ setup. It is worth remarking that in nonlinear runs only the modes with $n\in\{0,n_{min},2n_{min},...,(n_{k_y}-1)n_{min}\}$ are included. Results for the runs performed by setting $n_{min}=1$ (reference), 2, 3 and 5 are shown in figure \ref{figure6}. Figure \ref{figure6} (a) and (b) show that the drastic ion flux reduction due to $\gamma_{E\times B}$ generation persists even when the number of modes is reduced by 4/5 ($n_{min}=5$). This is further explained by the results reported in figure \ref{figure6} (c). The plot shows that by progressively increasing $n_{min}$ we are removing modes that only give a minor contribution to the total ion heat flux. We see that in the setup with $n_{min}=5$ the $n=10$ mode is the main driver of turbulent fluxes. With higher $n_{min}$, several single-$n$ modes would self-interact and pile-up to generate the zonal structure. \\
In order to study how the $n=10$ mode drives the development of the $n=0$ one at $q=1$ when $n_{min}=5$, time traces of the electrostatic potential are plotted in figure \ref{figure6bis} at this position. The linear growth phase of the modes is highlighted. Growth rates $\gamma$ are computed in this time interval through a linear regression of the time trace. We obtain $\gamma_{n=0}=1.13\times 10^{-1}$ and $\gamma_{n=10}=0.63\times 10^{-1}$. Therefore, $\gamma_{n=0}\approx 2\gamma_{n=10}$ holds. The most unstable turbulent mode $n=10$ drives the zonal one through interaction with itself (i.e. another $n=10$ mode) at the rational surface. 

\vspace{-0.5pc}
\subsection{Fast ion dilution effect on zonal structures properties} \label{subsec:FI_dilution}
\vspace{-0.5pc} 

\begin{figure*}[ht!]
\noindent\begin{subfigure}[t]{0.33\textwidth}
\begin{center}
\includegraphics[width=\textwidth]{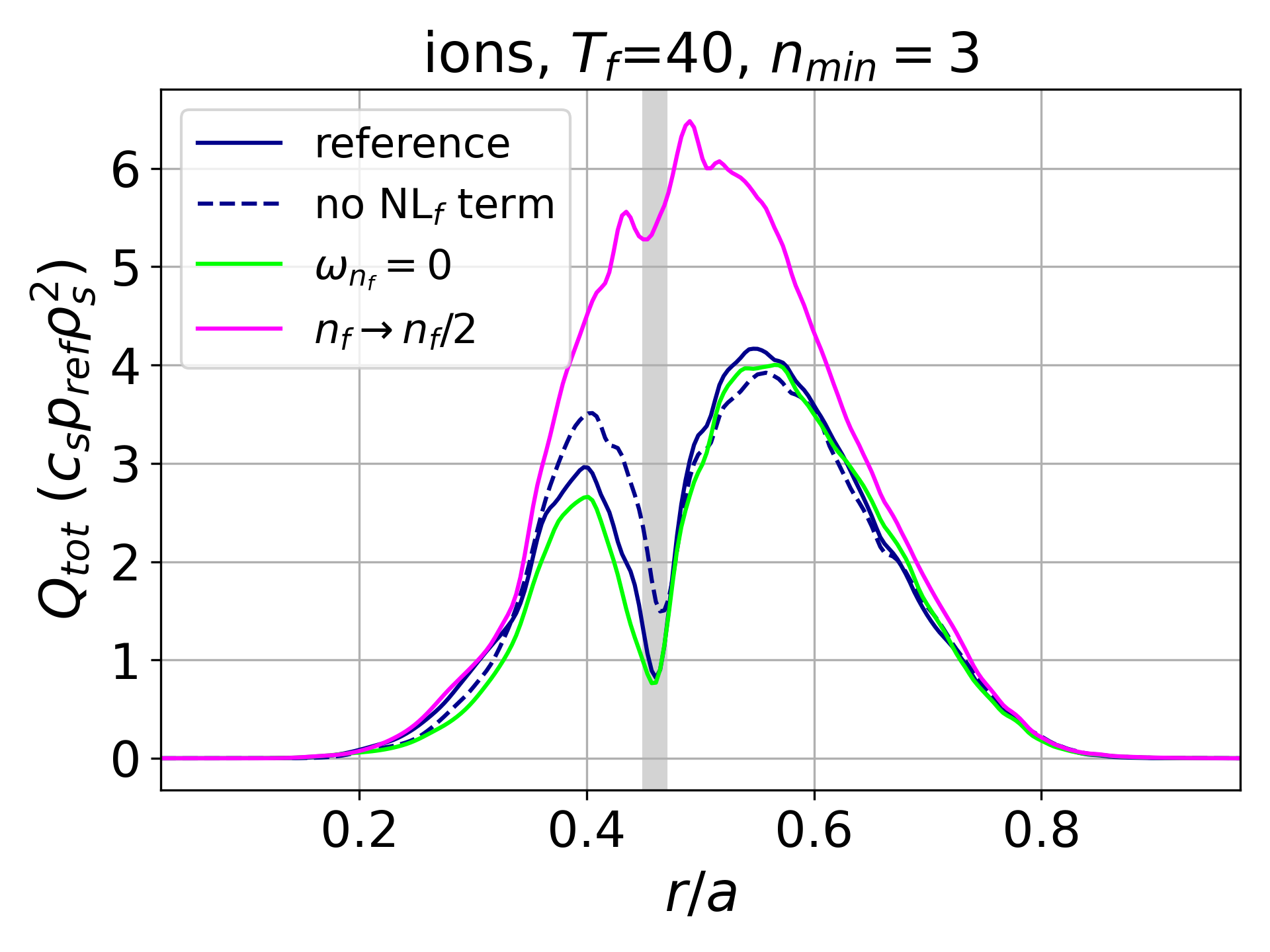}
\caption{}
\end{center}
\end{subfigure}
\noindent\begin{subfigure}[t]{0.33\textwidth}
\begin{center}
\includegraphics[width=\textwidth]{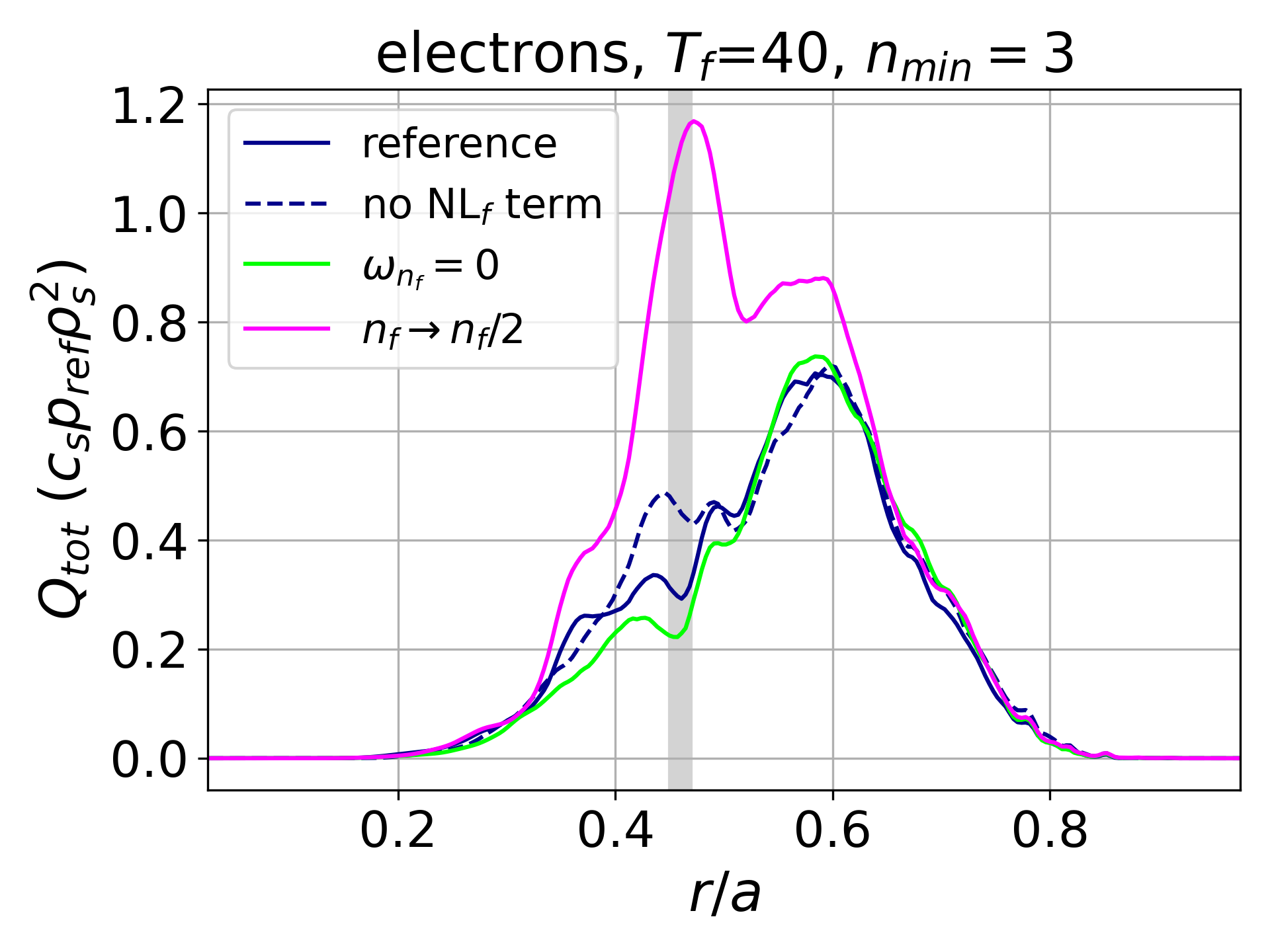}
\caption{}
\end{center}
\end{subfigure}
\noindent\begin{subfigure}[t]{0.33\textwidth}
\begin{center}
\includegraphics[width=\textwidth]{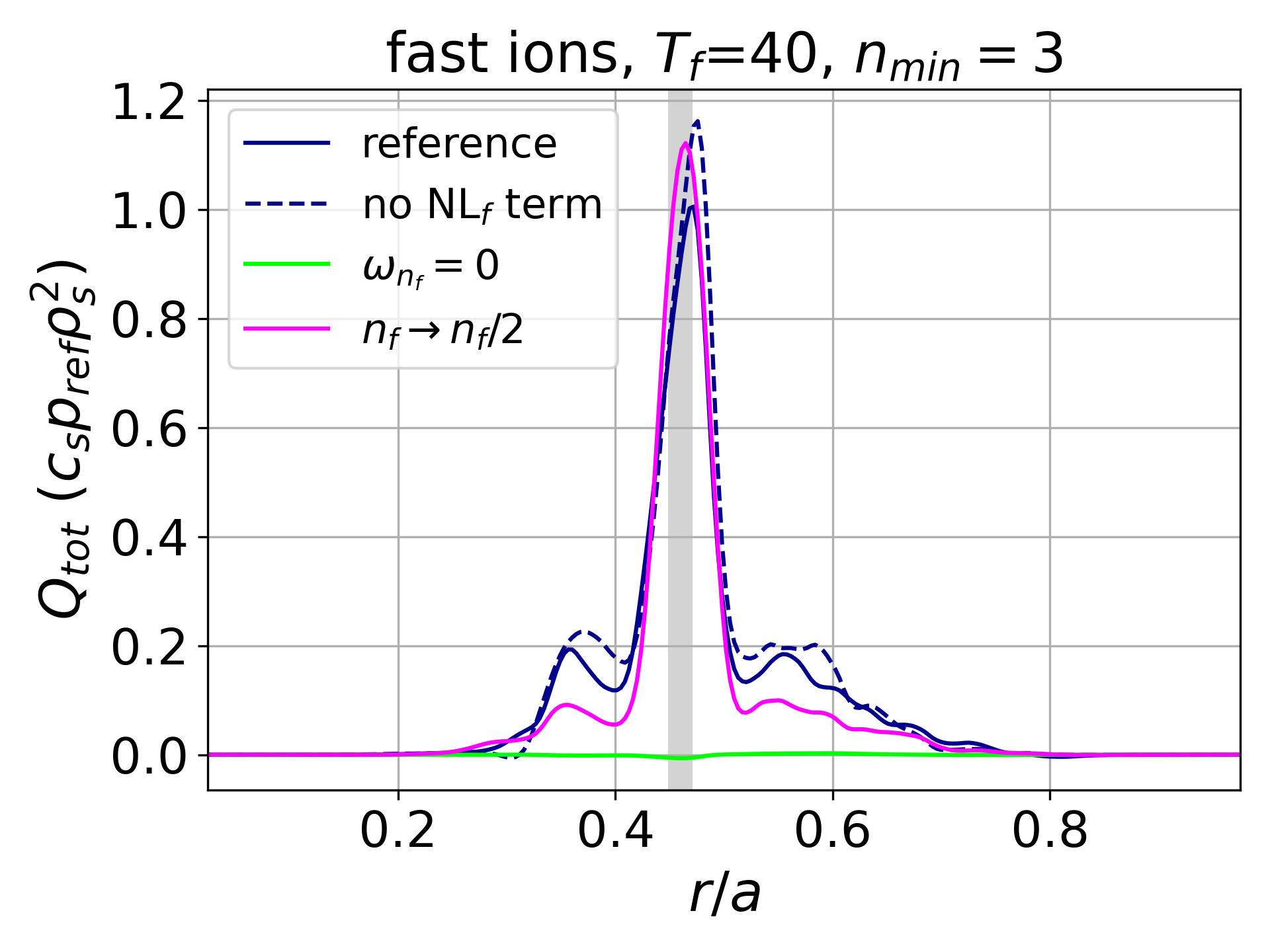}
\caption{}
\end{center}
\end{subfigure}
\caption{Turbulent total heat fluxes for (a) ions, (b) electrons and (c) fast ions species for different FI setups. We consider scenarios where (1) the nonlinear term in the Vlasov equation for fast ions is suppressed (no NL$_f$ term), (2) fast ion density is flat ($\omega_{n_f}=0$) and (3) fast ion density is halved ($n_f\rightarrow n_f/2$).}
\label{figure12-13-14}
\end{figure*}

\begin{figure*}[ht!]
\noindent\begin{subfigure}[t]{0.33\textwidth}
\begin{center}
\includegraphics[width=\textwidth]{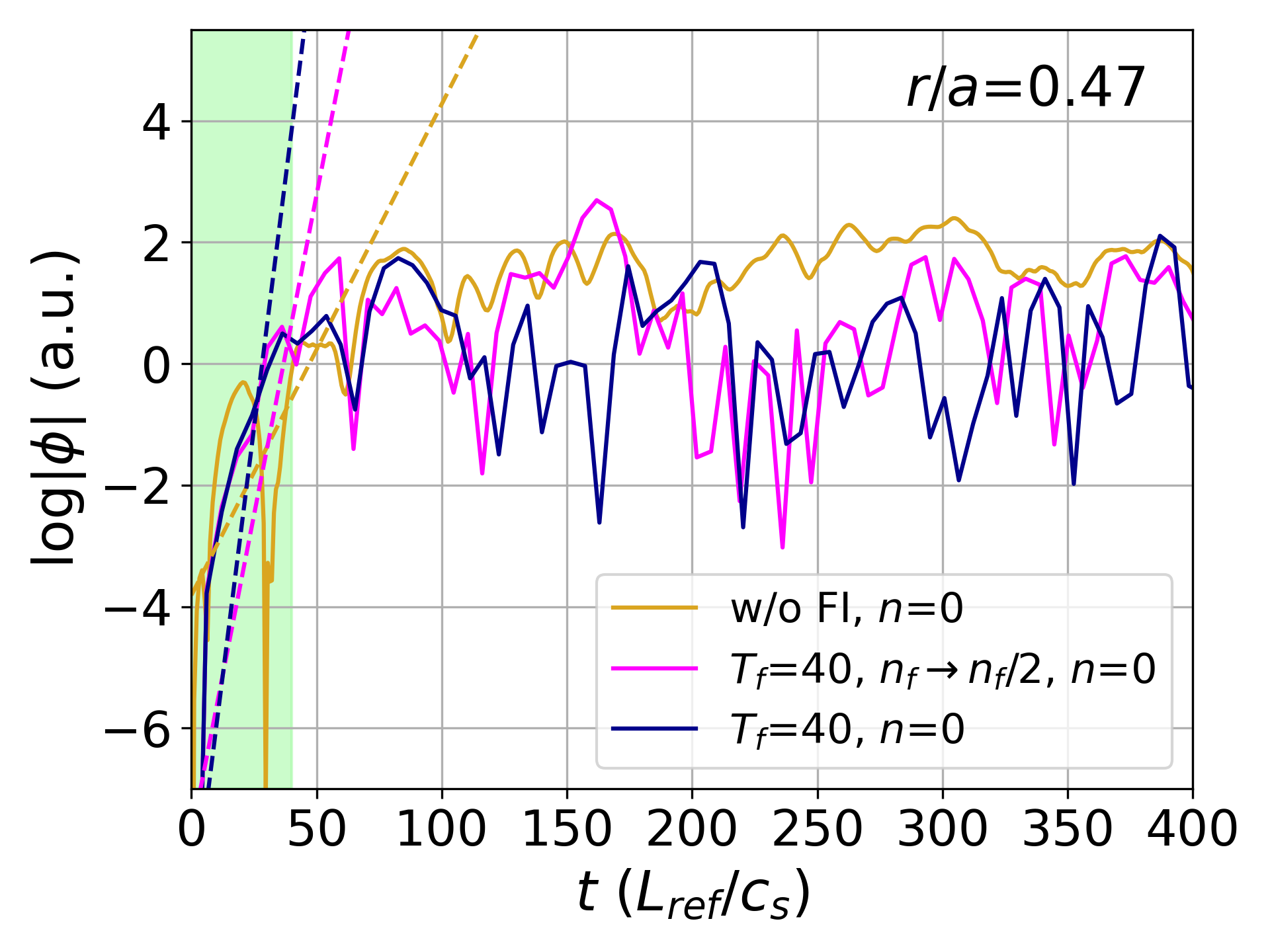}
\caption{}
\end{center}
\end{subfigure}
\noindent\begin{subfigure}[t]{0.33\textwidth}
\begin{center}
\includegraphics[width=\textwidth]{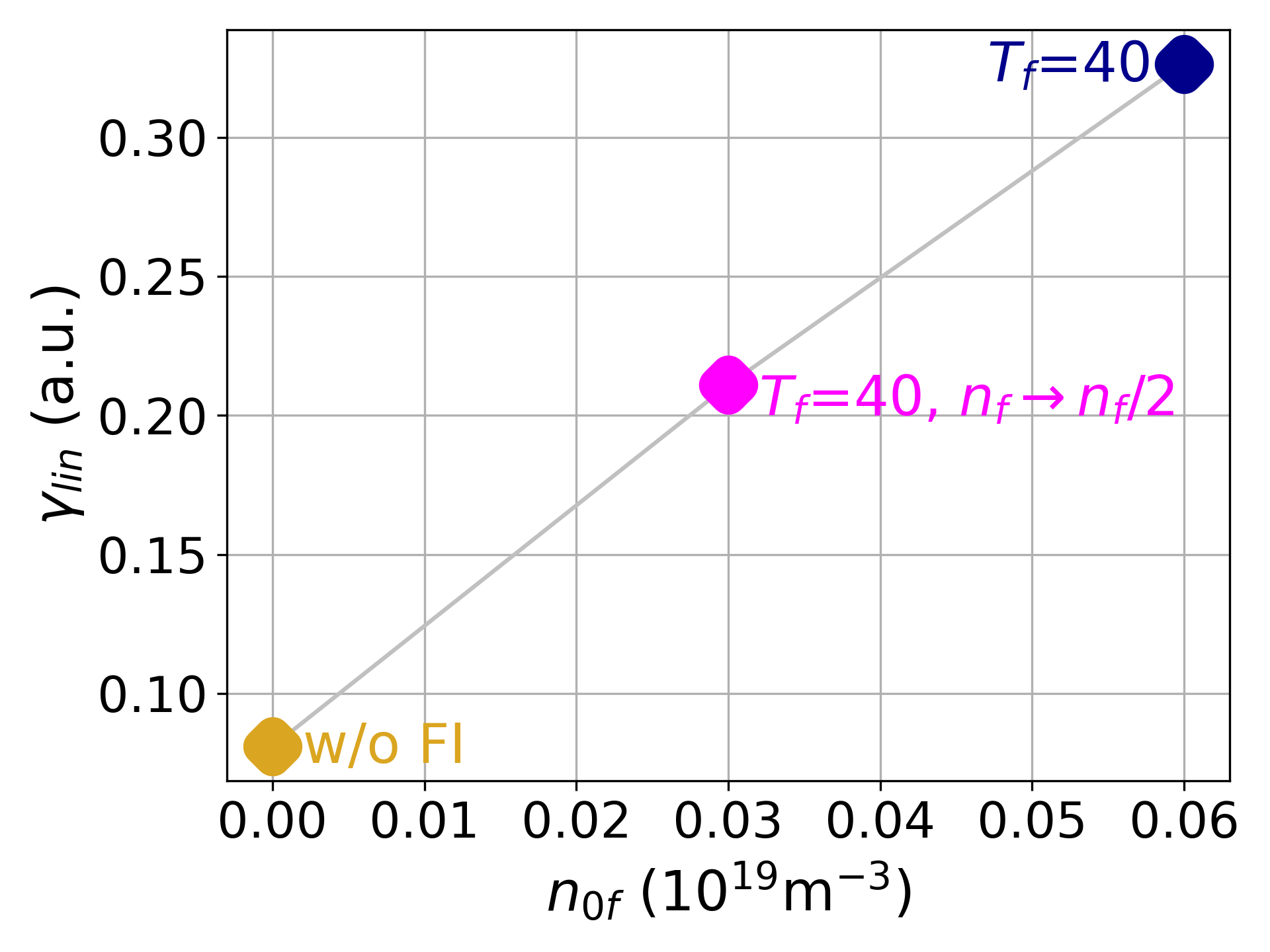}
\caption{}
\end{center}
\end{subfigure}
\caption{(a) Time traces of $\textrm{log}|\phi|_{n=0}$ for different $n_f$ values at $r/a=0.47$ ($q=1$). The shaded region represents the linear growths phase where the profiles are fitted with a linear regression. Results of the fit are reported as dashed lines. (b) Zonal potential linear growth rates computed in the shaded region as a function of $n_{0f}$ ($n_f$ at $q=1$).}
\label{figure16}
\end{figure*}

\begin{figure*}[ht!]
\noindent\begin{subfigure}[t]{0.33\textwidth}
\begin{center}
\includegraphics[width=\textwidth]{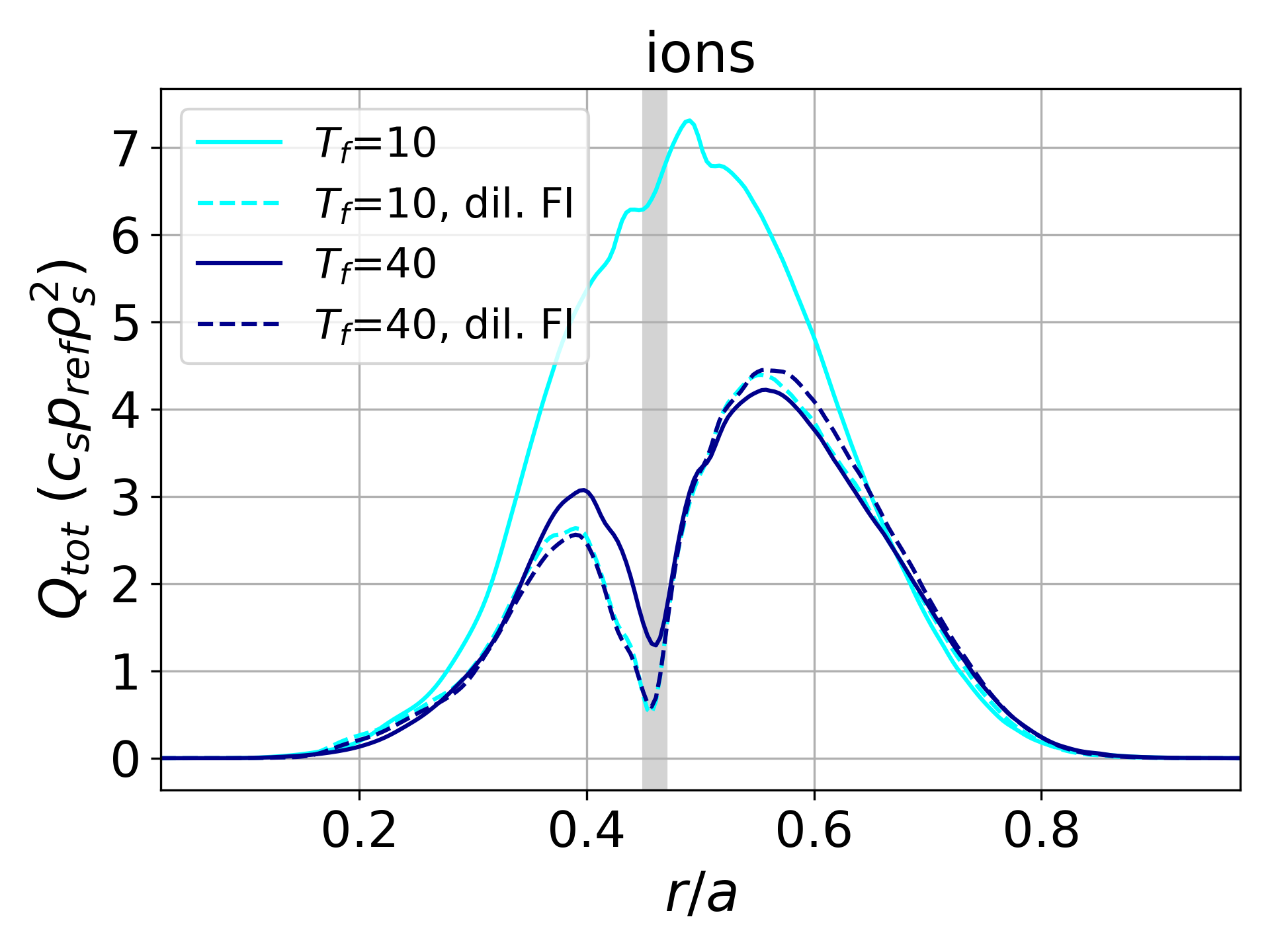}
\caption{}
\end{center}
\end{subfigure}
\noindent\begin{subfigure}[t]{0.33\textwidth}
\begin{center}
\includegraphics[width=\textwidth]{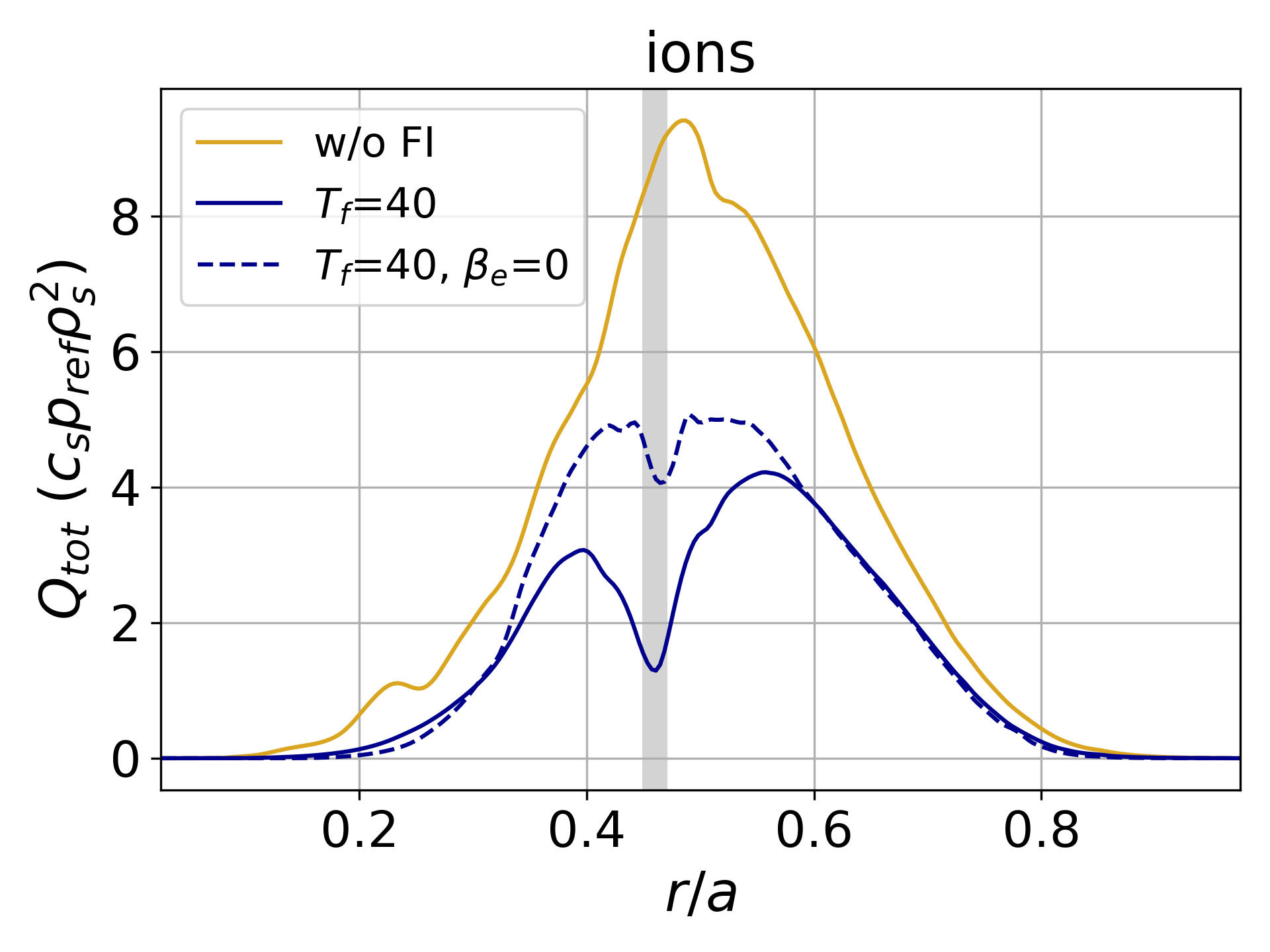}
\caption{}
\end{center}
\end{subfigure}
\noindent\begin{subfigure}[t]{0.33\textwidth}
\begin{center}
\includegraphics[width=\textwidth]{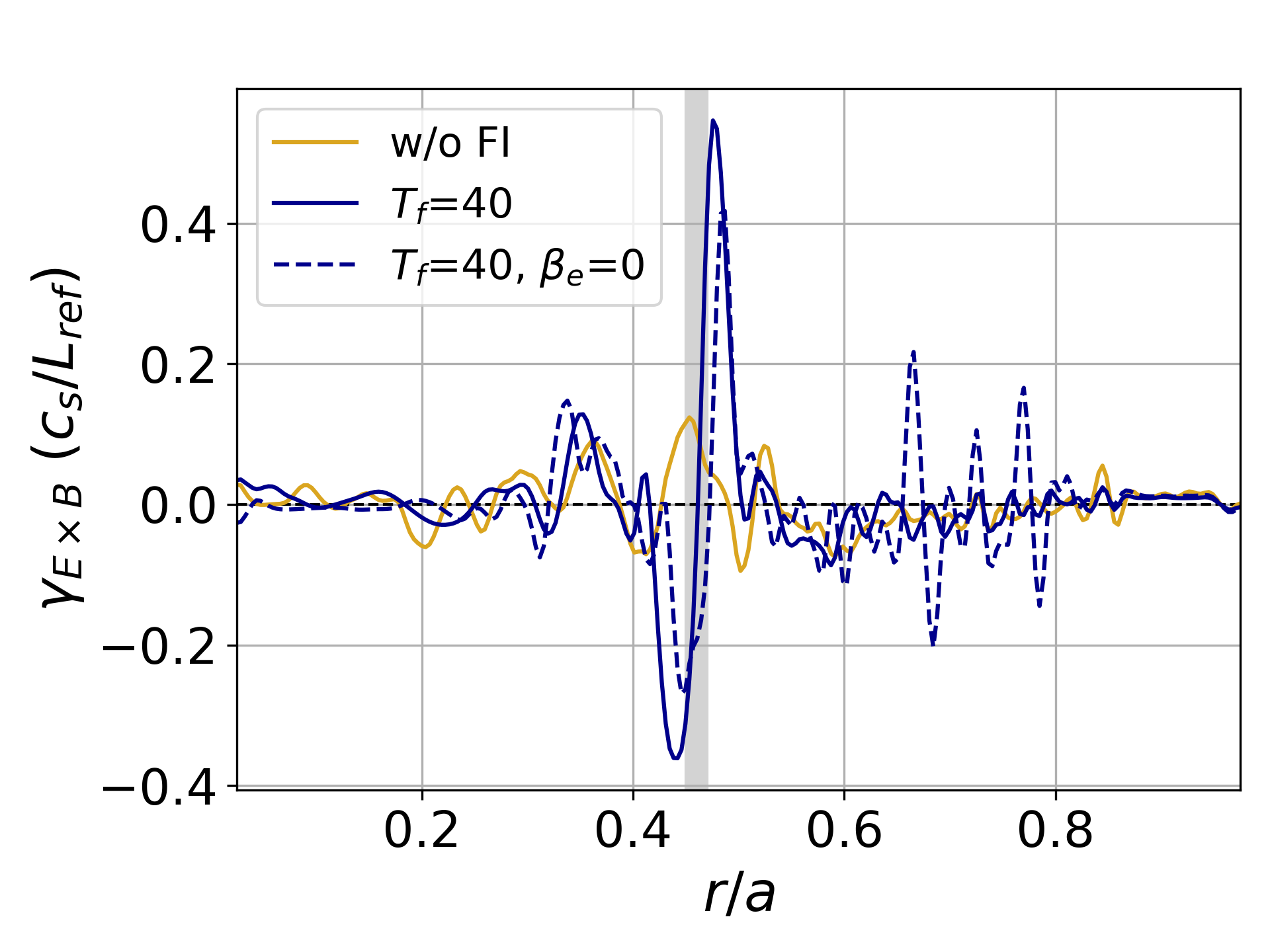}
\caption{}
\end{center}
\end{subfigure}
\caption{Turbulent total ion heat fluxes for (a) $T_f=10$, 40, with "standard" and "diluted" fast ions, and (b) $T_f=40$ with $n_{min}=3$ and $\beta_e=0$, compared with the nominal cases with and without FI. (c) Shearing rate $\gamma_{E\times B}$ radial profile for the same scenarios reported in (b).}
\label{figure15}
\end{figure*}  

\begin{figure*}[ht!]
\noindent\begin{subfigure}[t]{0.33\textwidth}
\begin{center}
\includegraphics[width=\textwidth]{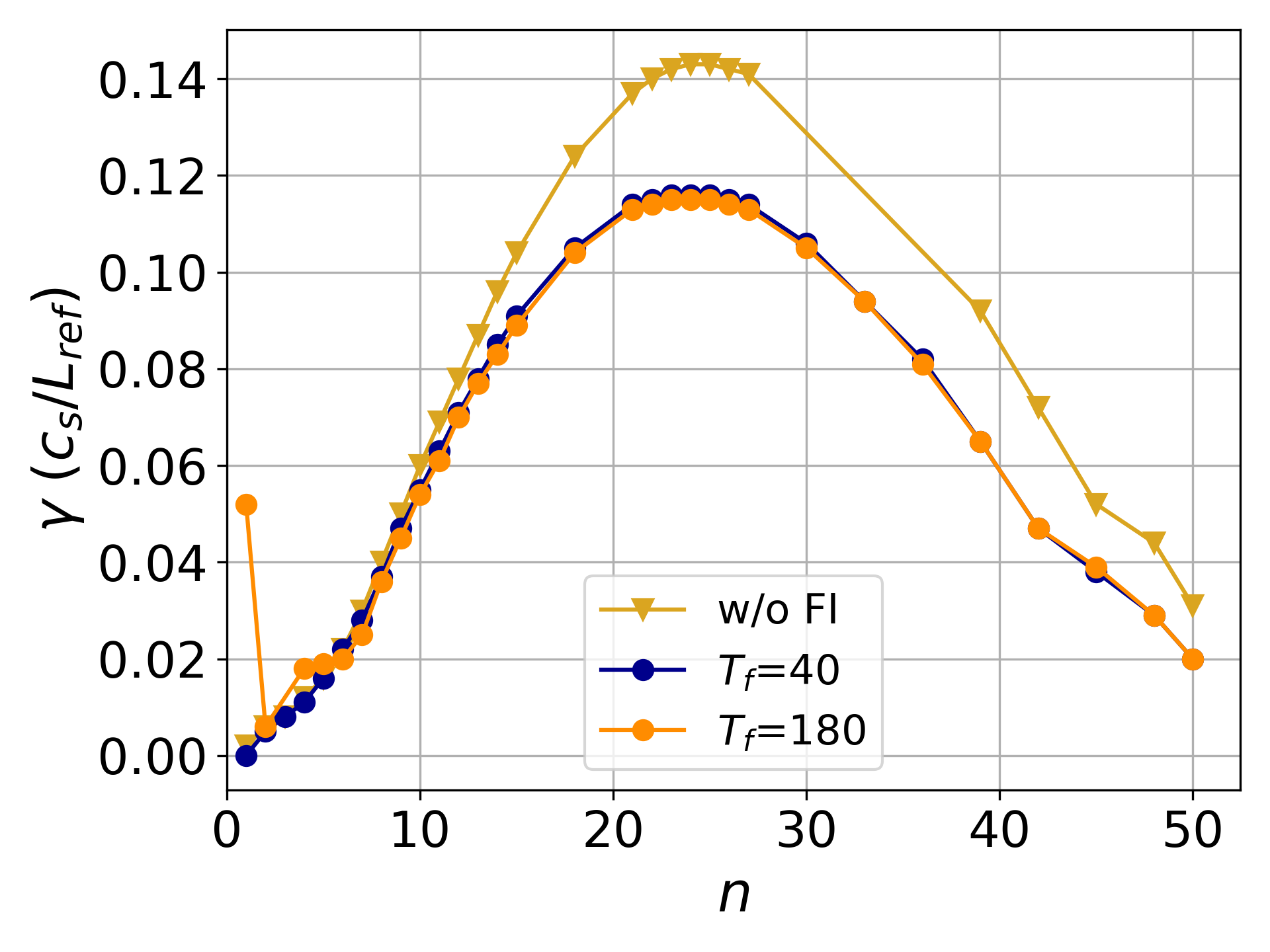}
\caption{}
\end{center}
\end{subfigure}
\noindent\begin{subfigure}[t]{0.33\textwidth}
\begin{center}
\includegraphics[width=\textwidth]{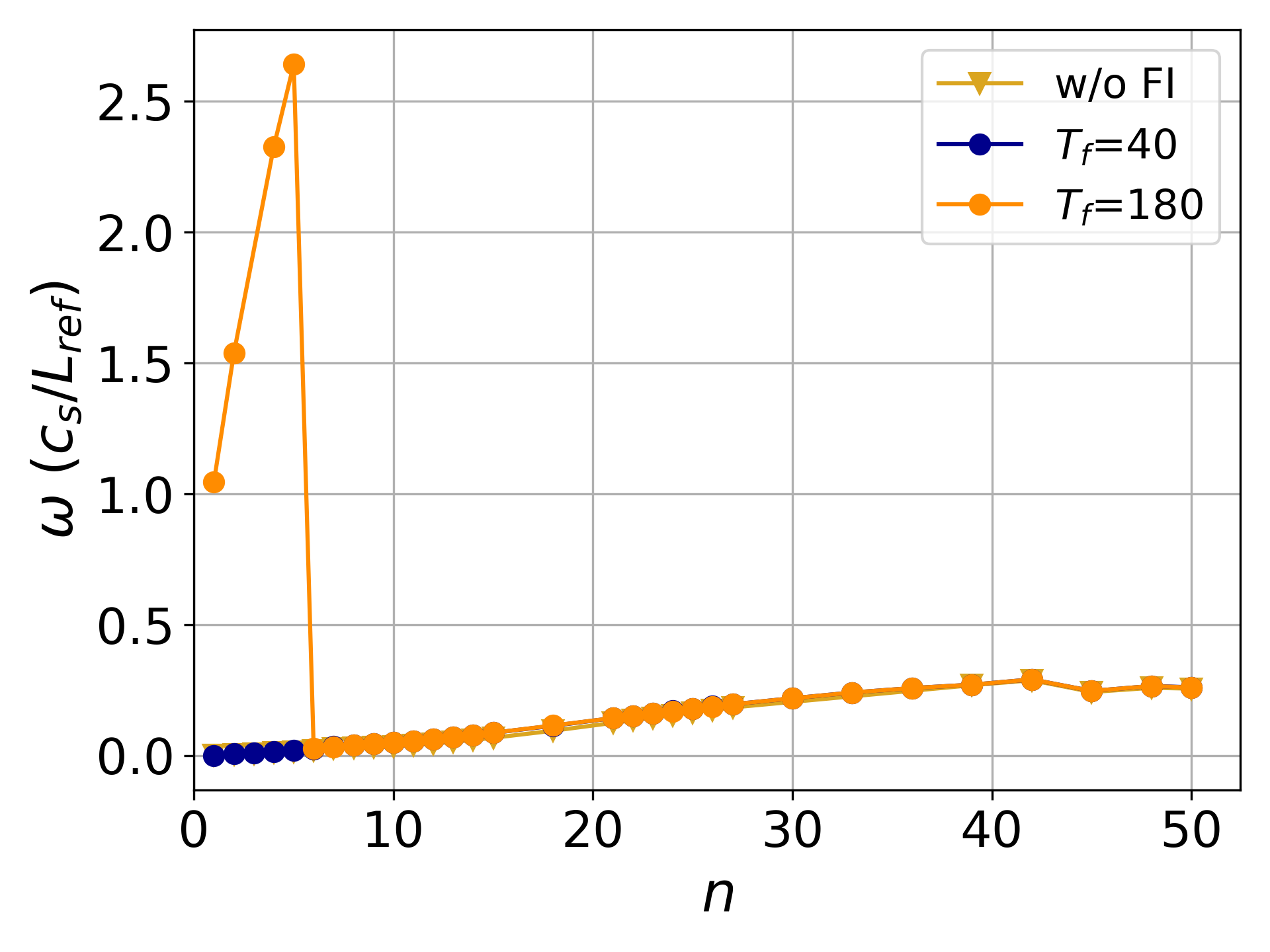}
\caption{}
\end{center}
\end{subfigure}
\noindent\begin{subfigure}[t]{0.33\textwidth}
\begin{center}
\includegraphics[width=\textwidth]{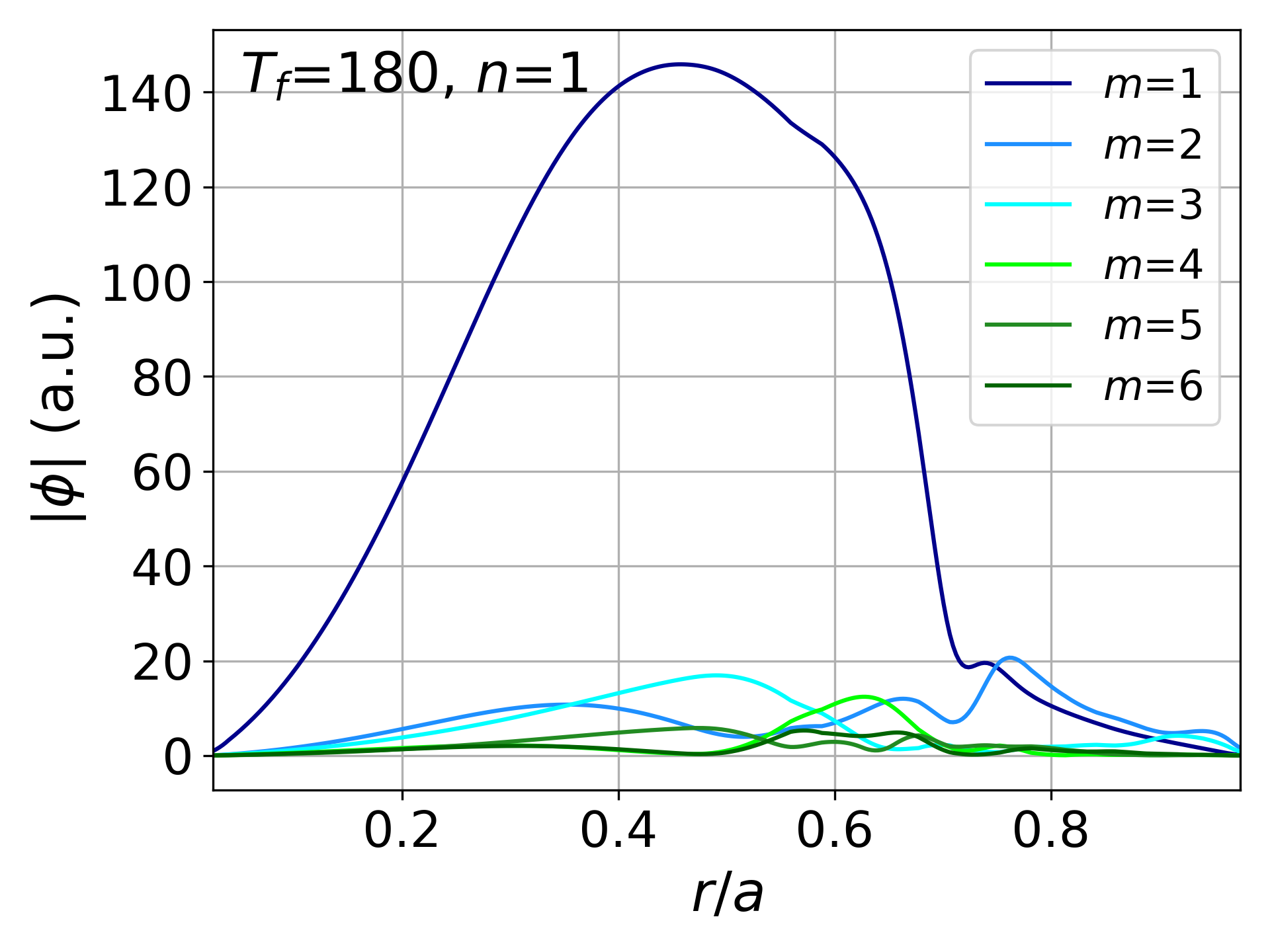}
\caption{}
\end{center}
\end{subfigure}
\caption{Normalized (a) growth rate and (b) frequency radial average for the setup without fast particles and the ones with $T_f=40$, $T_f=180$. (c) Different poloidal mode number ($m$) contributions to the $n=1$ mode profile for $T_f=180$.}
\label{figure17}
\end{figure*}

\begin{figure*}[ht!]
\noindent\begin{subfigure}[t]{0.33\textwidth}
\begin{center}
\includegraphics[width=\textwidth]{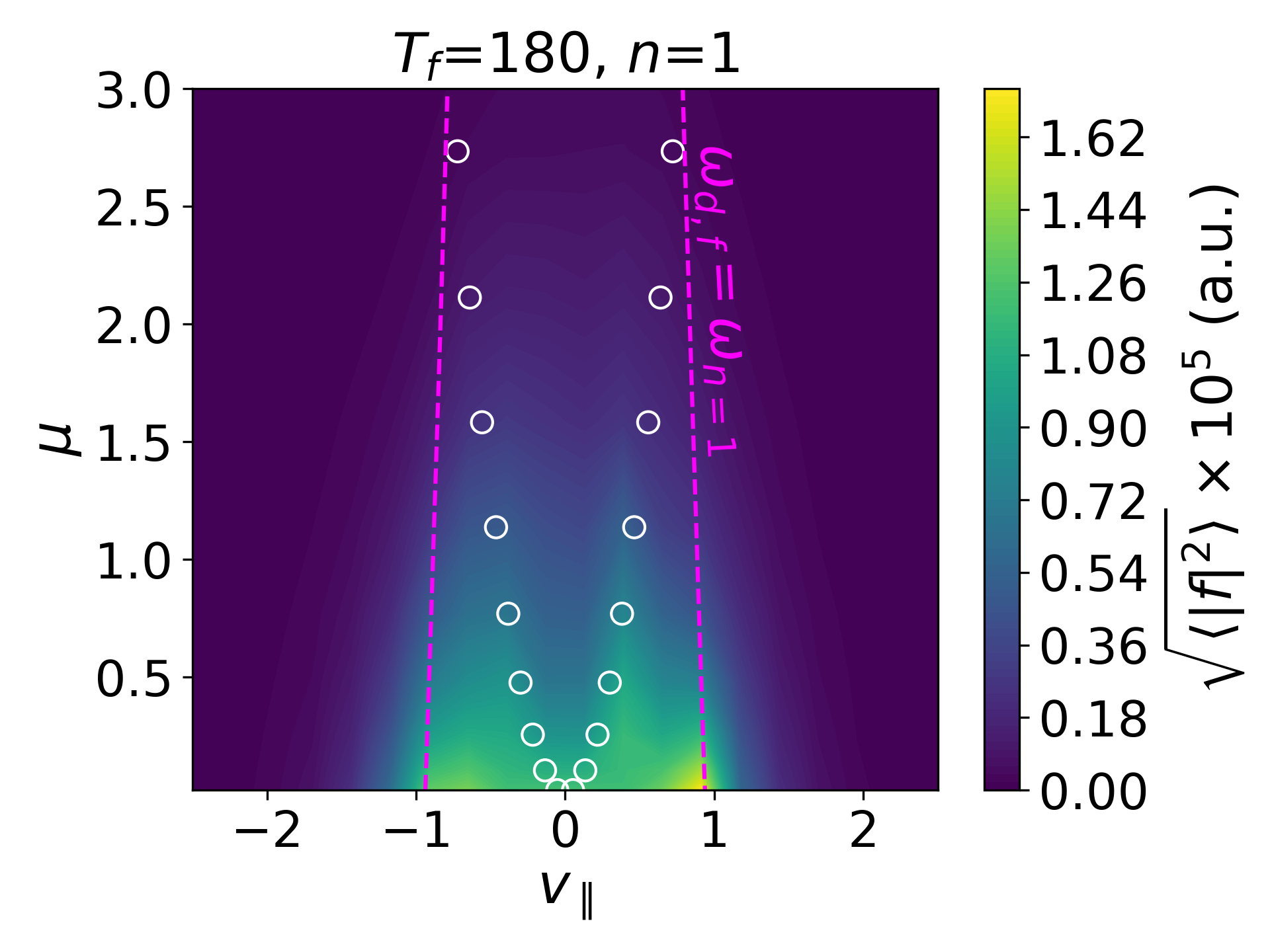}
%\caption{}
\end{center}
\end{subfigure}
\caption{Velocity space plot of the $T_f=180$ fast ions distribution function absolute value. White circles indicate the trapping region boundary at $z=0$. Magenta line corresponds to the $\omega_{d,f}=\omega_{n=1}$ resonance. Quantities are averaged in $r/a$ and $z$.}
\label{figure17bis}
\end{figure*}

Results from the previous sections show that the zonal structure generation at rational surfaces in the setup with fast ions is due to turbulence self-interaction. In this section, the role fast particles play in the development of the zonal mode through this mechanism is investigated. This is done by analyzing the dependence of the turbulence stabilization at $q=1$ on the fast ions characteristics. To this scope, we perform simulations in different fast ions setups. A simulation is performed by removing the nonlinear term from the Vlasov equation for the suprathermal species only. This allows to evaluate the contribution fast ions give to the zonal structure generation through nonlinear physics (e.g. by Reynolds stress generation). Furthermore, runs with flat fast ions density profile (i.e. $\omega_{n_f}=0$) and halved fast ions nominal density ($n_{0f}$ changed to $n_{0f}/2$ in Eq.(\ref{Eq1})) are performed. Results from these simulations are reported in figure \ref{figure12-13-14}. In order to lower computational costs, all these runs are performed with $n_{min}=3$. As discussed in section \ref{subsec:n_min_5}, this choice does not influence the zonal structure characteristics at $q=1$. From figure \ref{figure12-13-14} (a), we see that neither the density profile steepness (quantified by $\omega_{n_f}$) or the nonlinear fast particles term influence the turbulence suppression. Therefore, fast ions do not enter directly the zonal structure generation mechanism. Moreover, halving the fast ions density leads to a complete disappearance of the stabilizing effect on the ion channel. We conclude that a threshold in fast ions density exists in order for the zonal structure to develop. Starting from these conclusions, the role fast ions play in the zonal structure generation is explained by the dilution mechanism introduced in \cite{hahm} and \cite{choi}. This theory treats fast ions as a completely passive species in respect to the drift-wave turbulence. Therefore, the equation for the zonal potential $\langle\phi\rangle$ evolution is modified by the inclusion of a dilution factor $(1-f)$. Here, $f\equiv Z_fn_{0f}/n_{0e}$ is the ratio between fast ions and electron density. It follows that
the nonlinear threshold for the destabilization of the zonal mode decreases with increasing $f$, leading to a more effective zonal mode development when FI are present. Before proceeding, some remarks are needed. The theory presented in \cite{hahm} relies on the modulational instability formalism to describe the zonal flow generation by turbulence. Two regimes are considered: one where $k_y^2\rho_{s}^2\ll k_y^2\rho_{T_f}^2\ll 1$ and one where $k_y^2\rho_{s}^2\ll 1\ll k_y^2\rho_{T_f}^2$. The second one holds for high-$T_f$ setups considered here, with $k_y^2\rho_{T_f}^2\approx 1$ for $T_f=10$. Conversely, \cite{choi} describes the same dilution effect within a more general treatment of the drift wave + zonal mode system, which would be suitable for the case analyzed here. However, the limit considered in the work for the fast ions orbit $k_y^2\rho_{T_f}^2\ll 1$ does not hold for the cases we are studying. Results in \cite{hahm} show the same dependence of the zonal mode growth rate on $f$ in both limits when $\omega_{n_e}/\omega_{n_i}>1$ and the drift wave amplitude is low enough. Relying then on the fact that the same conclusions for the zonal mode development hold in the $k_y^2\rho_{s}^2\ll 1\ll k_y^2\rho_{T_f}^2$ limit, we decide to follow the description presented in \cite{choi}. Here, the zonal mode nonlinear growth rate $\Gamma$ is computed by coupling the modified evolution equation with the wave-kinetic equation \cite{diamond} for turbulence. One obtains
\begin{equation} \label{Eq10}
    \Gamma\approx\sqrt{\gamma_{mod}^2-(\lambda v_{gx})^2}.
\end{equation}
Here, 
\begin{equation} \label{Eq11} 
    \gamma_{mod}^2 \approx -(1-f)\lambda^2\sum_\mathbf{k}\frac{k_xk_y^2\eta_n\omega_*}{\bigl(1+(1-f)k_\perp^2\bigl)^2}\partial_{k_x}\bar{N}   
\end{equation}
and 
\begin{equation} \label{Eq12} 
    v_{gx} = -(1-f)^2\frac{2k_x\eta_n\omega_*}{\bigl(1+(1-f)k_\perp^2\bigl)^2}  
\end{equation}
quantify the turbulent drive of the zonal mode and the drift-wave group velocity, respectively. In Eq.(\ref{Eq11})-(\ref{Eq12}), $\lambda$ is the zonal mode wave vector, $\omega_*=k_y\rho_sc_s/L_{n_e}$ the diamagnetic frequency, $\eta_n=L_{n_e}/L_{n_i}$ and $k_\perp^2=k_x^2+k_y^2$. Moreover, $\bar{N}$ is the mean drift-wave wave action density. From Eq.(\ref{Eq10}) we see that the threshold for the zonal mode destabilization ($\Gamma>0$) lowers as the fast particles fraction increases, since the argument of the square root scales as $1-(1-f)^2$. This translates into a fast ions-induced zonal flow enhancement. Furthermore, \cite{choi} presents a modification of the predator-prey model of the zonal mode-regulated turbulence dynamics which takes into account these fast ions dilution effects. The drift-wave energy $\varepsilon_{NZ}$ at the saturation is computed, and the scaling $\varepsilon_{NZ}\sim(1-f)^3$ is found. These results describe an enhancement in the zonal flow level, leading to a strong reduction of the turbulence energy, when fast ions are included. This fits the picture drawn so far by the results of our simulations. \\
The results obtained for halved $n_f$ and reported in figure \ref{figure12-13-14} confirm our description. Indeed, only high-enough values of $f$ (i.e. of $n_f$) can lead to a relevant destabilization threshold reduction for the zonal mode. Thus, when $n_f$ is halved, the enhancing effect is too small to generate a zonal structure able to effectively reduce turbulence levels. We further explore the relation between fast ions density and zonal mode enhancement by comparing the linear growth phase of the zonal potential for different $n_f$ values. In particular, the reference setup without fast particles and the one with $T_f=40$ and $n_{min}=3$, with both nominal and halved $n_f$, are compared. Results are reported in figure \ref{figure16}. Figure \ref{figure16} (a) shows the time traces of the $n=0$ potential in the considered setups, along with the results from the linear regression performed in the linear phase. By using the linear growth rate $\gamma_{lin}$ of the zonal mode as a proxy for its destabilization efficiency, from figure \ref{figure16} (b) we see that the increase of $n_f$ leads to an enhancement of the mode development. The hypothesis of fast ions dilution enhancement of the zonal flow is now verified with nonlinear runs for $T_f=10$ and $T_f=40$ setting fast ions as a dilution species. Results for the ion heat flux profiles are reported in figure \ref{figure15} (a). We see that by excluding the fast ions dynamics from the field equations we recover the strong stabilization obtained for the high-$T_f$ cases.\\
It is worth noticing that the computations reported in \cite{choi} are electrostatic, while all the results described so far are obtained with $\beta_e\neq0$. In order to check the validity of our arguments, a simulation with $\beta_e=0$ is performed in the reference $T_f=40$ setup. Results are reported in figure \ref{figure15} (b), compared with the reference scenario and with the one without fast particles. We see that the fast ions-enhanced turbulence suppression at $q=1$ appears also in this case. However, the effect seems reduced if compared with the flux reduction obtained in the reference $T_f=40$ scenario. This is linked to the generation of an $\gamma_{E\times B}$ layer whose amplitude is $\sim30$\% smaller than the finite-$\beta_e$ case one, as shown in figure \ref{figure15} (c). Electromagnetic descriptions of (residual) zonal flows exist in the literature \cite{catto}. They show an increase in zonal mode amplitudes as electromagnetic effects are taken into account. However, an extension of the results presented in \cite{choi} to an electromagnetic scenario is far from the purposes of this work. \\
Finally, a comprehensive picture of the fast ions-induced turbulence suppression at rational surfaces developed so far can be drawn. In the analyzed setup $E\times B$ shearing structures are generated by self-interaction of ion temperature gradient-driven turbulence. This self-interaction is maximized at lower integer rational surfaces due to the shorter length eddies have to span before "biting their own tail". Potentially, eddies can reconnect in every setup in which turbulence levels are high enough in correspondence of rational surface. However, the $n=0$ mode generation threshold is lowered by the presence of fast ions through a dilution mechanism when $n_f$ and $T_f$ are high enough. Therefore, only plasmas with fast particles develop a zonal structure around the $q=1$ surface. The presence of this structure leads to the suppression of turbulence at this position. Such a stabilizing effect competes with the destabilizing one given by the quasi-resonant interaction between fast particles and drift-wave turbulence. The description holds in the limit $1\ll k_y^2\rho_{T_f}^2$. This results in a progressive loss of the favorable effect of the zonal flow enhancement as $T_f$ values are decreased. Therefore, this leads to a complete disappearance of the turbulence suppression for $T_f=10$, where $k_y^2\rho_{T_f}^2\approx 1$. 

\vspace{-0.5pc}
\section{Zonal structure generation by the fishbone mode} \label{sec:FB}
\vspace{-0.5pc}

\vspace{-0.5pc}
\subsection{Linear characterization of the fishbone mode} \label{subsec:FB_lin}
\vspace{-0.5pc}

\begin{figure*}[ht!]
\noindent\begin{subfigure}[t]{0.33\textwidth}
\begin{center}
\includegraphics[width=\textwidth]{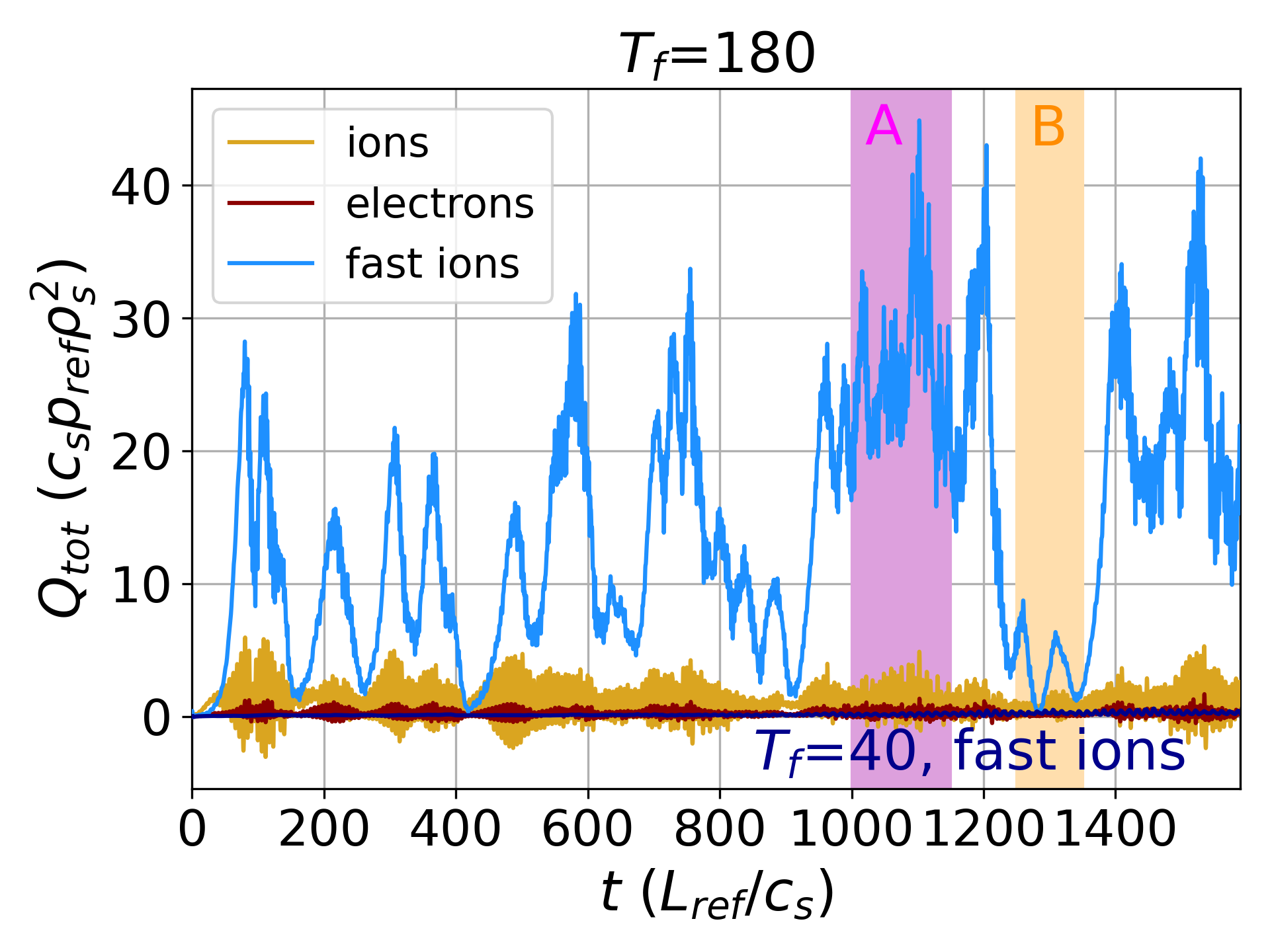}
\caption{}
\end{center}
\end{subfigure}
\noindent\begin{subfigure}[t]{0.33\textwidth}
\begin{center}
\includegraphics[width=\textwidth]{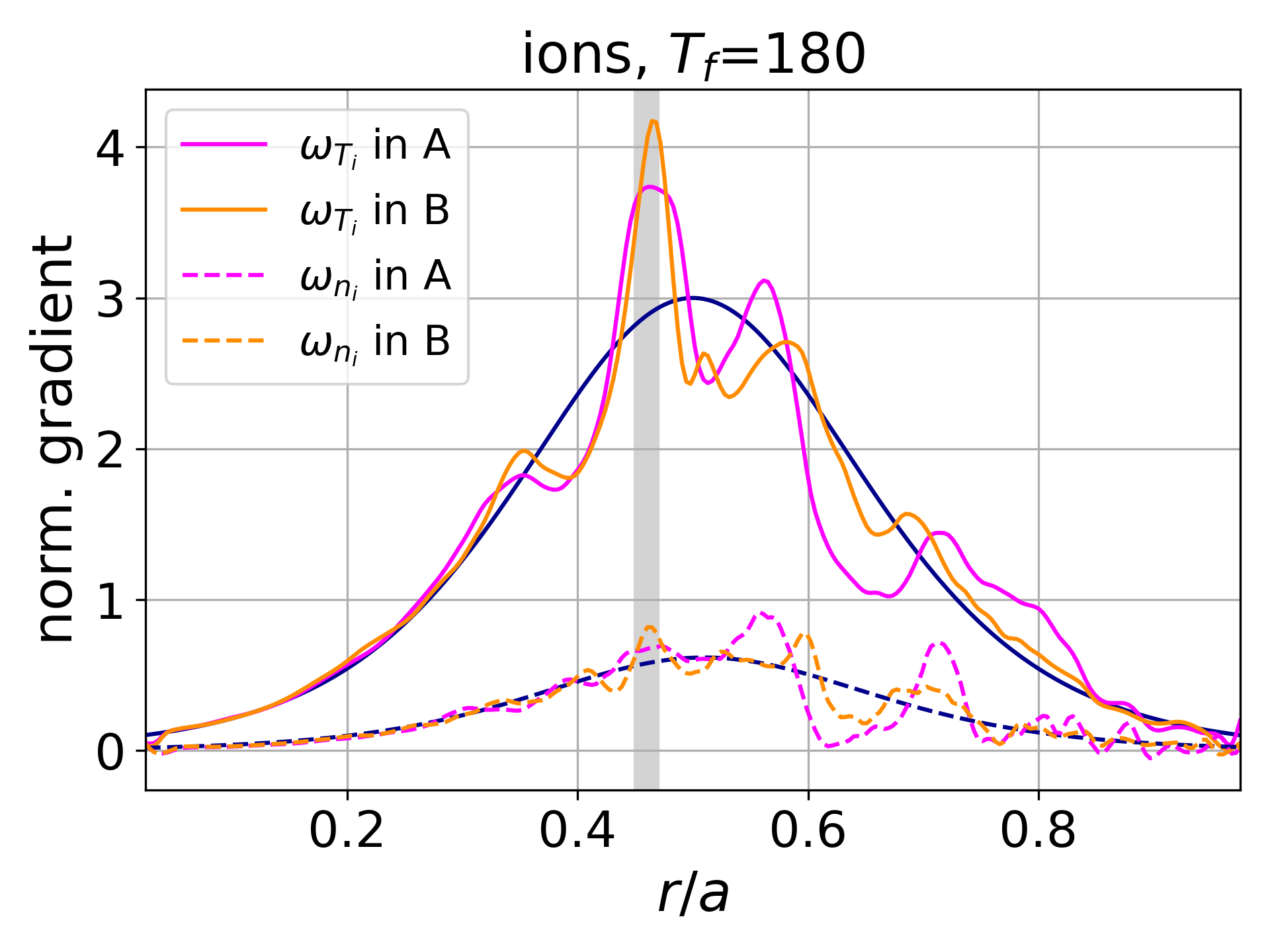}
\caption{}
\end{center}
\end{subfigure}
\noindent\begin{subfigure}[t]{0.33\textwidth}
\begin{center}
\includegraphics[width=\textwidth]{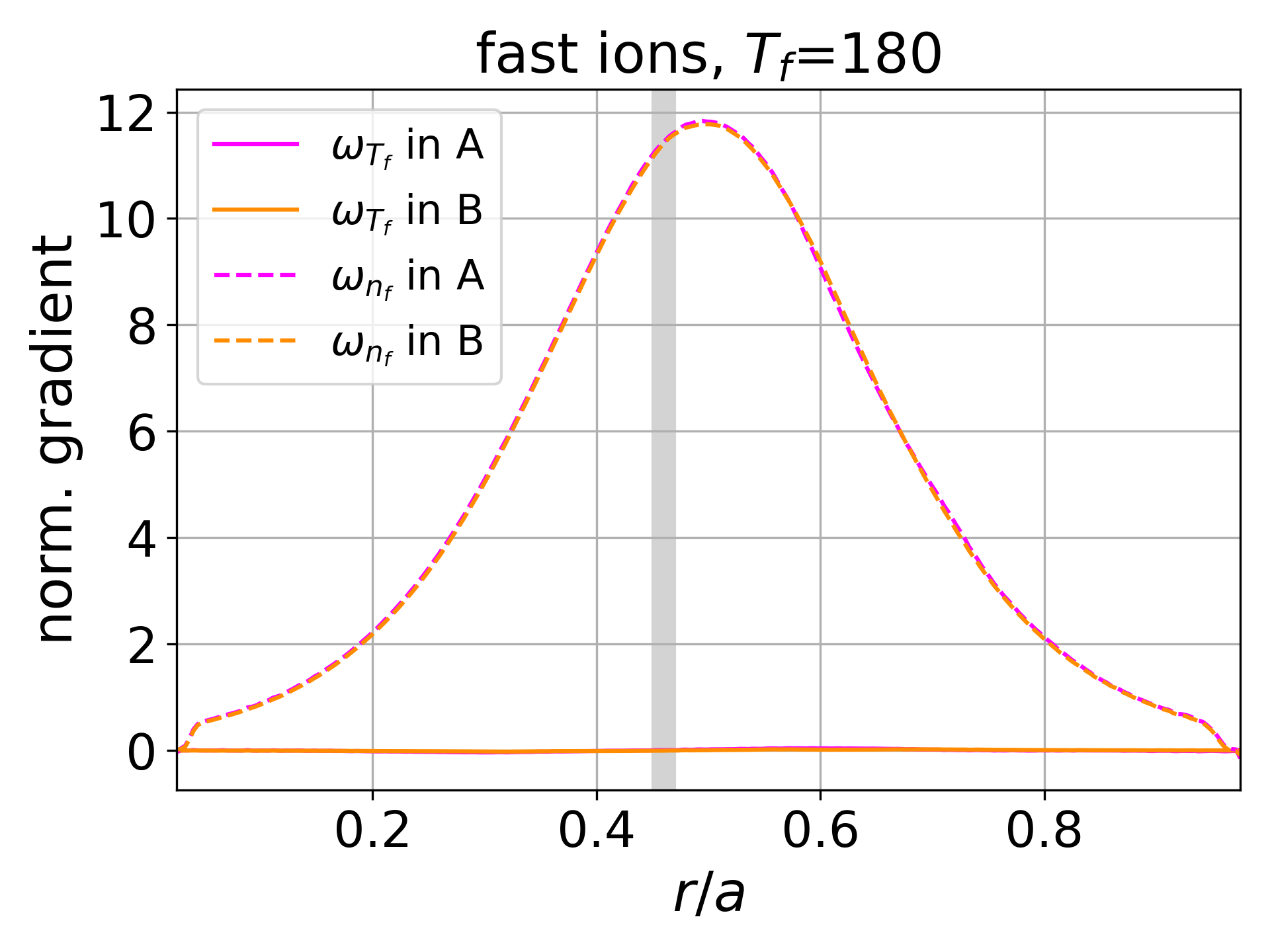}
\caption{}
\end{center}
\end{subfigure}
\caption{(a) Time traces of the total turbulent heat flux for $T_f=180$. Time trace for the fast ions heat flux in the $T_f=40$ setup is reported for comparison (dark blue line). (b) Ion temperature and density gradients (solid and dashed lines, respectively) for the two time intervals A and B (magenta and orange lines, respectively) compared with the initial ones (blue lines).(c) Same profiles for the FI species.}
\label{figure18}
\end{figure*}

\begin{figure*}[ht!]
\noindent\begin{subfigure}[t]{0.33\textwidth}
\begin{center}
\includegraphics[width=\textwidth]{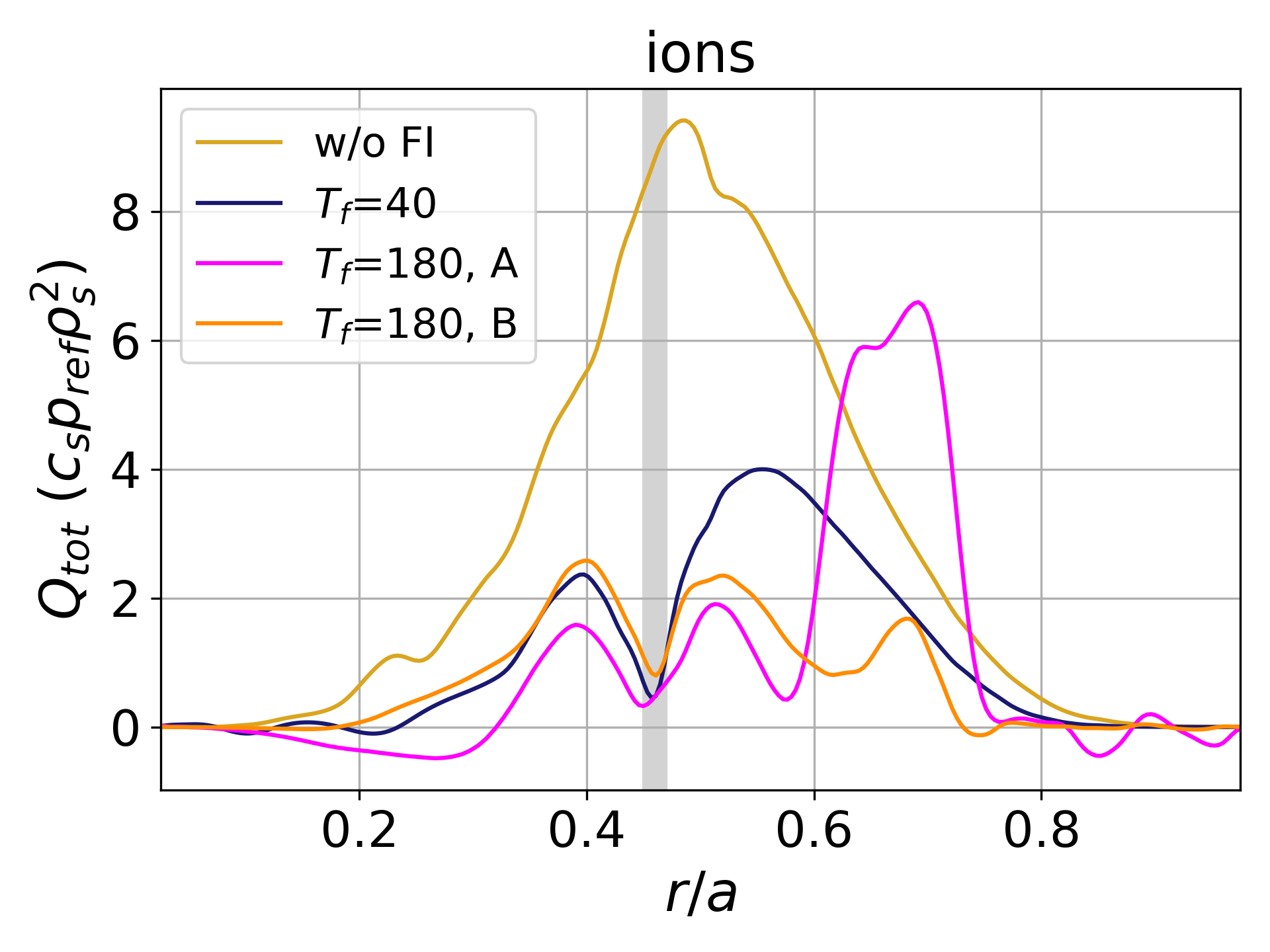}
\caption{}
\end{center}
\end{subfigure}
\noindent\begin{subfigure}[t]{0.33\textwidth}
\begin{center}
\includegraphics[width=\textwidth]{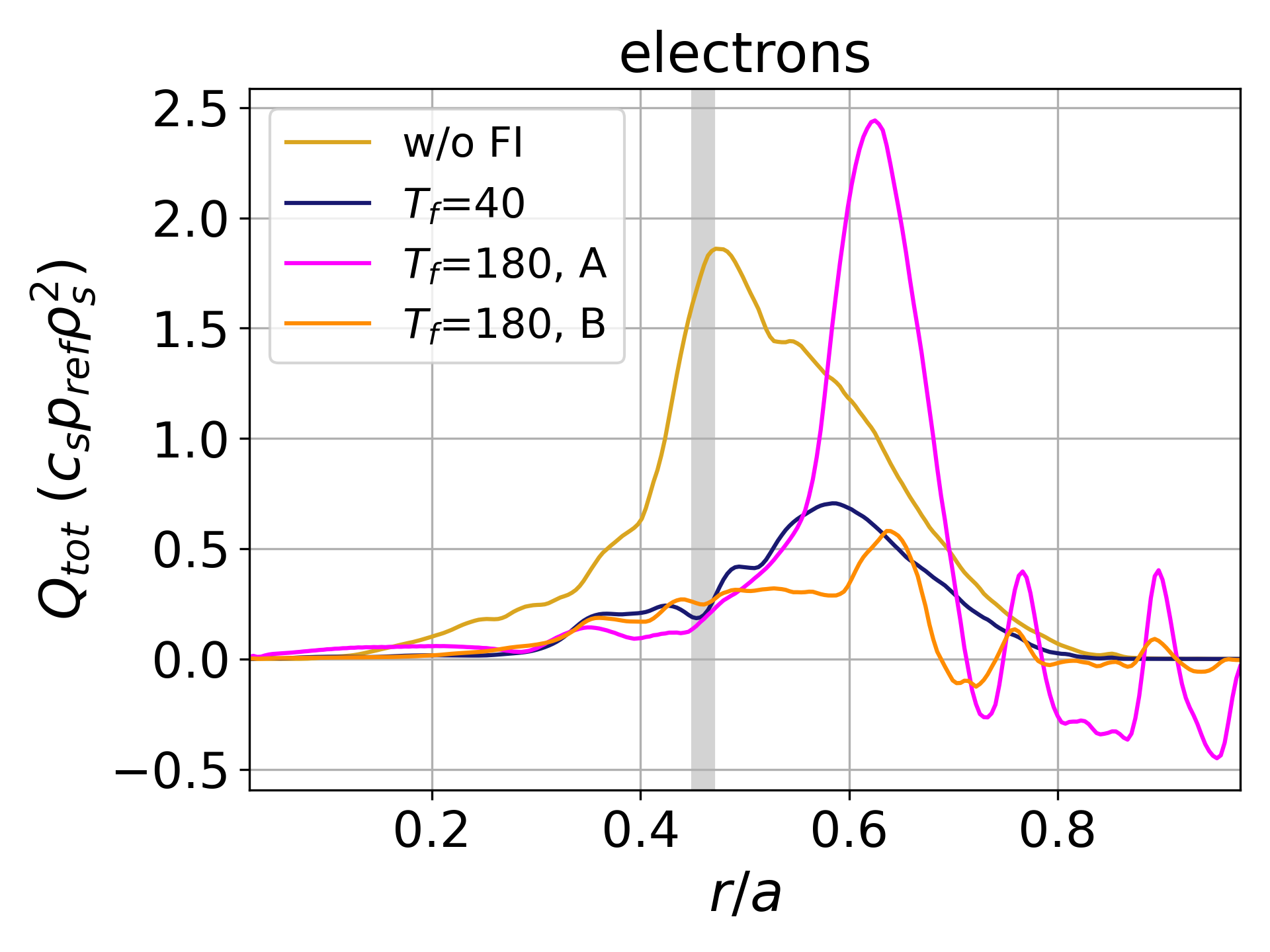}
\caption{}
\end{center}
\end{subfigure}
\noindent\begin{subfigure}[t]{0.33\textwidth}
\begin{center}
\includegraphics[width=\textwidth]{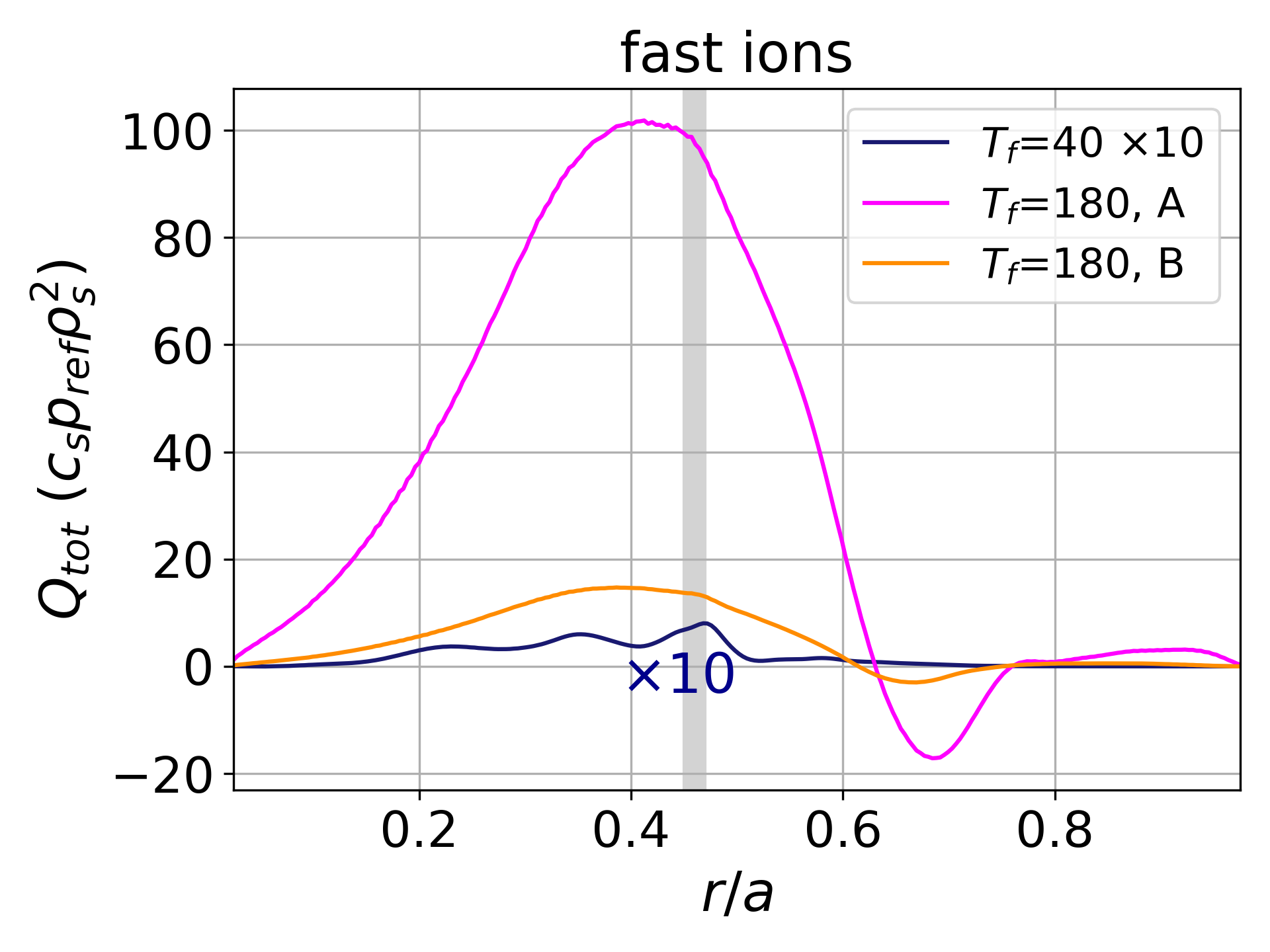}
\caption{}
\end{center}
\end{subfigure}
\caption{Turbulent total heat fluxes for (a) ions, (b) electrons and (c) fast ions species for different setups. The cases without fast particles, with $T_f=40$ and with $T_f=180$ averaged in A and in B are compared. Fast ions heat flux for $T_f=40$ is multiplied by a factor 10 to facilitate visualization.}
\label{figure19}
\end{figure*}

\begin{figure*}[ht!]
\noindent\begin{subfigure}[t]{0.33\textwidth}
\begin{center}
\includegraphics[width=\textwidth]{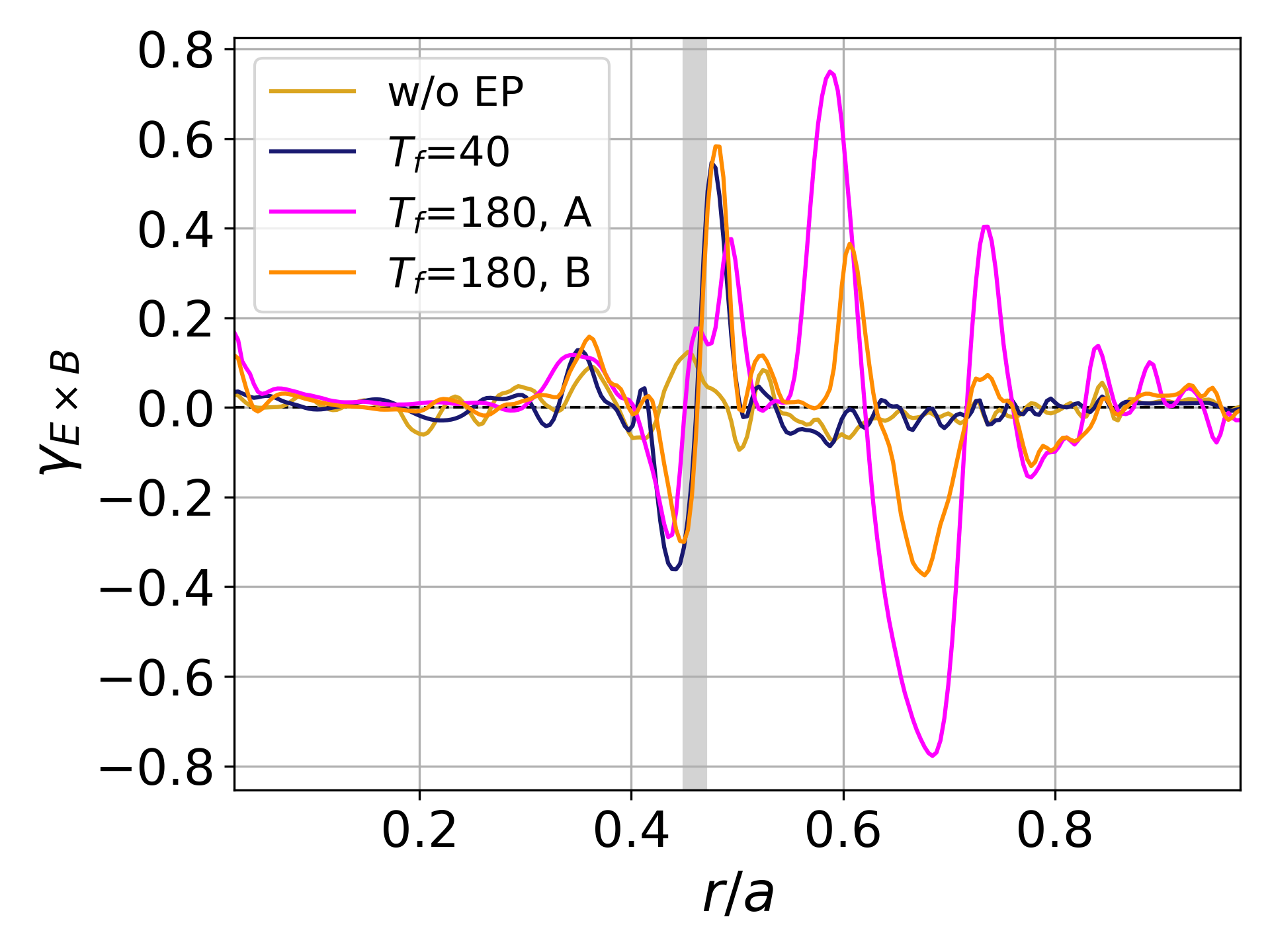}
\end{center}
\end{subfigure}
\caption{Shearing rate $\gamma_{E\times B}$ radial profile. The cases without fast particles, with $T_f=40$ and with $T_f=180$ averaged in A and in B are compared}
\label{figure20}
\end{figure*}

\begin{figure*}[ht!]
\noindent\begin{subfigure}[t]{0.33\textwidth}
\begin{center}
\includegraphics[width=\textwidth]{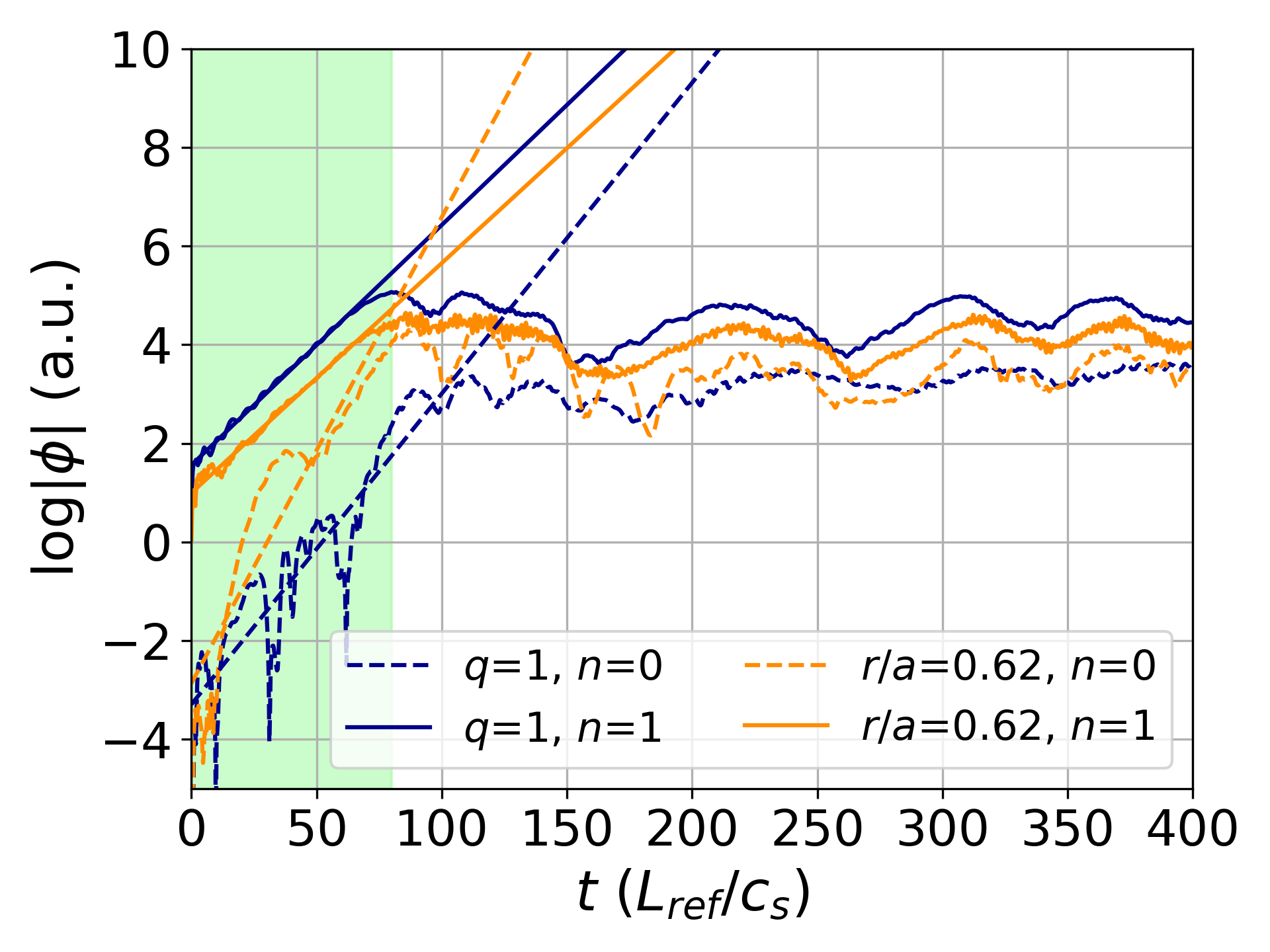}
\end{center}
\end{subfigure}
\caption{Time traces of $\textrm{log}|\phi|_{n=0}$, $\textrm{log}|\phi|_{n=1}$ at $r/a=0.47$ ($q=1$) and $r/a=0.62$. The shaded region represents the linear growths phase where the profiles are fitted with a linear regression.}
\label{figure21}
\end{figure*}

In this section, fast ions temperature is raised to $T_f=180$. We will show that this leads to the linear destabilization of an $n=m=1$ mode peaking around the $q=1$ rational surface. Moreover, nonlinear simulations including this mode will develop a new shearing layer responsible for a further suppression of turbulent fluxes. Notice that the $T_f/T_e$ ratio we retain in this setup is much higher than the one usually achieved in modern-day experiments with fast particles from NBI heating. Figure \ref{figure17} (a) and (b) show the growth rates and frequencies obtained via linear, global runs with $T_f=180$ compared with the setup without fast ions and the one with $T_f=40$. Concerning the ITG branch ($n>5$), no differences are observed when increasing $T_f$. This is expected, since from section \ref{sec:linear} we know the stabilizing effect induced by fast particles far enough from the $T_f\approx 2$ resonance is independent on their temperature. Focusing on lower toroidal mode numbers, we observe that the $n=1$, 2, 4 and 5 modes are driven unstable when the fast particle temperature is increased. Comparing their frequencies with the ones corresponding to the ITG branch, we see that they can be identified as EP-driven modes. Furthermore, the $n=1$ mode is characterized by a growth rate at least three times higher than the ones computed for the other low-$n$ modes. In order to investigate the nature of this mode, in figure \ref{figure17} (c) we plot the different $m$ contributions to the $n=1$ potential radial structure. The $m=1$ one is the dominant contribution to the envelope, presenting a very broad structure peaking around the $q=1$ surface. Moreover, by computing the beta-induced Alfven eigenmode (BAE) typical frequency \cite{lauber} for the setup, we obtain in normalized units $\omega_{BAE}=q\omega_{th,i}\sqrt{7/4+T_e/T_i}\approx 1.48$ ($\sim$73 kHz), being $\omega_{th,i}$ the thermal ion passing frequency. Thus, the $n=1$ mode frequency $\omega_{n=1}=1.05$ ($\sim$52 kHz) is comparable to the BAE one in this regime. According to this observations, we identify the mode as an high-frequency fishbone of the same nature as the one studied in \cite{zonca_FB}. Therefore, we expect the mode to oscillate in-phase with the fast particles precessional periodic motion. In order to verify this in our setup, we compare the real frequency of the $n=1$ mode with the fast ions precessional frequency. By looking at the trapping cone at $z=0$ shown in figure \ref{figure17bis}, we see that the passing particles are the main contributors to the fast ions physics in this setup. The precessional frequency for these particles is computed at position $r_0$ as \cite{nguyen}
\begin{equation} \label{Eq5}
    \omega_{d,f}=\dfrac{q(r_0)\lambda E}{ZeB_0r_0R}\Biggl(2y^2\biggl(\dfrac{\mathbb{E}(1/y^2)}{\mathbb{K}(1/y^2)}-1\biggl)+1+4\hat{s}(r_0)y^2\dfrac{\mathbb{E}(1/y^2)}{\mathbb{K}(1/y^2)}\Biggl).
\end{equation}
In the equation, $\lambda=\mu B_0/E$ is the pitch angle, $E=mv_\parallel^2/2+\mu B_0$ the particle energy, $y^2=(1-\lambda+\epsilon\lambda)/2\epsilon\lambda$ and $\mathbb{K}$, $\mathbb{E}$ elliptic integrals of the first and second kind, respectively. The relation $\omega_{n=1}=\omega_{d,f}$ is verified to hold in our setup by plotting its solutions in the velocity space. This is done in figure \ref{figure17bis}. We see that the maxima of the fast particles distribution function correspond to the fulfillment of the resonance condition described above. The hypothesis on the nature of the $n=1$ mode is then confirmed. 

\vspace{-0.5pc}
\subsection{Turbulent transport modifications and zonal mode beat-driven destabilization} \label{subsec:FB_nl}
\vspace{-0.5pc}

We now perform an analysis of the interaction between the turbulence described in section \ref{sec:transp_supp} and the spectrum of the low-$n$ EP-driven modes dominated by the $n=1$ one. In figure \ref{figure18} (a) we report the time traces for the total turbulent heat fluxes for the three plasma species during a nonlinear run in the $T_f=180$ setup. Turbulent fast particle fluxes are much higher than the thermal ones and present an oscillatory behavior, alternating between phases with higher and lower $Q_{tot}$ values. Two representatives for these phases are selected: $t\in[1000,1150]$ (labeled with "A") and $t\in[1250,1350]$ (labeled with "B") for "high" and "low" fluxes, respectively. This oscillation in the fast ions fluxes can be linked to a modification of the thermal species profiles, as shown in figure \ref{figure18} (b). Here, radial profiles of the ion temperature (solid line) and density (dashed) gradient at the beginning of the simulation (blue lines) are compared to the same quantities time-averaged in A (magenta) and B (orange). A strong flattening of the temperature and density profiles can be observed at $q=1$ and around $r/a=0.62$. For the second position, the feature is more pronounced for phase A. Therefore, higher fast particles fluxes are associated to a stronger flattening of the thermal species profiles. No drastic differences between A and B are observed for what concerns the fast ions profiles, as reported in figure \ref{figure18} (c). This is even more evident when we compare the turbulent fluxes radial profiles by distinguishing between phases A and B. Figure \ref{figure19} reports this quantity for all the species, comparing the setup without fast ions with the nominal one with FI and the one with $T_f=180$. Focusing on the ion species at position $q=1$, we see that the drastic flux reduction studied in section \ref{sec:transp_supp} manifests in the same fashion also in the nonlinear run where the $n=1$ mode is driven unstable. This is expected, since we know this turbulence suppression effect only depends on the presence of fast particles with temperature high enough to satisfy the limit $k_y^2\rho_s^2\ll 1 \ll k_y^2\rho_{T_f}^2$. A further reduction of the heat flux is observed between $r/a=0.5$ and 0.6, approximately, when moving from $T_f=40$ to $T_f=180$. This feature persists during phase B also inside $r/a\approx 0.75$, while in phase A a high contribution to $Q_{tot}$ appears in $0.6\lesssim r/a\lesssim 0.75$. The observations hold also for the electronic fluxes. When compared to the ones obtained for $T_f=40$, fast particle fluxes increase by one and two orders of magnitude during phase B and phase A, respectively. This difference between phases A and B can be simply quantified by evaluating the radial average of the ion total heat fluxes, $\bar{Q}_i$. For the four cases compared in figure \ref{figure19}, we have $\bar{Q}_i$(w/o FI)=2.76, $\bar{Q}_i$($T_f=40$)=1.07, $\bar{Q}_i$($T_f=180$, A)=0.97 and $\bar{Q}_i$($T_f=180$, B)=0.72. To better understand the nature of this further stabilization, along with the detrimental flux increase during phase A, in figure \ref{figure19} we plot the radial profile of the shearing rate. As anticipated, the zonal structure developing around $q=1$ has the same characteristics reported for the one which is driven unstable for $T_f=40$. Moreover, a new shearing layer develops around $r/a=0.62$ for $T_f=180$. This structure is responsible for the strong turbulent fluxes reduction observed in this position when considering phase B (orange lines in figure \ref{figure19}). However, the higher shearing rate detected in phase A is found to be detrimental, leading to the extreme profile relaxation in figure \ref{figure18} (b), which on its turn leads to the drastic increase of the thermal fluxes outside $r/a=0.62$. \\
The relation between the development of the zonal structure around $r/a=0.62$ and the $n=1$ mode is investigated by comparing their growths in the turbulence linear phase. Time traces for the $n=0$ and $n=1$ components of the electrostatic potential are studied for the $T_f=180$ case at $q=1$ and $r/a=0.62$ in figure \ref{figure21}. By interpolating with a linear function the time traces in the green region, we compute the growth rates for the modes to be $\gamma_{n=0}=0.63\times 10^{-1}$, $\gamma_{n=1}=0.49\times 10^{-1}$ around $q=1$ and $\gamma_{n=0}=0.94\times 10^{-1}$, $\gamma_{n=1}=0.47\times 10^{-1}$ around $r/a=0.62$. We have that $\gamma_{n=0}=2\gamma_{n=1}$ holds where the new shearing layer develops. Therefore, the $n=m=1$ mode drives the zonal structure through a beat-drive-like mechanism which is typical of fast ion-driven modes \cite{chen_zonca_2,qiu,disiena_1}. The relation does not hold at $q=1$, where the mechanism underneath the generation of the zonal mode is of a different nature. \\
In this section we have detailed an important aspect of the interplay between rational surfaces and fast ions. We have shown that, when a fishbone mode is driven unstable by the presence of energetic particles, the zonal structure it generates through beat-drive contributes to the turbulence suppression. It is important to point out that the effect is beneficial provided that it does not drive excessive thermal profile flattening. This can by itself lead to strong outward turbulent fluxes (as shown in figure \ref{figure19} (a)), detrimental for the plasma confinement.

\vspace{-0.5pc}
\section{CONCLUSIONS} \label{sec:conclusions}
\vspace{-0.5pc}

In this work, we have studied the influence fast ions with a vast range of $T_f/T_e$ values have on turbulence development and saturation at rational surfaces. This has been done through gyrokinetic simulations with the GENE code. Plasma profiles have been designed in order to linearly destabilize only drift-wave modes even when fast particles with high $T_f/T_e$ are included. The $n=1$ fishbone has been subsequently driven unstable by drastically increasing fast ions temperature. The main linear instabilities have been characterized, and we have identified a first stabilizing effect of fast particles via dilution of the thermal plasma. Moreover, a destabilizing quasi-resonant effect of the fast ions population has been diagnosed. This mechanism competes with the dilution when $T_f$ values are close to the one which satisfies the resonant condition between drift-wave and suprathermal minority. Nonlinear, global simulations show the most prominent feature of the setup. This consists in a pronounced reduction of ion turbulent fluxes at the $q=1$ rational surface when fast particles with an high enough temperature are included. By looking at the $\gamma_{E\times B}$ shearing rate radial profile, we see that this marked stabilization is accompanied by the development of a large shearing structure at the same position. Furthermore, the amplitude of this layer grows with $T_f$. This feature has been investigated by shifting the safety factor profile, thus the rational surface where the zonal structure develops. We have shown that the shearing layer generation at the rational surface is due to a mechanism of self-interaction of the turbulent eddies, which gets maximized at low order rational surfaces. This has been clearly demonstrated by analyzing the mode interaction with itself along the magnetic field line studying the correlation function of the electrostatic potential. The mechanism of self-interaction has been analyzed in a simplified setup with $n_{min}=5$ showing the same turbulence suppression as the nominal one. We have demonstrated that in this setup the $n=0$ mode is driven only by the contribution given by $n=10$ at the $q=1$ surface. The role fast ions play in enhancing the zonal structure has been explained in terms of dilution effect in the paradigm describing the drift-wave + zonal flow system. This description provides a zonal flow destabilization threshold reduction when fast particles are present inside the system. We have verified this mechanism to be of fundamental importance by setting fast ions as a dilution species inside the code. The effect clearly competes with the linear quasi-resonant destabilization to modify the turbulence level around the rational surface where the shearing layer is produced. Therefore, this explains the increasing $\gamma_{E\times B}$ amplitude (and decreasing fast particles heat fluxes) observed with increasing $T_f$ values. By raising the fast ions temperature to $T_f=180$, an $n=1$ BAE-frequency fishbone has been driven unstable in the plasma around $q=1$. When the mode is present, the nonlinear simulation oscillates between a phase with high outwards fast particles fluxes and one with lower ones. This is linked to the beat-drive by the fishbone of a zonal structure which in the "low flux" phase reduces the turbulent fluxes without impacting drastically on the thermal profiles. However, this shearing layer amplitude is larger in the "high flux" phase and strongly flattens the profiles, consequently increasing the thermal fluxes. The fishbone presence is then revealed to be beneficial for the turbulence suppression only when it does not affect the redistribution of thermal species. \\
In conclusion, this paper has presented a numerical study of two effects fast particles induce at rational surface. Firstly, the fast ions dilution effect on zonal modes development has been studied. The considered shearing structure the fast particles act upon has been found to be generated by turbulence self interaction. Only recently such mechanism has become object of intensive study by the community. These observations altogether form a novel picture in which fast particles enhance the generation of favorable shearing structures. The development of this structures is granted by the presence of the $q=1$ surface in correspondence to the region in which the turbulence drive peaks. The drive of a fishbone mode at $q=1$ due to the presence of energetic particles has been shown to further reduce turbulent fluxes. This interplay between fast ions and rational surfaces shows that these are fundamental ingredients for turbulence suppression and development of high confinement scenarios in tokamaks.  

\vspace{-0.5pc}
\section*{ACKNOWLEDGEMENTS}
\vspace{-0.5pc}

The authors would like to thank X. Wang for the useful discussion and advices. The simulations presented in this work have been performed at the VIPER HPC from Max Planck Computation and Data Facility (Germany) and at the Leonardo HPC from CINECA consortium (Italy). This work was carried out within the framework of the EUROfusion Consortium, funded by the European Union via the Euratom Research and Training Programme (Grant Agreement No. 101052200--EUROfusion). Views and opinions expressed are, however, those of the authors only and do not necessarily reflect those of the European Union or the European Commission. Neither the European Union nor the European Commission can be held responsible for them.

%\section*{References}

\end{document}